\documentclass[12pt,a4paper]{article}

\usepackage{epsfig,graphicx,bbm}
\usepackage{amsmath,amsfonts,amssymb}
\usepackage{citesort}
\usepackage[footnotesize]{caption}

\newcommand{\unit}{\leavevmode\hbox{\small1\kern-3.6pt\normalsize1}}

\begin{document}

\begin{flushright}
FTUAM 06/11\\
IFT-UAM/CSIC-06-35\\
\vspace*{3mm}
{\today}
\end{flushright}

\vspace*{5mm}
\begin{center}
{\Large \textbf{Impact of $\theta_{13}$ on Lepton Flavour Violating
    processes \\[2mm] within SUSY Seesaw} }

\vspace{0.5cm} 
{\large
S.~Antusch, E.~Arganda, M.~J.~Herrero and A.~M.~Teixeira}\\[0.4cm]

{\textit{Departamento de F\'{\i }sica Te\'{o}rica C-XI 
and }}\\

\vspace*{0.2cm} 
{\textit{Instituto de F\'{\i }sica Te\'{o}rica C-XVI, \\[0.2cm]
Universidad Aut\'{o}noma de Madrid,
Cantoblanco, E-28049 Madrid, Spain}}\\[0pt]

\vspace*{0.3cm} 
\begin{abstract}
We study the impact of neutrino masses and mixings on 
LFV processes within the context of the supersymmetric 
seesaw scenario, where the CMSSM is extended by three 
right-handed (s)neutrinos. A hierarchical spectrum is considered for both
heavy and light neutrinos. We systematically analyse  
the interesting relation between the leptonic mixing angle 
$\theta_{13}$ and LFV muon and tau decays, namely $l_j \to l_i \, \gamma$ and
$l_j \to 3 \, l_i$, and discuss the interplay with the other relevant parameters.
We require compatibility with low energy neutrino data, 
bounds on both LFV decays and charged lepton electric 
dipole moments, and impose a successful baryogenesis via 
thermal leptogenesis. Particular emphasis is given to the 
implications that a future $\theta_{13}$ measurement can 
have on our knowledge of the heavy neutrino sector.
\end{abstract}
\end{center}

\newpage
 
 
\section{Introduction}\label{intro}
The impressive experimental evidence of neutrino
masses~\cite{neutrinodata} has lead to
the first clear signal of physics beyond the standard model (SM).
One of the simplest extensions of the SM that 
allows to naturally explain 
the smallness of the neutrino masses (without excessively tiny Yukawa
couplings) consists in incorporating right-handed
Majorana neutrinos, and imposing a seesaw mechanism for the neutrino
mass generation~\cite{seesaw:I,seesaw:II}. 
The seesaw mechanism offers in
addition the interesting 
possibility of baryogenesis via leptogenesis~\cite{Fukugita:1986hr}.
Within the framework of leptogenesis, 
the observed baryon asymmetry of the Universe (BAU) is explained 
by the out-of-equilibrium decays of the same heavy right-handed
neutrinos
which are responsible for the suppression of the light neutrino masses.
The scale of new physics is naturally introduced
by the heavy right-handed neutrino masses which, for the simplest case of
just one right-handed neutrino, and assuming neutrino Yukawa couplings, $Y_\nu$,
of $\mathcal{O}(1)$, typically lies close to $10^{14}$ GeV.

Supersymmetric (SUSY) extensions of the SM, including three
right-handed neutrino superfields, are well motivated
models which can accommodate a seesaw mechanism,
and at the same time stabilise the hierarchy between the scale of new
physics and the electroweak (EW) scale. 
One of the most striking phenomenological 
implications of SUSY seesaw models
is the prediction of sizable rates for lepton flavour violating (LFV)
processes~\cite{Borzumati:1986qx}.  
Assuming $Y_\nu \approx \mathcal{O}(1)$ and that the scale of
soft-SUSY breaking is of the order (or below) 1 TeV, 
the radiatively induced SUSY corrections 
driven by the neutrino Yukawa couplings 
lead to rates for the LFV observables which are
many orders of magnitude larger than those expected from the SM seesaw.
Even though this holds irrespective of the chosen mass pattern for the
right-handed neutrinos, it has been shown that when compared to the
degenerate case~\cite{Hisano:1995cp,Hisano:1995nq}
the hierarchical scenario leads to larger LFV
rates, which may even be within the reach 
of current experimental 
bounds~\cite{Hisano:1998fj,Buchmuller:1999gd,Casas:2001sr,Lavignac:2001vp,Bi:2001tb,Ellis:2002fe,Fukuyama:2003hn,Masiero:2004js,Fukuyama:2005bh,Petcov:2005jh,Arganda:2005ji,Deppisch:2005rv,Calibbi:2006nq}. 
In this sense, the
$l_j \to l_i\,\gamma$ and $l_j \to 3\,l_i$ ($i \neq j)$ 
lepton decay channels, as well as $\mu - e$ conversion in
heavy nuclei, are among the most interesting
processes~\cite{Kuno:1999jp}. Experimentally, the
most promising decay is the $\mu \to e\, \gamma$ process, which
exhibits the most stringent present bounds, and offers a
significant improvement regarding the future sensitivity. 
Furthermore, in the presence of complex neutrino Yukawa
couplings, one can also construct from the latter LFV decays interesting
observables, which are sensitive to CP violation in the neutrino
sector. For instance, one can build 
T- and P-odd asymmetries in $\mu \to e\,\gamma$ and
$\mu \to 3\,e$ decays, which were addressed
in~\cite{Okada:1999zk,Ellis:2001xt}. 

In addition to the large number of parameters of the minimal
supersymmetric standard model (MSSM), the seesaw mechanism introduces
18 new parameters in the neutrino sector. 
As a first step to simplify the analysis of the LFV
rates in a SUSY seesaw model, we choose to work in the so-called constrained MSSM
(CMSSM), assuming universality of the soft-SUSY breaking parameters at
the scale of gauge coupling unification, $M_X$. 
This allows to reduce the unknown parameters in the SUSY sector 
to the five usual parameters of a minimal supergravity (mSUGRA) framework.   
Moreover, regarding the neutrino sector, we will assume a 
hierarchical spectrum for both light and heavy neutrinos.

Among the various seesaw parameters, those connected to 
the light neutrino sector can
in principle be derived from low-energy neutrino data,
while those associated with the heavy neutrino sector, 
in particular the values of the masses, are \`a priori clearly unreachable.
However, and given the fact that both light and heavy neutrinos enter in the
determination of the LFV rates (via the Yukawa interactions),
a powerful link between the low- and 
high-energy neutrino parameters can be obtained from these LFV processes. 
From the requirement of compatibility with both current LFV bounds and 
low-energy neutrino data, one can then 
extract information on the heavy neutrino sector, thus providing an
indirect access to the heavy neutrino parameters.  
In the presence of
additional CP phases (other than those associated to the light
sector), and assuming that BAU is generated from thermal leptogenesis, one
can obtain a further insight on the heavy neutrino parameters. 
More specifically, 
one can obtain a lower bound on the mass of the lightest right-handed
neutrino, which in turn translates into lower mass bounds for the other
heavy states.

Here we address the subject of how to extract information 
on the unknown SUSY-seesaw
parameters from both the analysis of LFV decays and the requirement 
of successful BAU. As already said, we restrict 
ourselves to the scenario of hierarchical heavy neutrinos, 
which leads to the most interesting predictions. 
There are studies
also addressing the same subject, within similar
SUSY seesaw frameworks and  hierarchical neutrino scenarios. 
In particular, some constraints on the 
heavy neutrino and SUSY sectors have been considered
by~\cite{Ellis:2002fe,Fukuyama:2003hn,Masiero:2004js,Fukuyama:2005bh,Petcov:2005jh,Arganda:2005ji,Deppisch:2005rv,Calibbi:2006nq} and 
further implications 
regarding low-energy CP violating observables were studied
in~\cite{Okada:1999zk,Ellis:2001xt,Ellis:2001yz,Ellis:2002xg}. 
In addition, it has been noticed that not only are 
the LFV branching ratios (BRs) 
sensitive to the heavy neutrino parameters, but there is
also a potentially relevant role being played by 
the yet undetermined low-energy neutrino parameters.
The latter include the
overall light neutrino mass scale and CP violating phases. 
Concretely, the effects of this 
scale and of the two Majorana phases on the radiative LFV 
decays have been analysed in~\cite{Petcov:2006pc}.

An even more intriguing situation occurs regarding the sensitivity of the 
LFV rates to $\theta_{13}$, which is our main interest here. 
Although it is not expected to be of relevance
in the context of the LFV phenomenology, 
the few studies regarding the dependence of LFV rates on 
$\theta_{13}$~\cite{Masiero:2004js,Arganda:2005ji}
have revealed that, for some specific seesaw cases, the 
LFV muon decays indeed exhibit a 
strong sensitivity to this parameter. In particular,
the dependence of the BR($\mu \to e\,\gamma$) on $\theta_{13}$ 
was noticed~\cite{Masiero:2004js} in the context of SUSY grand unified
theories (SUSY-GUTs), and  
in~\cite{Arganda:2005ji} within the CMSSM. In the latter, it was further
emphasised that in addition to $\mu \to e\,\gamma$ other LFV decays 
as $\mu \to 3\,e$ are also very sensitive to $\theta_{13}$. 
Both these 
studies assumed a simple scenario where no additional 
mixing, other than that induced from the low-energy 
mixing angles $\theta_{ij}$, was present in the neutrino sector.

In this work we systematically explore the sensitivity of LFV
processes to $\theta_{13}$ in a broader class of SUSY seesaw
scenarios, with different possibilities for the mixing in the neutrino sector, 
and we incorporate in our analysis the requirement of BAU via
thermal leptogenesis. Specifically, we conduct a comprehensive and comparative
study of the dependence on $\theta_{13}$ in all the following
decay channels: 
$\mu^- \to e^-\, \gamma$, $\mu^- \to e^-\,e^-\,e^+$, 
$\tau^- \to e^-\, \gamma$, $\tau^- \to e^-\,e^-\,e^+$, 
$\tau^- \to \mu^-\, \gamma$ and $\tau^- \to \mu^-\,\mu^-\,\mu^+$. 
We will show here that various of these channels indeed offer 
interesting expectations regarding the sensitivity to  $\theta_{13}$.
In our analysis, we work in the context of a CMSSM
extended by three right-handed neutrinos and their SUSY partners,
and use the requirement of generating a successful BAU in order 
to constrain the explored seesaw parameter space.
Our main motivation to perform the present study has been triggered by 
the potential measurement of $\theta_{13}$, as suggested by 
the experimental program of MINOS and OPERA,
which claim a future experimental sensitive of 
$\theta_{13} \lesssim 8^\circ$~\cite{Ables:MINOS:2004} 
and $\theta_{13} \lesssim 7^\circ$~\cite{Komatsu:2002sz,Migliozzi:2003pw}, respectively.
With the advent of other experiments, like Double Chooz and T2K, 
the expected sensitivity will be further improved to
$\theta_{13} \lesssim 4^\circ$~\cite{Huber:2006vr} 
and $\theta_{13} \lesssim 2^\circ$~\cite{Itow:2001ee}. An ambitious program to
push the sensitivity to less than $1^\circ$ is envisioned by Neutrino
Factory~\cite{Blondel:2006su,Huber:2006wb} and/or Beta
Beam~\cite{Burguet-Castell:2005pa,Campagne:2006yx} facilities.

Our ultimate goal is to explore the impact of a 
potential $\theta_{13}$ measurement on the LFV branching ratios,
which together with the current and future experimental bounds (measurements)
on the latter ratios, could lead to a better knowledge (determination) of 
the heavy neutrino parameters.

Our work is organised as follows. In Section~\ref{model}, we present
the SUSY seesaw scenario, describing the seesaw
mechanism for the neutrino mass generation, and discussing how flavour
mixing in the slepton and sneutrino sectors arises in this context.
We further address the constraints on the seesaw parameters
from the requirement of generating a successful BAU via thermal
leptogenesis, and from imposing compatibility with experimental data
on charged lepton electric dipole moments (EDMs). In Section~\ref{results},
we explore in detail how the several parameters affect the theoretical
predictions for the LFV rates, and whether the former can modify the
sensitivity of a given LFV observable to $\theta_{13}$. 
Section~\ref{prospects:hints} is devoted to the discussion of the
hints on SUSY and seesaw parameters which can be derived 
from a potential measurement of $\theta_{13}$
and LFV branching ratios.
Finally, our conclusions are summarised in Section~\ref{conclusions}.

\section{LFV within the SUSY seesaw model}\label{model}

In what follows, we first present the SUSY seesaw scenario within the CMSSM,
then proceed to describe how LFV processes arise in this
framework and finally discuss the implications regarding BAU and charged
lepton EDMs. 

\subsection{The SUSY seesaw scenario}\label{model:susyseesaw}
The leptonic superpotential containing the relevant terms to describe a 
type-I SUSY seesaw is given by
\begin{equation}\label{W:Hl:def}
W\,=\,\hat N^c\,Y_\nu\,\hat L \, \hat H_2 \,+\,
\hat E^c\,Y_l\,\hat L \, \hat H_1 \,+\,
\frac{1}{2}\,\hat N^c\,m_M\,\hat N^c\,,
\end{equation}
where $\hat N^c$ is the additional superfield that contains the three right-handed 
neutrinos $\nu_{R_i}$ and their scalar partners $\tilde \nu_{R_i}$.
The lepton Yukawa couplings $Y_{l,\nu}$ and the
Majorana mass $m_M$ are $3\times 3$ matrices in lepton flavour
space. From now on, we will assume that we are in a basis where 
$Y_l$ and $m_M$ are diagonal.

After EW symmetry breaking, the charged lepton and 
Dirac neutrino mass matrices
can be written as
\begin{equation}
m_l\,=\,Y_l\,\,v_1\,, \quad \quad
m_D\,=\,Y_\nu\,v_2\,,
\end{equation}
where $v_i$ are the vacuum expectation values (VEVs) of the neutral Higgs
scalars, with $v_{1(2)}= \,v\,\cos (\sin) \beta$ and $v=174$
GeV.

The $6\times 6$ neutrino mass matrix is given by
\begin{equation}\label{seesaw:def}
M^\nu\,=\,\left(
\begin{array}{cc}
0 & m_D^T \\
m_D & m_M
\end{array} \right)\,. 
\end{equation}
The eigenvalues of $M^\nu$ are the masses of the six physical 
Majorana neutrinos. In the seesaw limit, the three right-handed 
masses are much heavier than the EW scale,  
$m_{M_i}\,\gg\,v$, and one obtains three light and three heavy
states, $\nu_i$ and $N_i$, respectively. 

Block-diagonalisation of the neutrino mass matrix of
Eq.~(\ref{seesaw:def}), leads (at lowest order in the $(m_D/m_M)^n$
expansion) to the standard seesaw equation 
for the light neutrino mass matrix,
\begin{equation}\label{seesaw:light}
m_\nu\,=\, - m_D^T m_M^{-1} m_D  \,, 
\end{equation}
as well as the simpler relation for the heavy mass eigenstates,
$m_{N}\,=\,m_M$.
Since we are working in a basis where $m_M$ is diagonal, the heavy
eigenstates are then given by 
\begin{equation}\label{def:Ndiag}
m_N^\text{diag}\,=\,m_N\,=\, \text{diag}\,(m_{N_1},m_{N_2},m_{N_3})\,.
\end{equation}
The matrix $m_\nu$ can be diagonalised by the
Maki-Nakagawa-Sakata unitary matrix
$U_{\text{MNS}}$~\cite{Maki:1962mu,Pontecorvo:1957cp}, leading
to the following masses for the light physical states
\begin{align}\label{physicalmasses}
m_{\nu}^\text{diag}&
\,=\,U_\text{MNS}^T \,m_{\nu}\, U_\text{MNS} 
\,=\, \text{diag}\,(m_{\nu_1},m_{\nu_2},m_{\nu_3})\,.
\end{align}
Here we use the standard parameterisation for $U_\mathrm{MNS}$ 
given by
\begin{equation}
U_\text{MNS}=
\left( 
\begin{array}{ccc} 
c_{12} \,c_{13} & s_{12} \,c_{13} & s_{13} \, e^{-i \delta} \\ 
-s_{12}\, c_{23}\,-\,c_{12}\,s_{23}\,s_{13}\,e^{i \delta} 
& c_{12} \,c_{23}\,-\,s_{12}\,s_{23}\,s_{13}\,e^{i \delta} 
& s_{23}\,c_{13} \\ 
s_{12}\, s_{23}\,-\,c_{12}\,c_{23}\,s_{13}\,e^{i \delta} 
& -c_{12}\, s_{23}\,-\,s_{12}\,c_{23}\,s_{13}\,e^{i \delta} 
& c_{23}\,c_{13}
\end{array} \right) \,.\, V\,,
\label{Umns}
\end{equation}
with
\begin{equation}
 V\,=\,\text{diag}\,(e^{-i\frac{\phi_1}{2}},e^{-i\frac{\phi_2}{2}},1)\,,
\end{equation}
and $c_{ij} \equiv \cos \theta_{ij}$, $s_{ij} \equiv \sin \theta_{ij}$.
$\theta_{ij}$ are the neutrino flavour mixing angles, $\delta$ is the Dirac
phase and $\phi_{1,2}$ are the Majorana phases. 

In view of the above, 
the seesaw equation (\ref{seesaw:light}) can be solved for $m_D$
as~\cite{Casas:2001sr} 
\begin{equation}\label{seesaw:casas}
m_D\,=\, i \sqrt{m^\text{diag}_N}\, R \,
\sqrt{m^\text{diag}_\nu}\,  U^\dagger_{\text{MNS}}\,,
\end{equation}
where $R$ is a generic complex orthogonal $3 \times 3$ matrix that
encodes the possible extra neutrino mixings (associated with the
right-handed sector) in addition to the ones in
$U_{\text{MNS}}$. $R$ can be parameterised 
in terms of three complex angles, $\theta_i$ $(i=1,2,3)$ as
\begin{equation}\label{Rcasas}
R\, =\, 
\left( 
\begin{array}{ccc} 
c_{2}\, c_{3} & -c_{1}\, s_{3}\,-\,s_1\, s_2\, c_3
& s_{1}\, s_3\,-\, c_1\, s_2\, c_3 \\ 
c_{2}\, s_{3} & c_{1}\, c_{3}\,-\,s_{1}\,s_{2}\,s_{3} 
& -s_{1}\,c_{3}\,-\,c_1\, s_2\, s_3 \\ 
s_{2}  & s_{1}\, c_{2} & c_{1}\,c_{2}
\end{array} 
\right)\,,
\end{equation}
with $c_i\equiv \cos \theta_i$, $s_i\equiv \sin\theta_i$. 
Eq.~(\ref{seesaw:casas}) is a convenient means of parameterising our ignorance 
of the full neutrino Yukawa couplings, while at the same time allowing
to accommodate the experimental data.
Notice that it is only valid at the 
right-handed neutrino scales $m_M$, so that the quantities appearing
in Eq.~(\ref{seesaw:casas}) are the renormalised ones, 
$m^\text{diag}_\nu\,(m_M)$ and $U_{\text{MNS}}\,(m_M)$.

We shall focus on the simplest scenario, where both the heavy and 
light neutrinos are hierarchical, and in particular we will 
assume a normal hierarchy,
\begin{align}
&m_{N_1}\, \ll\, m_{N_2}\, \ll \,m_{N_3}\,, \nonumber \\
&m_{\nu_1}\, \ll\, m_{\nu_2}\, \ll\, m_{\nu_3}\,. 
\end{align}  
The masses $m_{\nu_{2,3}}$ 
can be written in terms of the lightest mass $m_{\nu_{1}}$, and of  
the solar and atmospheric mass-squared differences as
\begin{align}
& m_{\nu_2}^2\,=\, \Delta m_\text{sol}^2 \, + \, m_{\nu_1}^2\,, \nonumber\\  
& m_{\nu_3}^2\,=\, \Delta m_\text{atm}^2 \, + \, m_{\nu_1}^2\,.
\end{align}

Moving now to the s-spectrum, we encounter an enlarged slepton sector
due to the inclusion of the 
SUSY partners of $\nu_{R_i}$, namely 
${\tilde \nu}_{R_i}$. The soft-SUSY breaking
Lagrangian for the slepton sector will also include new terms and parameters.
In addition to the 
left- and right-handed soft-breaking masses, $m_{\tilde L}$, 
$m_{\tilde E}$, and trilinear couplings, $A_l$, for the charged sleptons,
one now includes 
soft-breaking sneutrino masses, $m_{\tilde M}$, sneutrino trilinear
couplings $A_\nu$, and the new bilinear parameter, $B_M$.

The universality conditions of the soft-SUSY breaking parameters 
at the high-energy scale $M_X$
(with $M_X \gg m_M$) for the full slepton sector then read as follows:
\begin{align}\label{CMSSM:univcond}
&
\left(m_{\tilde L}\right)^2_{ij}\,=\,
\left(m_{\tilde E}\right)^2_{ij}\,=\,
\left(m_{\tilde M}\right)^2_{ij}\,=\, M_0^2\,\delta_{ij}\,,\,\,
\nonumber \\
& \left(A_l\right)_{ij}\,=\, A_0 \, \left(Y_l \right)_{ij}\,,\,\,
\left(A_\nu \right)_{ij}\,=\, A_0 \, \left(Y_\nu \right)_{ij}\,,
\end{align}
where $M_0$ and $A_0$ are the universal scalar soft-breaking mass and
trilinear coupling of the CMSSM, and $i,j$ denote lepton flavour indices, with
$i,j=1,2,3$. 
Here, we choose $M_X$ to be the gauge coupling unification scale.
The extended CMSSM is further specified by the
universal gaugino mass, $M_{1/2}$, the ratio of the Higgs VEVs, $\tan \beta$, 
and the sign of the bilinear $\mu$-parameter ($\text{sign}\,\mu$). 

The CMSSM predictions for the low-energy parameters are
obtained by solving the full renormalisation
group equations (RGEs), which must now include the appropriate   
equations and extra terms for the extended neutrino and sneutrino 
sectors.
Due to the existence of intermediate scales $m_M$ introduced by the seesaw
mechanism, the running must be carried in two steps.
The full set of equations is first run down from $M_X$ to
$m_M$. At the seesaw scales, 
the right-handed neutrinos as well as their SUSY partners
decouple, and the new RGEs (without the equations and terms for 
$\nu_R$ and $\tilde \nu_R$) are then run down from $m_M$ to the EW scale, where the
couplings and mass matrices are finally computed.

\subsection{Lepton flavour violating decays}\label{model:lfv}

In the present study, and since we work within the CMSSM, 
all LFV originates solely from the neutrino Yukawa couplings.
For the LFV process that we are interested in, 
the flavour mixing in the neutrino sector is transmitted to the charged
lepton sector via radiative corrections involving $Y_\nu$.
These corrections can be important since, due to the Majorana nature of
the neutrinos, the Yukawa couplings may be sizable (as large as
$\mathcal{O}(1)$). 
In particular, we will consider here
the following LFV muon and tau decays:
$\mu \to e \gamma$, $\tau \to \mu \gamma$, $\tau \to e \gamma$, 
$\mu \to 3\,e $, $\tau \to 3 \,e$ and $\tau \to 3\,\mu$.

Under the requirement that at the seesaw scales ($m_M$) $Y_\nu$ satisfies
Eq.~(\ref{seesaw:casas}), 
the running from $M_X$ down to the EW scale will induce flavour mixing 
in the low-energy charged slepton squared mass matrix,
$M_{\tilde{l}}^2$, whose $LL$, $RR$, $LR$ and $RL$ elements are given by 
\begin{align}
M_{LL}^{ij \, 2} & \,=  \,
m_{\tilde{L}, ij}^2 \, + \, v_1^2  \,\left( Y_l^{\dagger} \, Y_l 
\right)_{ij} \, + \, 
m_Z^2  \,\cos 2 \beta \, \left(-\frac{1}{2} \,+ \, \sin^2 \theta_{W}
\right)  \,  \delta_{ij} \,, \nonumber \\
M_{RR}^{ij \, 2} & \,=  \,
m_{\tilde{E}, ij}^2 \, + \, v_1^2 \, \left( Y_l^{\dagger} \, Y_l
\right)_{ij} \, -  \, 
m_Z^2 \, \cos 2 \beta  \,\sin^2 \theta_{W}  \,\delta_{ij} \,, \nonumber \\
M_{LR}^{ij \, 2} & \,=  \,
v_1  \,\left(A_l^{ij}\right)^{*}  \,- \,\mu \, Y_l^{ij} \, v_2 \,, \nonumber \\
M_{RL}^{ij \, 2} & \,=  \,\left(M_{LR}^{ji \, 2}\right)^{*} \, ,
\end{align}
with $m_Z$ the $Z$-boson mass and $\theta_W$ the weak mixing angle.
Below $m_M$, the right-handed sneutrinos decouple, and
the low-energy sneutrino mass eigenstates are dominated by the 
$\tilde \nu_L$ components~\cite{Grossman:1997is}. 
Thus, the sneutrino flavour
mixing is confined to the left-handed sector, and described by the
following $3 \times 3 $ matrix:
\begin{equation}
M_{\tilde{\nu}}^{ij \, 2}
\,=\,
m_{\tilde{L}, ij}^2  + \frac{1}{2}\, m_Z^2 \,\cos 2 \beta \, 
\delta_{ij}\,.
\end{equation} 

The physical masses and states are obtained by diagonalising the
previous mass matrices, leading to
\begin{align}
{M_{\tilde l}^2}^\text{diag} & \,=\, 
R^{l}  \,M_{\tilde l}^2  \,R^{l\,\dagger} \, = \,
\text{diag} \,(m_{\tilde l_1}^2,..,m_{\tilde l_6}^2) \,,
\nonumber \\
{M_{\tilde \nu}^2}^\text{diag}  & \,=  \,
R^{\nu}  \,M_{\tilde \nu}^2  \,R^{\nu\,\dagger} \, = \,
\text{diag}\,(m_{\tilde \nu_1}^2, \,
m_{\tilde \nu_2}^2, \,m_{\tilde \nu_3}^2)\,,
\end{align}
where $R^{l,\nu}$ are unitary rotation matrices.

\vspace*{5mm}
The LFV ratios for the decay processes of our interest are 
obtained here via a full one-loop computation 
(and in terms of physical eigenstates), 
including all relevant SUSY diagrams.

For the radiative decays $l_j \to l_i \, \gamma$ ($i
\neq j$), the branching ratios are given by
\begin{equation}
\text{BR}(l_j \to l_i\, \gamma)\,  = \, \frac{e^2}{16 \, \pi}\, 
\frac{m^5_{l_j}}{\Gamma_{l_j}}\, \left(|A_2^L|^2\, +\, |A_2^R|^2\right)\,, 
\end{equation}
where ${\Gamma_{l_j}}$ is the total lepton width, 
and the form factors $A_2^{L,R}$ receive contributions from two types 
of diagrams, sneutrino-chargino loops and charged slepton-neutralino
loops.
These BRs were computed in
Ref.~\cite{Borzumati:1986qx,Hisano:1995cp}. 
We will use in our analysis the explicit formulae for the 
$|A_2^{L,R}|$ form factors as in Ref.~\cite{Arganda:2005ji}.

Regarding the LFV decays into
three leptons, $l_j \to 3 \, l_i$, the one-loop computation was
presented in~\cite{Hisano:1995cp}, and
later revised and completed in~\cite{Arganda:2005ji}. 
The latter work included the
full set of SUSY one-loop contributing diagrams, namely photon-,
$Z$-, and Higgs-penguins, as well as box diagrams.
As shown by the explicit computation of~\cite{Arganda:2005ji},
the dominant contribution is clearly coming
from the photon-penguin diagrams and, more specifically, from the same form
factors $A_2^{L,R}$ as in the case of the radiative 
decays\footnote{This has also been concluded in a generic, non-seesaw,
MSSM scenario~\cite{Brignole:2004ah}.}.
This is valid even in the case of very large $\tan \beta$ where the Higgs-penguin diagrams,
although enhanced, induce contributions
which are still many orders of magnitude below those associated with the
photon-penguins\footnote{Notice that the Higgs-penguin contribution could only be
  relevant in a generic MSSM 
  framework~\cite{Babu:2002et,Paradisi:2005tk}.}.
Therefore, the BR for the $l_j \to 3 \, l_i$ decays can be 
approximated by the simple expression,
\begin{equation}
 \text{BR}(l_j \to 3 l_i)\,=\,
 \frac{\alpha}{3\,\pi}\,
\left(\log\frac{m_{l_j}^2}{m_{l_i}^2}\,-\,\frac{11}{4}\right)\,
 \times\,
\text{ BR}(l_j \to l_i\, \gamma)\,,
\end{equation}           
where $\alpha$ is the electromagnetic coupling constant.
However, and although we have verified that this is indeed a very good
approximation, we use in the present analysis the full one-loop
formulae for the 
BR($l_j \to 3 l_i$) from Ref.~\cite{Arganda:2005ji}.

Finally, and regarding the estimation of the low-energy parameters, we
consider the full 2-loop RGE
running, except for the neutrino sector which is treated at the
1-loop level. Nevertheless, for the forthcoming discussion,
it will be clarifying and interesting to 
compare the full results
with the simplified estimation which is obtained within the leading 
logarithmic approximation (LLog).
In the latter framework, 
the RGE generated flavour mixing in the slepton
sector is summarised by the following logarithmic contributions,
\begin{align}\label{misalignment_sleptons}
(\Delta m_{\tilde{L}}^2)_{ij}&\,=\,
-\frac{1}{8\, \pi^2}\, (3\, M_0^2+ A_0^2)\, (Y_{\nu}^\dagger\, 
L\, Y_{\nu})_{ij} 
\,,\nonumber \\
(\Delta A_l)_{ij}&\,=\,
- \frac{3}{16 \,\pi^2}\, A_0\, Y_{l_i}\, (Y_{\nu}^\dagger\, L\, Y_{\nu})_{ij}
\,,\nonumber \\
(\Delta m_{\tilde{E}}^2)_{ij}&\,=\,
0\,\,;\, L_{kl}\, \equiv \,\log \left( \frac{M_X}{m_{M_k}}\right) \,
\delta_{kl}\,,
\end{align}
which are originated by the running from $M_X$ to the right handed
mass scales $m_{M_i}$, $i=1,2,3$. 
The matrix elements $( Y_\nu^\dagger L Y_\nu )_{ij}$ in
Eq.~(\ref{misalignment_sleptons}) can be simply written in terms of
the parameterisation of
Eqs.~(\ref{def:Ndiag}-\ref{Rcasas}). In particular, we obtain
\begin{align}\label{Y21:LLog}
\lefteqn{v_2^2 \,( Y_\nu^\dagger \,L\, Y_\nu )_{21}\, = \,
} \nonumber \\
&  
{L_{33}}\,{m_{N_3}}\,\left[ {c_{13}}\,
      \left( e^{\frac{i }{2}\,{{\phi }_1}}\,{\sqrt{{m_{\nu_1}}}}\,
         {c_{12}}\,{s_2} + 
        e^{\frac{i }{2}\,{{\phi }_2}}\,{\sqrt{{m_{\nu_2}}}}\,{c_2}\,
         {s_1}\,{s_{12}} \right)  + 
     e^{i \,\delta }\,{\sqrt{{m_{\nu_3}}}}\,{c_1}\,{c_2}\,{s_{13}}
     \right] \,\nonumber \\
& \qquad \left[ {\sqrt{{m_{\nu_3}}}}\,{c_1}\,{c_2}\,{c_{13}}\,
      {s_{23}} - e^{-\frac{i }{2}\,{{\phi }_1}}{{\sqrt{{m_{\nu_1}}}}\,{s_2}\,
        \left( {c_{23}}\,{s_{12}} + 
          e^{i \,\delta }\,{c_{12}}\,{s_{13}}\,{s_{23}}
          \right) } \right.\nonumber \\ 
	  &\qquad + \left. 
     e^{-\frac{i }{2}\,{{\phi }_2}}{{\sqrt{{m_{\nu_2}}}}\,{c_2}\,{s_1}\,
        \left( {c_{12}}\,{c_{23}} - 
          e^{i \,\delta }\,{s_{12}}\,{s_{13}}\,{s_{23}}
          \right) } \right] \nonumber \\
& 
+ {L_{22}}\,m_{N_2}\,\left[ e^{\frac{i }{2}\,{{\phi }_1}}\,
      {\sqrt{{m_{\nu_1}}}}\,{c_2}\,{c_{12}}\,{c_{13}}\,{s_3} + 
     e^{\frac{i}{2}\,{{\phi }_2}}\,{\sqrt{{m_{\nu_2}}}}\,{c_{13}}\,
      \left( {c_1}\,{c_3} - {s_1}\,{s_2}\,{s_3} \right) \,
{s_{12}}\right.\nonumber \\
& \left. \qquad     - e^{i \,\delta }\,{\sqrt{{m_{\nu_3}}}}\,
      \left( {c_3}\,{s_1} + {c_1}\,{s_2}\,{s_3} \right) \,{s_{13}}
     \right] \, \nonumber \\
& \qquad\left[ - {\sqrt{{m_{\nu_3}}}}\,{c_{13}}\,
        \left( {c_3}\,{s_1} + {c_1}\,{s_2}\,{s_3} \right) \,
        {s_{23}}  - 
     e^{-\frac{i }{2}\,{{\phi }_1}}{{\sqrt{{m_{\nu_1}}}}\,{c_2}\,{s_3}\,
        \left( {c_{23}}\,{s_{12}} + 
          e^{i \,\delta }\,{c_{12}}\,{s_{13}}\,{s_{23}}
          \right) } \right. \nonumber \\
& \left. \qquad+ 
     e^{-\frac{i }{2}\,{{\phi }_2}}{{\sqrt{{m_{\nu_2}}}}\,
        \left( {c_1}\,{c_3} - {s_1}\,{s_2}\,{s_3} \right) \,
        \left( {c_{12}}\,{c_{23}} - 
          e^{i \,\delta }\,{s_{12}}\,{s_{13}}\,{s_{23}}
          \right) } \right]\nonumber \\
& + {L_{11}}\,m_{N_1}\,\left[ e^{\frac{i }{2}\,{{\phi }_1}}\,
      {\sqrt{{m_{\nu_1}}}}\,{c_2}\,{c_3}\,{c_{12}}\,{c_{13}} - 
     e^{\frac{i }{2}\,{{\phi }_2}}\,{\sqrt{{m_{\nu_2}}}}\,{c_{13}}\,
      \left( {c_3}\,{s_1}\,{s_2} + {c_1}\,{s_3} \right) \,{s_{12}}  \right.\nonumber \\
&  \qquad \left.   + e^{i \,\delta }\,{\sqrt{{m_{\nu_3}}}}\,
      \left( -\left( {c_1}\,{c_3}\,{s_2} \right)  + 
        {s_1}\,{s_3} \right) \,{s_{13}} \right] \,\nonumber \\
&  \qquad \left[ {\sqrt{{m_{\nu_3}}}}\,{c_{13}}\,
      \left( -\left( {c_1}\,{c_3}\,{s_2} \right)  + 
        {s_1}\,{s_3} \right) \,{s_{23}} - 
     e^{-\frac{i }{2}\,{{\phi }_1}}{{\sqrt{{m_{\nu_1}}}}\,{c_2}\,{c_3}\,
        \left( {c_{23}}\,{s_{12}} + 
          e^{i \,\delta }\,{c_{12}}\,{s_{13}}\,{s_{23}}
          \right) } \right.\nonumber \\
&	\left. \qquad  - 
     e^{-\frac{i }{2}\,{{\phi }_2}}{{\sqrt{{m_{\nu_2}}}}\,
        \left( {c_3}\,{s_1}\,{s_2} + {c_1}\,{s_3} \right) \,
        \left( {c_{12}}\,{c_{23}} - 
          e^{i \,\delta }\,{s_{12}}\,{s_{13}}\,{s_{23}}
          \right) } \right]\,.  	   
\end{align}
The above is the relevant matrix element for the 
$\mu \to e \gamma$ and $\mu \to 3e$
decays, which will be the most emphasised in the present work. 
Correspondingly, the expression for the $(Y_\nu^\dagger \,L\, Y_\nu )_{32}$  
($(Y_\nu^\dagger \,L\, Y_\nu )_{31}$) matrix element, omitted here 
for brevity,
is the relevant one with respect to $\tau \to \mu \gamma$ and $\tau \to
3\,\mu$ ($\tau \to e \gamma$ and $\tau \to 3\,e$) decays.

Since the dominant contribution to the $\mu \to e \gamma$ decay stems
from the RGE induced flavour mixing in $(\Delta
m_{\tilde{L}}^2)_{21}$, within the framework of the mass insertion and
leading logarithmic approximations, 
one then obtains a simple formula given by
\begin{equation}\label{BR:MIA:LL}
\text{BR}(\mu \to e \,\gamma)\,=\, 
\frac{\alpha^3\, \tan^2 \beta}{G_F^2\, m_\text{SUSY}^8}\,
\left|
\frac{1}{8\,\pi^2}\, \left(3\, M_0^2+ A_0^2\right)\, \left(Y_{\nu}^\dagger\, 
L\, Y_{\nu}\right)_{21} 
\right|^2\,,
\end{equation}
where $G_F$ is the Fermi constant, $(Y_{\nu}^\dagger\, 
L\, Y_{\nu})_{21}$ has been given in Eqs.~(\ref{Y21:LLog}, \ref{Rcasas}),
and $m_\text{SUSY}$ represents a generic SUSY mass.

\subsection{Thermal leptogenesis and gravitino constraints}\label{bau-theory}

In our analysis, we will take into account constraints on LFV from the
requirement of successfully generating the baryon asymmetry of the
Universe via thermal leptogenesis~\cite{Fukugita:1986hr}. 
In this scenario, the BAU is explained by the out-of-equilibrium decay of the 
same heavy right-handed neutrinos which are responsible for the suppression of 
light neutrino masses in the seesaw mechanism. 
The needed CP asymmetry for BAU is obtained from the CP violating
phases in the complex angles $\theta_i$ (see Eqs.~(\ref{seesaw:casas}), (\ref{Rcasas})),
which also have a clear impact on the LFV rates. 
Here we assume that 
the necessary population of right-handed neutrinos 
emerges via processes in the thermal bath of the early Universe. 
We will furthermore consider cosmological constraints on the reheat 
temperature after inflation
associated with thermally produced gravitinos. 
The reheat temperature, $T_{\mathrm{RH}}$, has a 
strong impact on thermal leptogenesis 
since the thermal production of right-handed neutrinos $N_1$ is suppressed
if $T_{\mathrm{RH}}\ll m_{N_1}$.

\subsubsection{Gravitino problems and the reheat 
temperature}\label{Sec:GravitinoProblem}
Thermally produced gravitinos can lead to two generic 
constraints on the reheat temperature~\cite{Kohri:2005wn}.  
Both are associated with the fact that 
in the scenarios under consideration, and assuming R-parity conservation, 
the gravitinos will ultimately decay in the lightest 
supersymmetric particle (LSP). 
Firstly, they can decay late, after the Big Bang nucleosynthesis (BBN) 
epoch, and potentially spoil the success of BBN.  
This leads to upper bounds on the reheat temperature which depend on 
the specific supersymmetric model as well as on the mass of the 
gravitino. In particular, with a heavy gravitino (roughly above $100$
TeV), the BBN constraints can be nearly avoided. In our study, we 
will consider the gravitino mass as a free parameter, 
so that we can safely avoid the latter constraints for any given 
reheat temperature. 
Secondly, the decay of a gravitino produces one LSP, which has an impact 
on the relic density of the latter. 
The number of thermally 
produced gravitinos increases with the reheat temperature, and  
we can estimate the contribution to the dark matter (DM) 
relic density arising from
non-thermally produced LSPs via gravitino decay as~\cite{Kohri:2005wn}
\begin{equation}
\Omega^{\mathrm{non-th}}_\mathrm{LSP} h^2 \approx 0.054 
\left( \frac{m_\mathrm{LSP}}{100 \, \text{GeV}}\right) 
\left( \frac{T_\mathrm{RH}}{10^{10}\,\text{GeV}} \right) ,
\end{equation}
which depends on the LSP mass, $m_\mathrm{LSP}$, as well as on the reheat 
temperature $T_\mathrm{RH}$.     
Taking the bound 
$\Omega^{\mathrm{non-th}}_\mathrm{LSP} h^2 \le
\Omega_\mathrm{DM} h^2 \lesssim 0.13$ from the Wilkinson Microwave
Anisotropy Probe (WMAP)~\cite{Spergel:2006hy}, 
we are led to an upper bound on the reheat temperature of
\begin{equation}\label{TRHbound}
T_{\mathrm{RH}} \lesssim 2.4 \times 10^{10} \, \text{GeV} \,
\left(  \frac{100 \: \text{GeV}}{m_\mathrm{LSP} }\right)  .
\end{equation}
For the considered SUSY scenarios, the mass of the LSP (which is the
lightest neutralino) is in the range 
100 GeV$\, - \, 150$ GeV,
resulting in an estimated upper bound on the reheat temperature of
approximately
$T_{\mathrm{RH}} \lesssim 2 \,\times\, 10^{10} \, \text{GeV}$.
In the following subsection, we will consider the constraints on the
$R$-matrix parameters from the 
requirement of generating the BAU via 
thermal leptogenesis, while taking into account the latter bound on the 
reheat temperature.

\subsubsection{Thermal leptogenesis}
In the chosen scenario of hierarchical right-handed neutrinos, 
the baryon asymmetry arises from the out-of-equilibrium 
decay of the lightest right-handed neutrino $N_1$. 
The produced lepton asymmetry is then partially transformed into a baryon 
asymmetry via sphaleron conversion. In the MSSM, the resulting baryon 
to photon ratio from thermal leptogenesis 
can be written as~\cite{Giudice:2003jh}
\begin{equation}\label{eq:BAU_epsilon_eta}
  \frac{n_\mathrm{B}}{n_\gamma}
 \approx - \,1.04\,\times\,10^{-2}\,\varepsilon_{1}
 \, \eta\; ,
\end{equation}
where  
$\varepsilon_1$ is the decay asymmetry of $N_1$ into Higgs and 
lepton doublets and $\eta$ is an efficiency factor for thermal 
leptogenesis, which can be estimated by solving the Boltzmann
equations. 
The efficiency strongly depends on the ratio  
$m_{N_1}/T_\mathrm{RH}$ as well as on the parameter 
$\widetilde m_1$~\cite{Buchmuller:2000as}, which is defined as 
\begin{equation}\label{Eq:m1tilde}
\widetilde m_1 = \frac{\sum_{f} (Y_\nu)_{1f} (Y^\dagger_\nu)_{f1} \, 
v_2^2 }{ m_{{N_1}} } \; .
\end{equation}

In the following, regarding the estimation of the efficiency 
$\eta (\widetilde m_1,m_{N_1}/T_\mathrm{RH})$, we will use the 
numerical results of Ref.~\cite{Giudice:2003jh} for 
$10^{-7}$~eV~$ \leq \widetilde m_1 \leq 1$~eV and 
$0.1 \leq m_{N_1}/T_\mathrm{RH} \leq 100$ 
(under the assumption of a zero initial population of $N_1$). 
As presented in \cite{Giudice:2003jh}, the efficiency dramatically drops 
if either $m_{N_1} \gg T_\mathrm{RH}$ or if 
$\widetilde m_1$  strongly deviates from its optimal value 
$\widetilde m_1 \approx 10^{-3}$ eV. Thus, the optimisation of this
efficiency factor, to obtain a successful BAU, 
suggests that $m_{N_1}\,\lesssim \,T_\mathrm{RH}$, which we will assume 
for the forthcoming LFV analysis.

With respect to the decay asymmetries we will use the 1-loop 
results~\cite{Covi:1996wh}
\begin{equation}\label{eq:DecayAss}
\varepsilon_{1} = \frac{1}{8 \pi} 
 \frac{\sum_{j\not=1} \mbox{Im}\,\{ [(Y_\nu Y^\dagger_\nu)_{1j}]^2\}}{
 \sum_f \, |(Y_\nu)_{1f}|^2} 
 \, \sqrt{x_j} \,
 \left[ \frac{2}{1-x_j} -  \ln \left( \frac{x_j +1}{x_j}\right) \right] ,
\end{equation}
with $x_j = m^2_{N_j} / m^2_{N_1}$, for $j \not= 1$.

Since in our analysis we use the $R$-matrix parameterisation of 
Eq.~(\ref{seesaw:casas}), it is convenient to rewrite both the decay 
asymmetry $\varepsilon_1$ (in the limit of hierarchical right-handed 
neutrinos) and the washout parameter $\widetilde m_1$, in terms of 
the $R$-matrix parameters~\cite{Davidson:2002qv},
\begin{eqnarray}
\label{eq:BAU_R}  
\varepsilon_1 \approx -\frac{3}{8\pi} \frac{m_{N_1}}{v_2^2} 
\frac{\sum_j m_{\nu_j}^2 \mbox{Im}(R^2_{1j})}{\sum_i m_{\nu_i} 
|R_{1i}|^2}\; , \quad
\label{Eq:m1tilde_R}
\widetilde m_1 = \sum_j m_{\nu_j} |R_{1j}|^2\; .
\end{eqnarray}
As seen from the previous equation, a successful leptogenesis requires 
complex values of the $R$-matrix entries in order to 
generate a non-zero decay asymmetry. 

The BAU estimate in Eq.~(\ref{eq:BAU_epsilon_eta}) should be compared with 
the reported WMAP 68\% confidence range for the baryon-to-photon 
ratio~\cite{Spergel:2006hy} 
\begin{equation}\label{wmapBAU}
\frac{n_\text{B}}{n_\gamma}\,=\,(6.0965\,\pm\,0.2055)\,\times\,10^{-10}\,.
\end{equation}

\begin{figure}[t]
  \begin{center} 
	\psfig{file=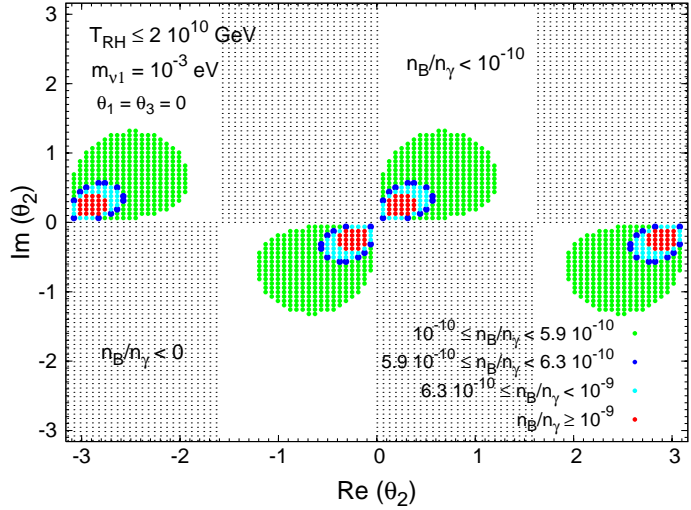,width=70mm,angle=270,clip=} 
    \caption{Constraints on the $R$-matrix angle $\theta_{2}$ (in
    radians) defined in Eq.~(\ref{Rcasas}), 
    from the requirement of a successful BAU via thermal leptogenesis 
    (compatible with the constraint on $T_{\mathrm{RH}}$ from
    Eq.~(\ref{TRHbound})). From out- to inner-most rings, the regions
    are associated with the following BAU ranges: $n_\text{B}/n_\gamma
    \in [10^{-10}, \,5.9 \times 10^{-10}]$, $n_\text{B}/n_\gamma
    \in [5.9 \times 10^{-10}, \,6.3 \times 10^{-10}]$, $n_\text{B}/n_\gamma
    \in [6.3 \times 10^{-10}, \,10^{-9}]$ and 
    $n_\text{B}/n_\gamma \gtrsim 10^{-9}$.
    }\label{fig:bau:modt2:argt2} 
  \end{center}
\end{figure}

\begin{figure}[t]
  \begin{center} 
	\psfig{file=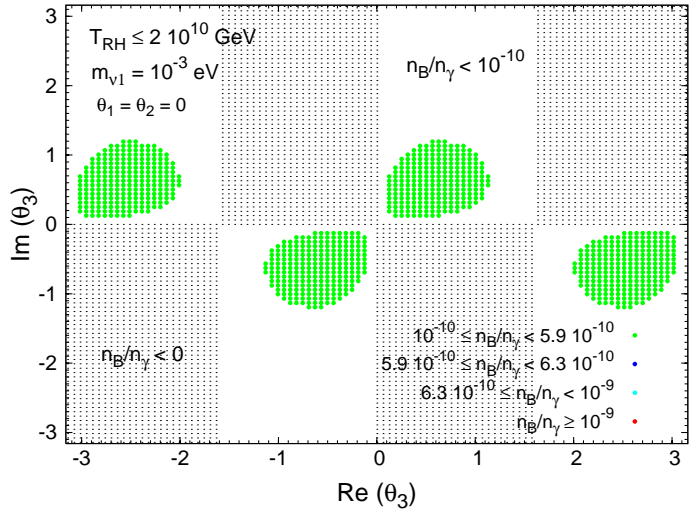,width=70mm,angle=270,clip=} 
\caption{\label{fig:bau:modt3:argt3}
  Constraints on the $R$-matrix angle $\theta_{3}$ defined in 
  Eq.\ (\ref{Rcasas}), from the requirement of successful BAU via thermal 
  leptogenesis with the constraint 
  $T_{\mathrm{RH}} \lesssim 2 \,\times\, 10^{10} \,\text{GeV}$. Colour
  code as in Fig.~\ref{fig:bau:modt2:argt2} (in this case only the
  region $n_\text{B}/n_\gamma
    \in [10^{-10}, \,5.9 \times 10^{-10}]$ is present).} 
  \end{center}
\end{figure}

Finally, the constraints on the $R$-matrix parameters  
from the requirement of a successful BAU
compatible with the upper bound 
$T_{\mathrm{RH}} \lesssim 2 \,\times\, 10^{10} \: \text{GeV}$
are summarised in Figs.~\ref{fig:bau:modt2:argt2} 
and~\ref{fig:bau:modt3:argt3}.
Figure~\ref{fig:bau:modt2:argt2} illustrates the impact of $\theta_2$
(with $\theta_1=\theta_3=0$)
on the estimated BAU. As one can see, the 68\% WMAP confidence range of
Eq.~(\ref{wmapBAU}) corresponds to a
very narrow ring (represented by the darkest region in
Fig.~\ref{fig:bau:modt2:argt2})
in the Re($\theta_2$)-Im($\theta_2$) plane. 
Notice also that values of either Re($\theta_2$) or Im($\theta_2$)
larger than 1.2 radians (mod $\pi$) 
lead to very small values of the BAU,namely $n_\text{B}/n_\gamma <
10^{-10}$. On the other hand, the analogous study of 
Fig.~\ref{fig:bau:modt3:argt3} shows
that with just $\theta_3$ ($\theta_1=\theta_2=0$)
one cannot accommodate the WMAP range. 
Similarly, values of Re($\theta_3$) or Im($\theta_3$)
larger than 1.2 radians (mod $\pi$) also lead to excessively  
small $n_\text{B}/n_\gamma$ ($< 10^{-10}$). Similar results regarding the
constraints on $\theta_2$ and $\theta_3$ from successfull BAU via leptogenesis
have been found in~\cite{Petcov:2005jh} and~\cite{Deppisch:2005rv}. We also see from     
Figs.~\ref{fig:bau:modt2:argt2} and~\ref{fig:bau:modt3:argt3} 
that a significant part of the parameter space is excluded since 
the baryon asymmetry is produced with the wrong sign, 
$n_\mathrm{B}/n_\gamma < 0$, which contradicts observation. 

Regarding $\theta_1$, and even though it cannot independently account
for a successful BAU, it may have an impact on leptogenesis if 
$\theta_2$ and/or $\theta_3$ 
are non-zero, as can be inferred from Eq.~(\ref{eq:BAU_R}). 

For the present study of LFV, we adopt a conservative approach, and we only 
require the  
estimated baryon-to-photon ratio to be within the range
\begin{eqnarray}\label{BAUeffrange}
\frac{n_\mathrm{B}}{n_\gamma} \in \left[10^{-10},10^{-9}\right] .
\end{eqnarray}  
This broad range for $n_\text{B}/n_\gamma$ reflects
the theoretical uncertainties in our 
estimate which may come, 
for instance, from flavour effects in the Boltzmann 
equations~\cite{Abada:2006fw,Nardi:2006fx,Abada:2006ea} and,
more generally, from the approximations made 
in~\cite{Giudice:2003jh} in order to calculate the efficiency factor $\eta$. 
To accommodate the extended range of
Eq.~(\ref{BAUeffrange}), 
Figs.~\ref{fig:bau:modt2:argt2} and~\ref{fig:bau:modt3:argt3} 
suggest that one should take values of $\theta_2$ and $\theta_3$ 
not larger than
approximately 1 radian (mod $\pi$).
 
\subsection{Implications for charged lepton EDMs}

The presence of CP violating phases in the 
neutrino Yukawa couplings has further
implications on low-energy phenomenology. In particular, RGE running
will also induce, in addition to the LFV decays, 
contributions to 
flavour conserving CP violating observables, as is the case of the
charged lepton EDMs. Here, we also analyse the potential constraints
on the SUSY seesaw parameter space arising from the present experimental 
bounds~\cite{pdg2004} on the EDMs of the electron, muon and tau.

As argued in~\cite{Ellis:2001yz,Ellis:2002xg,Masina:2003wt,Farzan:2004qu}, 
the dominant contributions to
the EDMs arise from the renormalisation of the  
charged lepton soft-breaking parameters.
In particular, 
the EDMs are strongly sensitive to the non-degeneracy of the
heavy neutrinos, and to the several CP violating phases of the model
(in our case, the complex $R$-matrix angles).
In the present analysis we estimate the relevant contributions to
the charged lepton EDMs, taking into account the associated one-loop
diagrams (chargino-sneutrino and neutralino-slepton mediated),
working in the mass eigenstate basis, and closely following the
computation of~\cite{Ibrahim:1997gj,Abel:2001vy} and~\cite{Ellis:2001xt}. 
Instead of conducting a detailed survey, we only use the EDMs as a
viability constraint, and postpone a more complete study, including
all phases, to a forthcoming work.
The discussion of
other potential CP violating effects, as for instance CP asymmetries
in lepton decays, is also postponed to a future study.

\section{Results and discussion}\label{results}

In this section we present the numerical results for the LFV branching
ratios arising in the SUSY seesaw scenario previously described. 
In particular, we aim at investigating the dependence of
the BRs on the several input parameters, namely on 
$\theta_i$, $m_{N_i}$, and $m_{\nu_1}$,
and how the results would reflect the impact of a potential $\theta_{13}$
measurement. 
In all cases, we further discuss how 
the requirement of a viable BAU would affect the allowed parameter
range, and in turn the BR predictions.

Regarding the dependence of the BRs on the CMSSM parameters, 
and instead of scanning over the full ($M_{1/2},\,M_0,\,A_0,\,\tan
\beta,\,\text{sign}\,\mu$) parameter space, we study specific
points, each exhibiting distinct characteristics from
the low-energy phenomenology point of view. We specify these
parameters by means of the ``Snowmass Points and Slopes''
(SPS) cases~\cite{Allanach:2002nj} 
listed in Table~\ref{SPS:def:15}. 
\begin{center}
\begin{table}\hspace*{25mm}
\begin{tabular}{|c|c|c|c|c|c|}
\hline
SPS & $M_{1/2}$ (GeV) & $M_0$ (GeV) & $A_0$ (GeV) & $\tan \beta$ & 
 $\mu$ \\\hline
 1\,a & 250 & 100 & -100 & 10 &  $>\,0 $ \\
 1\,b & 400 & 200 & 0 & 30 &   $>\,0 $ \\
 2 &  300 & 1450 & 0 & 10 &  $>\,0 $ \\
 3 &  400 & 90 & 0 & 10 &    $>\,0 $\\
 4 &  300 & 400 & 0 & 50 &   $>\,0 $ \\
 5 &  300 & 150 & -1000 & 5 &   $>\,0 $\\\hline
\end{tabular} 
\caption{Values of $M_{1/2}$, $M_0$, $A_0$, $\tan \beta$, 
and sign $\mu$ for the SPS points considered in the analysis.}
\label{SPS:def:15}
\end{table}
\end{center}

\noindent
These points are benchmark scenarios for an mSUGRA SUSY
breaking mechanism. Points 1a and 1b are ``typical'' mSUGRA points
(with intermediate and large $\tan \beta$, respectively), 
lying on the so-called
bulk of the cosmological region. The focus-point region for the
relic abundance is represented by SPS 2, also characterised by a
fairly light gaugino spectrum. SPS 3 is directed towards the
coannihilation region, accordingly displaying a very small
slepton-neutralino mass difference. Finally, SPS 4 and 5 are extreme
$\tan \beta$ cases, with very large and small values, respectively.
Since the LFV rates are very sensitive to $\tan \beta$, we will also
display the BR predictions as a function of this parameter,
with $M_{1/2},\,M_0,\,A_0$ and $\text{sign}\, \mu$ as fixed by the SPS points.

To obtain the low-energy parameters of the model (and thus compute the
relevant physical masses and couplings), 
the full RGEs (including relevant terms and equations for the 
neutrinos and sneutrinos) are firstly run down from $M_X$ to $m_{M}$.
At the seesaw scale\footnote{In our
  analysis we do not take into account the effect of the heavy
  neutrino thresholds~\cite{Antusch:2002rr}. We have verified that, within the LLog
  approximation, these thresholds effects are in general negligible in our
  analysis.} 
(in particular at $m_{N_3}$), we 
impose the boundary condition of
Eq.~(\ref{seesaw:casas}). After the decoupling of the heavy neutrinos
and sneutrinos, the new RGEs are then run down from $m_{N_1}$ to 
the EW scale, at which the observables are computed.

The numerical implementation of the above procedure is achieved by
means of the public Fortran code {\tt SPheno2.2.2}~\cite{Porod:2003um}. The
value of $M_X$ is derived from the unification condition of the
$SU(2)$ and $U(1)$ gauge couplings (systematically leading 
to a value of $M_X$ very close to $2 \times 10^{16}$ GeV
throughout the numerical analysis), while $|\mu|$ is derived from  
the requirement of obtaining the correct radiative EW symmetry
breaking.
The code {\tt SPheno2.2.2} has been adapted in order to fully 
incorporate the right-handed
neutrino (and sneutrino) sectors, as well as the full lepton flavour 
structure~\cite{Arganda:2005ji}. The computation of the LFV 
branching ratios (for all channels) has
been implemented into the code with additional subroutines~\cite{Arganda:2005ji}. 
Likewise, the code has been enlarged with two other subroutines which
estimate the value of the BAU, and evaluate the contributions to the
charged lepton EDMs.  

The input values used regarding the light neutrino masses
and the $U_\text{MNS}$ matrix elements are
\begin{align}
& 
\Delta\, m^2_\text{sol} \,=\,8\,\times 10^{-5}\,\,\text{eV}^2\,,
\quad \quad 
\Delta \, m^2_\text{atm} \,=\,2.5\,\times 10^{-3}\,\,\text{eV}^2\,,
\nonumber \\
& 
\theta_{12}\,=\,30^\circ\,, 
\quad 
\theta_{23}\,=\,45^\circ\,,
\quad 
\theta_{13}\,\lesssim\,10^\circ\,,
\quad \quad 
\delta\,=\,\phi_1\,=\,\phi_2\,=\,0\,,
\end{align}
which are compatible with present experimental data (see, for
instance, the analysis of~\cite{Gonzalez-Garcia:2003qf,Maltoni:2004ei,Fogli:2005cq}).
As previously mentioned, we do not address the impact of
non-vanishing $U_\text{MNS}$ phases (Dirac or Majorana) in the LFV
branching ratios. The effects of Majorana phases on the BRs have been
discussed in Ref.~\cite{Petcov:2006pc}.

Regarding charged lepton EDMs, we require compatibility with the 
current experimental bounds~\cite{pdg2004}
\begin{align}\label{EDM:exp}
&|d_e|\,\lesssim (6.9\pm7.4)\,\times\,10^{-28}\,\,\text{e.cm}\,,
\quad
|d_\mu| \,\lesssim (3.7\pm3.4)\,\times\,10^{-19}\,\,\text{e.cm}\,,
\quad \nonumber \\
&\hspace*{10mm}
-2.2\,\times\,10^{-17}\,\lesssim \,d_\tau \,\lesssim
4.5\,\times\,10^{-17}\,\,\text{e.cm}\,. 
\end{align}

Finally, and before beginning the numerical analysis 
and discussion of the 
results, we briefly summarise\footnote{In
  Table~\ref{LFV:bounds:future}, the future
  prospects should be understood as order of magnitude conservative 
  estimates of the projected sensitivities.} 
in Table~\ref{LFV:bounds:future} the present LFV 
bounds~\cite{Brooks:1999pu,Aubert:2005wa,Aubert:2005ye,Bellgardt:1987du,Aubert:2003pc}, as well as the
future planned
sensitivities~\cite{mue:Ritt,Akeroyd:2004mj,Iijima,Aysto:2001zs},  
for the several channels under
consideration\footnote{
There are other LFV processes of interest, such as 
  $\tau \to \mu\,e\,e$, $\tau \to e\,\mu\,\mu$, semileptonic $\tau$
decays, and $\mu$-$e$ conversion in heavy nuclei, 
  which are not considered in the present work.
  With the advent of the PRISM/PRIME experiment at 
J-PARC~\cite{PRIME,Kuno:2005mm}, 
  $\mu$-$e$ conversion in heavy nuclei as Ti 
  may become the most stringent test for muon 
  flavour conservation.}.
\begin{center}
\begin{table}\hspace*{25mm}
\begin{tabular}{|c|c  c |}
\hline
LFV process & Present bound & Future sensitivity \\
\hline
BR($\mu \to e\,\gamma$) & $1.2 \times 10^{-11}$  & $1.3 \times
10^{-13}$  \\
BR($\tau \to e \,\gamma$) & $1.1 \times 10^{-7}$ & 
 $10^{-8}$ \\
BR($\tau \to \mu \,\gamma$) & $6.8 \times 10^{-8}$  &
$10^{-8}$  \\
BR($\mu \to 3\,e$) & $1.0 \times 10^{-12}$  & 
$10^{-13}$  \\
BR($\tau \to 3\,e$) & $2.0 \times 10^{-7}$  & 
$10^{-8}$  \\
BR($\tau \to 3\,\mu$) & $1.9 \times 10^{-7}$  & 
$10^{-8}$  \\\hline
\end{tabular}
\caption{Present bounds and future sensitivities for the LFV 
processes.}
\label{LFV:bounds:future}
\end{table}
\end{center}

\subsection{Sensitivity to $\theta_{13}$ in the case $R = \mathbbm{1}$}

We begin our study by revisiting the $R = \mathbbm{1}$ case which represents
the situation where there are no further neutrino mixings in the Yukawa
couplings other than those induced by the $U_\text{MNS}$.
In this case, the BR($\mu \to e \gamma$) dependence on $\theta_{13}$
was first observed in the context of SUSY GUTs~\cite{Masiero:2004js}.
In Ref.~\cite{Arganda:2005ji}, a comprehensive study of
all the leptonic decay channels was 
performed (in a full RGE approach), and it was noticed that  
$\mu \to e\, \gamma$ and $\mu \to 3\,e$ were the channels that
exhibited both a clear sensitivity to $\theta_{13}$ and promising prospects 
from the point of view of experimental detection.
Here, we complete the study of~\cite{Arganda:2005ji}, also analysing   
the other LFV channels. More concretely, 
we investigate how sensitive to $\theta_{13}$ the BR($l_j \to
l_i\,\gamma$) and BR($l_j \to 3\,l_i$) are. We also add some comments on the
comparison between the full and the LLog approximation results.

\begin{figure}[t]
  \begin{center} \hspace*{-10mm}
    \begin{tabular}{cc}
	\psfig{file=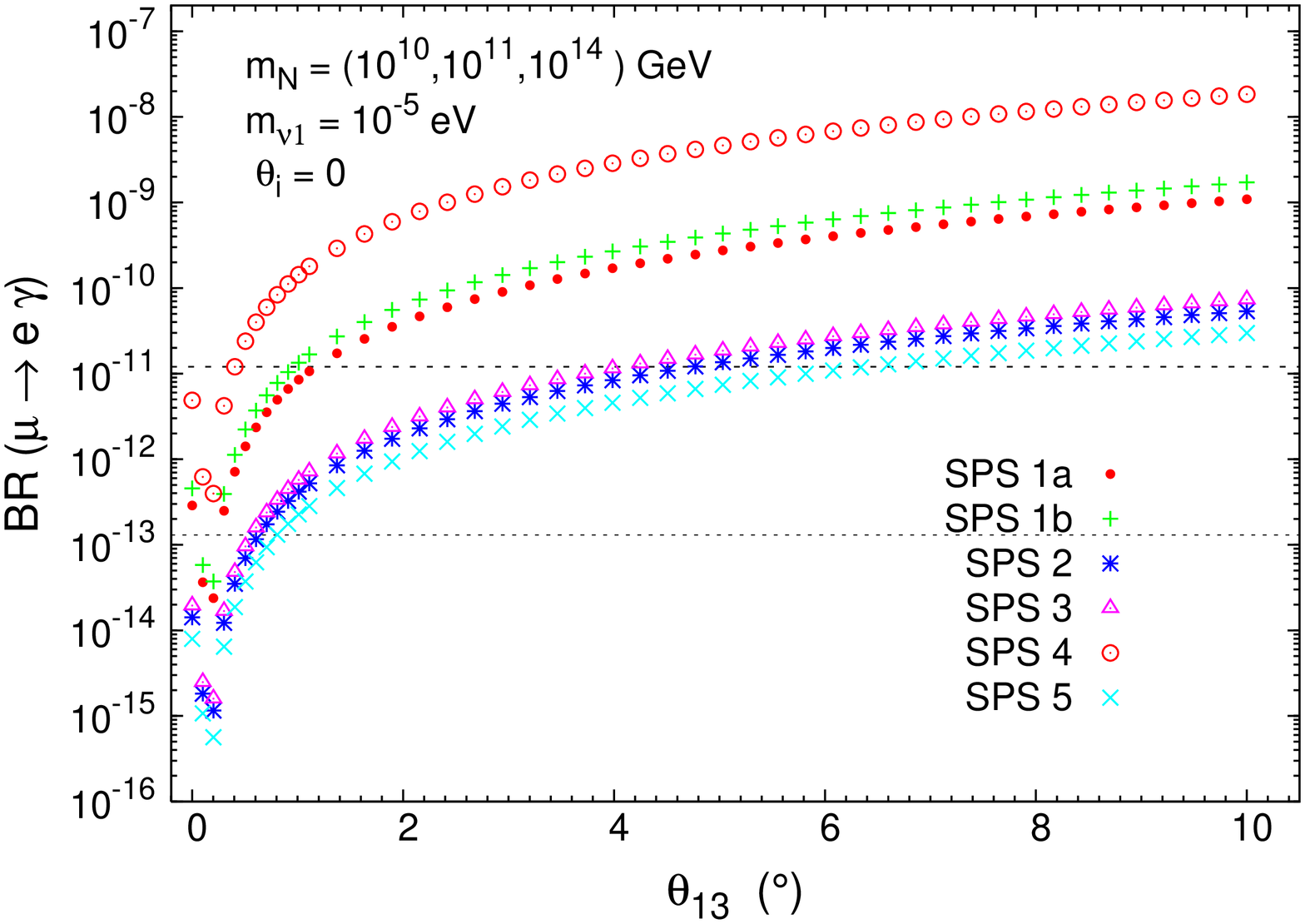,width=60mm,angle=270,clip=} &
	\psfig{file=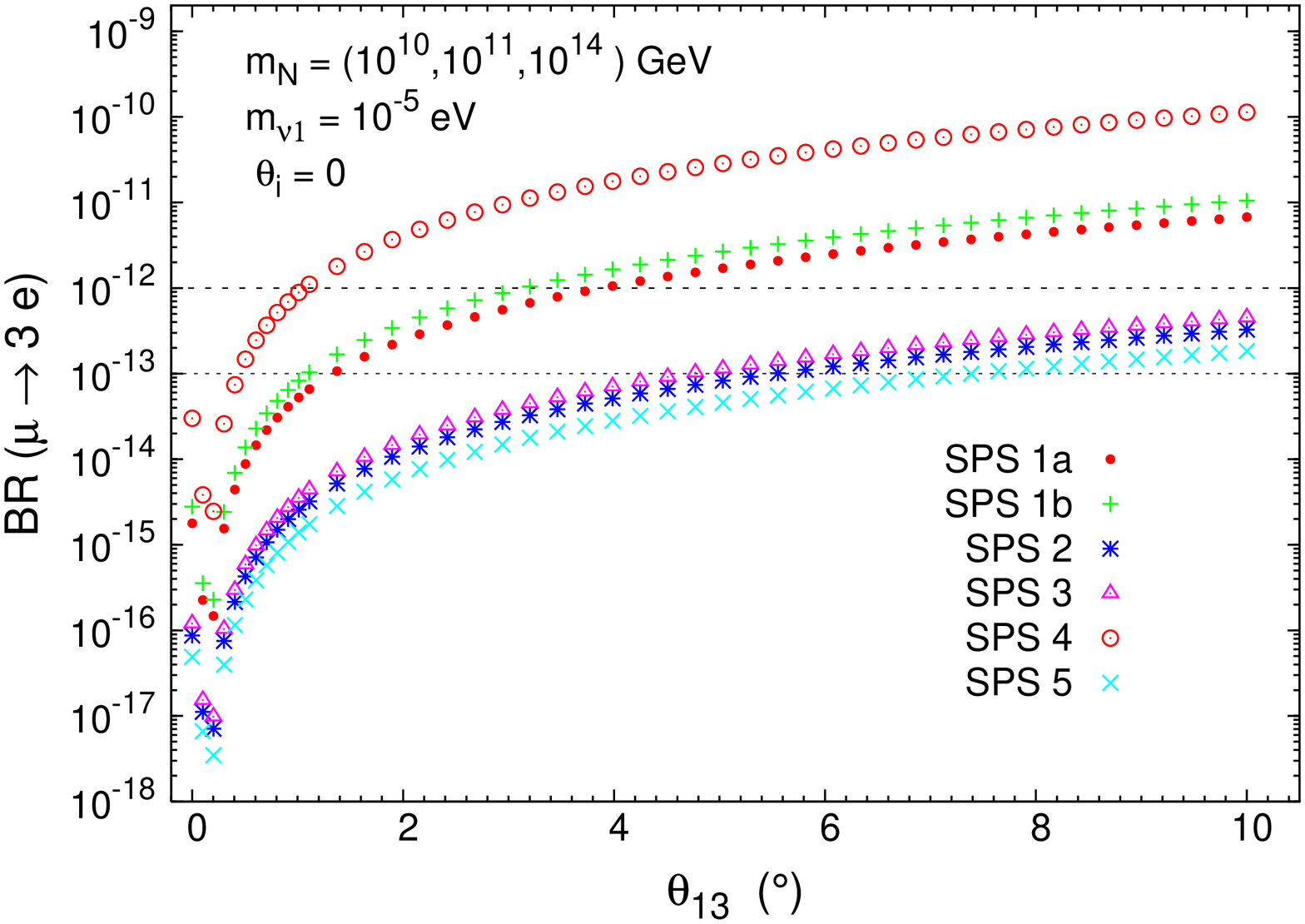,width=60mm,angle=270,clip=} 
    \end{tabular}
    \caption{BR($\mu \to e\, \gamma$) and BR($\mu \to
    3\,e $) as a function of $\theta_{13}$ (in degrees), for SPS 1a
    (dots), 1b (crosses), 2 (asterisks), 3 (triangles), 4 (circles) and 5
    (times). A dashed (dotted)
    horizontal line denotes the present experimental bound (future
    sensitivity).} 
    \label{fig:SPS:t13:ad}
  \end{center}
\end{figure}

\begin{figure}[t]
  \begin{center} \hspace*{-10mm}
    \begin{tabular}{cc}
	\psfig{file=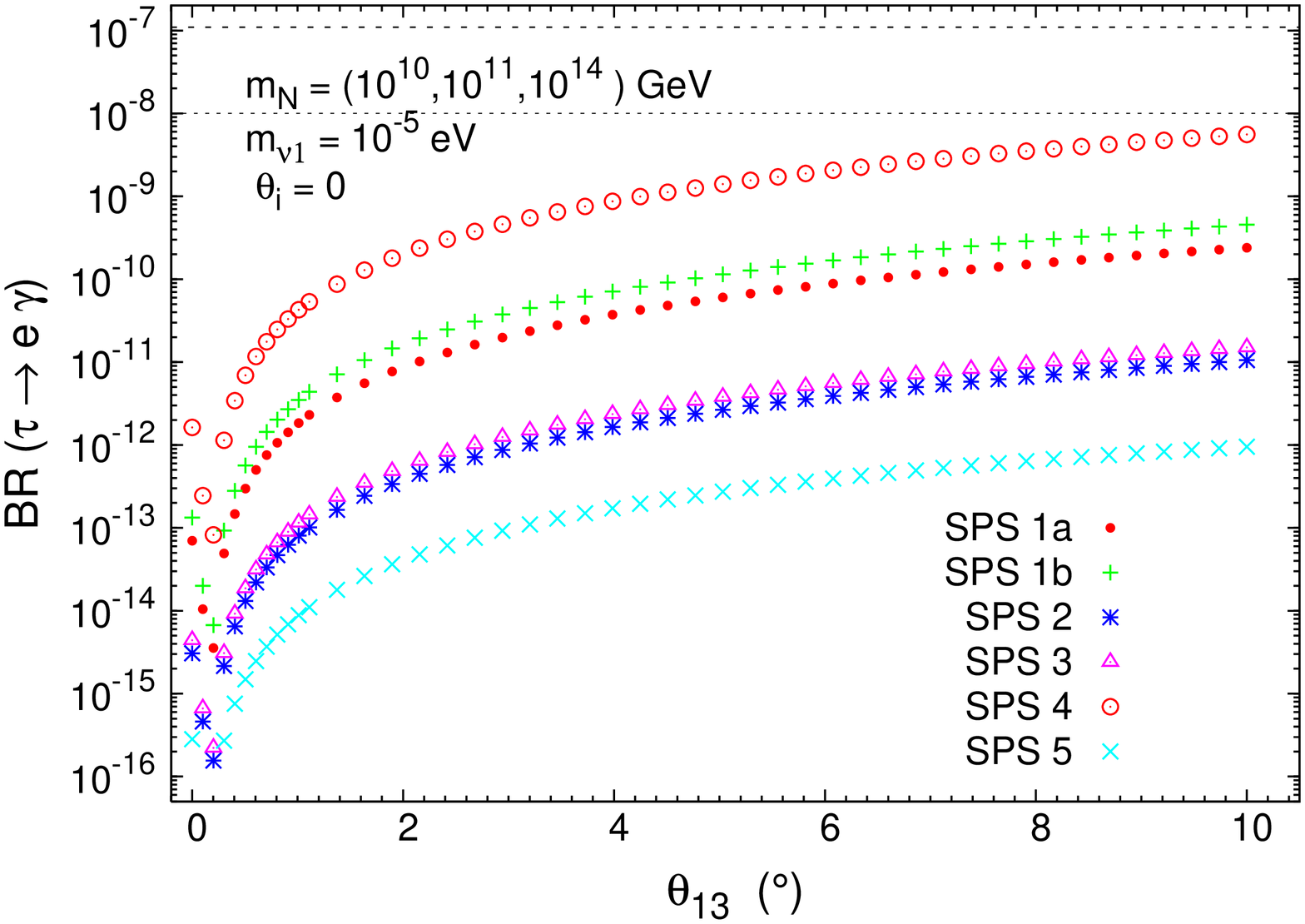,width=60mm,angle=270,clip=} &
	\psfig{file=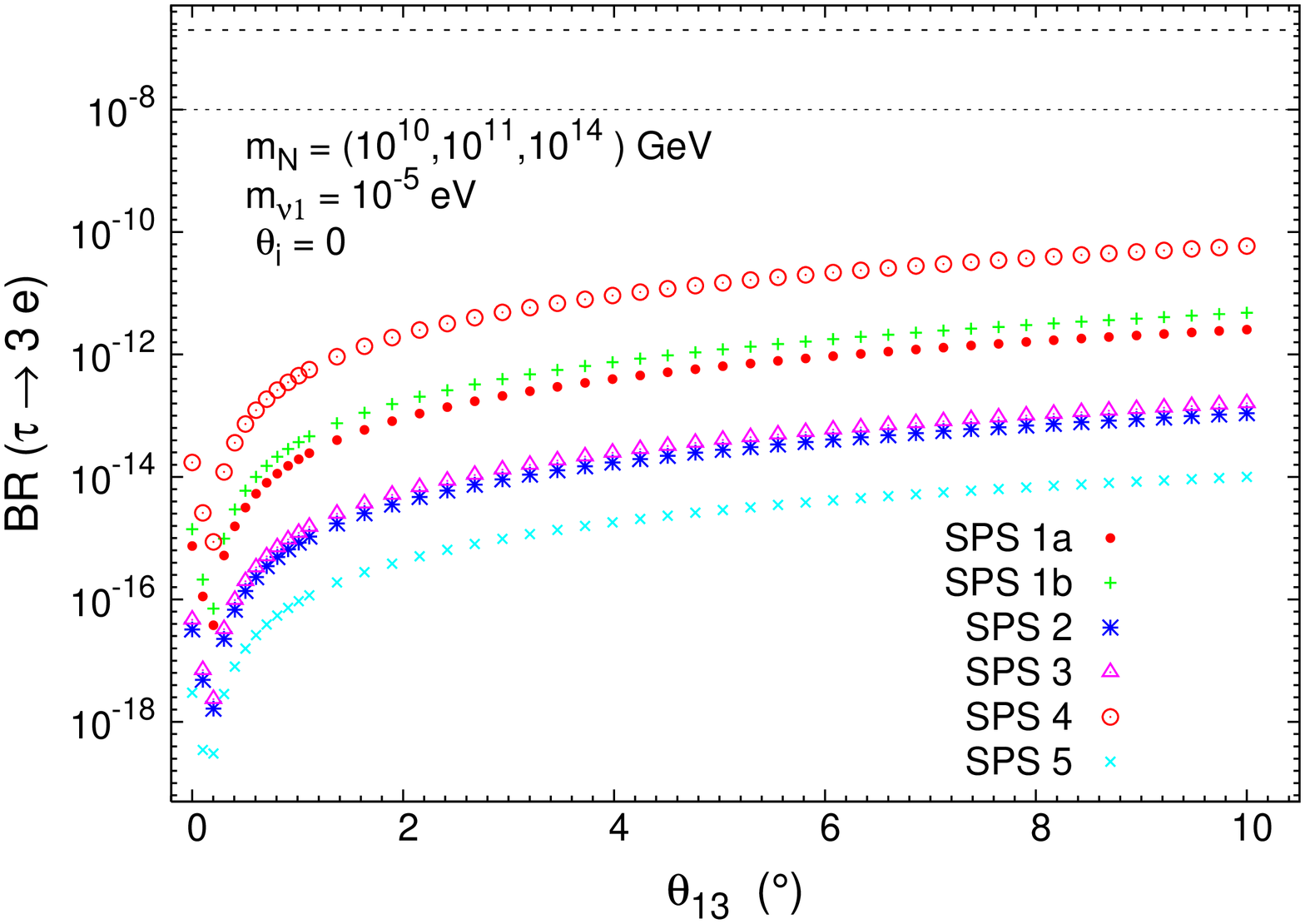,width=60mm,angle=270,clip=} 
    \end{tabular}
    \caption{BR($\tau \to e\, \gamma$) and BR($\tau \to
    3\,e $) as a function of $\theta_{13}$ (in degrees) for SPS 1a,
    1b, 2, 3, 4 and 5. Line and colour codes as in Fig.~\ref{fig:SPS:t13:ad}.} 
    \label{fig:SPS:t13:be}
  \end{center}
\end{figure}

In Figs.~\ref{fig:SPS:t13:ad} and~\ref{fig:SPS:t13:be}
we plot the
branching ratios of the decays $\mu \to e\, \gamma$,  $\mu \to 3\,e$,   
$\tau \to e\, \gamma$ and $\tau \to 3\,e$,  
as a function of $\theta_{13}$, which we vary\footnote{The scan
  step is purposely finer for small values of $\theta_{13}$.} in the range
$[0^\circ,\,10^\circ]$. We also display, for comparison, the lines
associated with the present experimental bounds and future sensitivities.   
In each case, we consider as input the six SPS points, and take
$\theta_i\,=\,0$, so that in this case no BAU is generated and 
there is no flavour mixing arising from the right-handed neutrino sector. 
Regarding the neutrino masses, we have assumed $m_{\nu_1}\,=\,
10^{-5}$ eV, while the masses of the heavy right-handed are set to
$m_{N}\, =\, (10^{10},\,10^{11},\,10^{14})$ GeV. 
In particular we have chosen $m_{N_1}$ to avoid the gravitino problem
in relation with non-thermal LSP production, as 
explained in Section~\ref{Sec:GravitinoProblem}. 
Notice that our choice of $m_{N_3}$ leads to large values for the 
third family Yukawa couplings~\footnote{Other approaches, for instance
in GUT-inspired frameworks, allow to derive the values of 
$m_{N_3}$ from unification of the Yukawa couplings of the third
family, and this may lead to even larger values of $(Y_\nu)_{33}$.
For example, an SO(10) GUT could lead 
to $m_{N_3}\approx 10^{15}$ GeV, as implied by $(Y_\nu)_{33}
\approx 1$ (see, for
example,~\cite{Masiero:2004js}).}, specifically 
$(Y_\nu)_{33} \approx (Y_\nu)_{32} \approx 
0.3$.


The first conclusion to be 
inferred from Fig.~\ref{fig:SPS:t13:ad} is
that, in agreement with~\cite{Arganda:2005ji},
the sensitivity to $\theta_{13}$ is clearly manifest in the 
$\mu \to e\, \gamma$ and $\mu \to3\,e $ channels. In addition, 
Fig.~\ref{fig:SPS:t13:be} shows that 
BR($\tau \to e\, \gamma$) and BR($\tau \to 3\,e $) also display a
strong dependence on $\theta_{13}$. Notice that in these tau decays
the BR predictions for the explored $\theta_{13}$ values lie below the
present and future experimental sensitivities\footnote{On the other hand, we remark
that compared to $\theta_{13}$, the uncertainties in the other neutrino
oscillation parameters, $\theta_{23}$, $\theta_{12}$, $\Delta m_{23}^{2}$ and
$\Delta m_{12}^{2}$, are expected to have only a smaller effect on the LFV
ratios (see e.g.~\cite{Deppisch:2002vz})}.

The observed qualitative behaviour with respect to $\theta_{13}$ 
can be easily understood from 
Eq.~(\ref{Y21:LLog}), 
which predicts that the dominant contribution proportional to
$(L_{33}\,m_{N_3}\,m_{\nu_3})^2$ should grow as
$(c_{13}\,s_{13})^2$. For small values of
$\theta_{13}$ , the ``dip'' exhibited by the BRs 
is a consequence of a shift in $\theta_{13}$ arising
from RGE running, changing it from $\theta_{13} \equiv \theta_{13}(m_Z)$ to 
$\theta_{13}(m_M)$. 
Renormalisation induces, in our example, that 
$\theta_{13}({m_M}) \approx \theta_{13} ({m_Z})- 0.2^\circ$, so that the minimum of the
BR is shifted from $\theta_{13}=0^\circ$ 
to $\theta_{13} \approx 0.2^\circ$ (which is consistent with 
analytical estimates~\cite{Antusch:2003kp}). 
More explicitly, even when starting with 
a value $\theta_{13}=0^\circ$  at the EW scale, RGE running leads to
the appearance of a negative value for $\theta_{13}(m_M)$ (or, equivalently, a
non-zero positive $\theta_{13}$ and $\delta = \pi$). 

Concerning the $\tau \to \mu\, \gamma$
and  $\tau \to 3\,\mu$ channels, the corresponding 
branching ratios do not exhibit any noticeable 
dependence on $\theta_{13}$, as expected from the
analytical expressions of the LLog approximation.
For the case $R = \mathbbm{1}$, and taking for example
$\theta_{13}=5^\circ$, these BRs 
are presented in Table~\ref{SPS:BRtau:15}.
\begin{table}[h]
\begin{tabular}{|l|c|c|c|c|c|c|}
\hline
BR & SPS 1a &SPS 1b  &SPS 2 &SPS 3 &SPS 4 &SPS 5\\\hline
$\tau \to \mu\, \gamma$ &
$ 4.2\times 10^{-9}$ &$7.9 \times 10^{-9}$ &$1.8 \times 10^{-10}$ &
$2.6 \times 10^{-10}$ &$9.7 \times 10^{-8}$ & $1.9 \times 10^{-11}$\\\hline
$\tau \to 3\,\mu$ &
$9.4 \times 10^{-12}$& $1.8 \times 10^{-11}$& $4.1 \times 10^{-13}$
 & $5.9 \times 10^{-13}$& $2.2 \times 10^{-10}$& $4.3 \times
10^{-14}$\\\hline
\end{tabular}
\caption{Predictions for the BR($\tau \to \mu\, \gamma$) and BR($\tau \to
  3\,\mu$) corresponding to the SPS points. The values of 
  $m_{N_i}$ and $m_{\nu_1}$ are as specified
  in Figs.~\ref{fig:SPS:t13:ad} and~\ref{fig:SPS:t13:be}. In each
  case, the predicted values should be compared with the present bounds
  (future prospects) BR($\tau \to
  \mu\, \gamma$) $<$ $6.8 \times 10^{-8}$ ($10^{-8}$) and BR($\tau \to
  3\,\mu$) $<$ $1.9 \times 10^{-7}$ ($10^{-8}$).}
\label{SPS:BRtau:15}
\end{table}

The conclusion to be 
inferred from Figs.~\ref{fig:SPS:t13:ad},~\ref{fig:SPS:t13:be}
and Table~\ref{SPS:BRtau:15} is that, for the assumed value of $m_{\nu_1}$,
and for the chosen seesaw scenario (which is specified by $\theta_i$ and
$m_{N_i}$), the experimental bounds for BR($\tau \to \mu\,\gamma$) 
already disfavour the CMSSM scenario of SPS 4 (for any value of
$\theta_{13}$). 
From the comparative analysis of the $\theta_{13}$-sensitive channels
it is also manifest that $\mu \to e\, \gamma$ and
$\mu \to 3\,e$ are the decays whose BRs are within the reach of 
present experiments, thus potentially allowing to constrain
the values of $\theta_{13}$. In fact, 
BR($\mu \to e\,\gamma$) suggests that SPS 
4, 1(a and b), 3, 2 and 5 are disfavoured for values of
$\theta_{13}$ larger than approximately 
$0.5^\circ$, $1^\circ$, $4^\circ$, $5^\circ$ 
and $6^\circ$, respectively, while a similar analysis of 
BR($\mu \to 3\,e$) would exclude $\theta_{13}$ values above 
$1^\circ$, $3^\circ$ and $4^\circ$ for SPS 4, 1a and 1b, correspondingly.
Nevertheless, it is crucial to notice that, as can be seen from 
Eqs.~(\ref{Y21:LLog}, \ref{BR:MIA:LL}), 
the value of $m_{N_3}$ plays a very relevant role. 
For instance, by lowering $m_{N_3}$
from $10^{14}$ GeV to $10^{13}$ GeV one could have compatibility
with the experimental bound on BR($\mu \to e\,\gamma$) for 
$\theta_{13} \lesssim 2^\circ$ for {\it all} SPS scenarios. Moreover, in this
case, even SPS 4 would be in agreement with the experimental
bound on BR($\tau \to \mu\,\gamma$).

The relative predictions for each of the SPS points can be easily understood
from the BRs dependence on the SUSY spectrum\footnote{For each
  SPS point, the associated spectrum can be found, for example, 
  in~\cite{Allanach:2002nj}.} and $\tan \beta$, 
which is approximately given by Eq.~(\ref{BR:MIA:LL}).
However, it is worth emphasising that although 
the several approximations leading to
Eq.~(\ref{BR:MIA:LL}) do provide a qualitative understanding of
the LFV rates, they are not sufficiently accurate, and do fail in
some regions of the CMSSM parameter space. 
In particular, for the SPS 5 scenario, we have verified that the LLog
predictions for the BRs arising from Eq.~(\ref{misalignment_sleptons}) 
differ from our results by several orders of
magnitude. We will return to this discussion at a later stage.

As already mentioned, in the context of SUSY GUTs, 
the dependence of the BR($\mu \to e\,\gamma$) on $\theta_{13}$ for the
same set of SPS points was presented in~\cite{Masiero:2004js}. 
Instead of the full
computations, the analysis was done using the LLog approximation, 
and the amount of slepton flavour violation was parameterised by means
of mass insertions.
In general, and even though a different seesaw scenario was considered,
the results are in fair agreement with Fig.~\ref{fig:SPS:t13:ad}, 
the only exception occurring for SPS 5. In fact,
while~\cite{Masiero:2004js} predicts the largest 
BR($\mu \to e\,\gamma$) for the SPS 5 case, our results of  
Fig.~\ref{fig:SPS:t13:ad} show that the rates for this point are
indeed the smallest ones. As already mentioned, this is due to the failure
of the LLog for SPS 5. 

Henceforth, and in view of the fact that not only is the
decay $\mu \to e\, \gamma$ one of the most sensitive to $\theta_{13}$,
but it is also the most promising regarding experimental
detection, we will mainly focus our discussion on the 
analysis of BR($\mu \to e\, \gamma$).

\subsection{Implications of a favourable BAU scenario 
on the sensitivity to $\theta_{13}$}

Motivated by the generation of a sufficient amount of CP asymmetry in the
decay of the right-handed heavy neutrinos, one has to depart from
the $R = \mathbbm{1}$ case, and this will naturally affect 
the predictions for the several BRs.
Nevertheless, it is worth stressing that the hierarchy of the SPS points
regarding the relative predictions to the distinct LFV
observables is not altered, and we also observe the same ordering
as that emerging from Figs.~\ref{fig:SPS:t13:ad},~\ref{fig:SPS:t13:be}
and Table~\ref{SPS:BRtau:15}, namely
BR$_{4}$~$>$~BR$_\text{1b}$~$\gtrsim$~BR$_\text{1a}$~$>$~BR$_{3}$~$\gtrsim$~BR$_{2}$~$>$~BR$_{5}$.  

As discussed in Section~\ref{bau-theory}, the $R$-matrix complex parameters 
$\theta_2$ and $\theta_3$ are instrumental in obtaining a value for
the baryon asymmetry in agreement with experimental
observation, while $\theta_1$ plays a comparatively less relevant
role. In what follows, we discuss how requiring a favourable BAU
scenario would constrain the $\theta_i$ ranges, and how this would
reflect on the BRs' sensitivity to $\theta_{13}$.

\subsubsection{Influence of $\theta_2$}

In view of the above, 
we begin by analysing the dependence of 
the BR($\mu \to e\, \gamma$) on $\theta_2$ and consider two
particular values of $\theta_{13}$,
$\theta_{13}=0^\circ\,,5^\circ$. 
We choose SPS 1a, and motivated from the discussion regarding 
Fig.~\ref{fig:bau:modt2:argt2}, take $0 \,\lesssim \, |\theta_2|
\,\lesssim \, \pi/4$, with $\arg \theta_2
\,=\,\{\pi/8\,,\,\pi/4\,,\,3\pi/8\}$. 

\begin{figure}[t]
  \begin{center} \hspace*{-10mm}
    \begin{tabular}{cc}
      \psfig{file=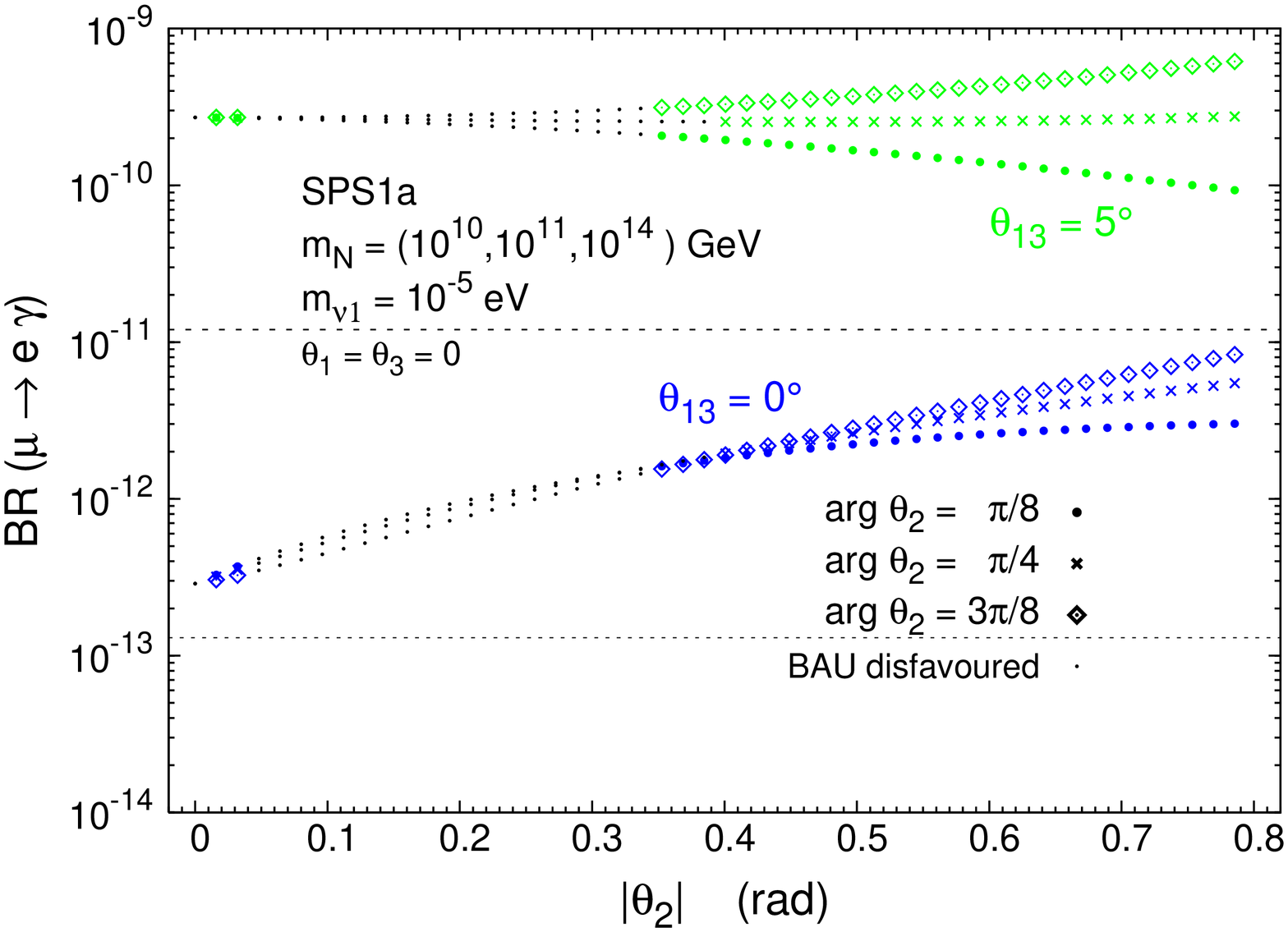,width=60mm,angle=270,clip=}
      &
      \psfig{file=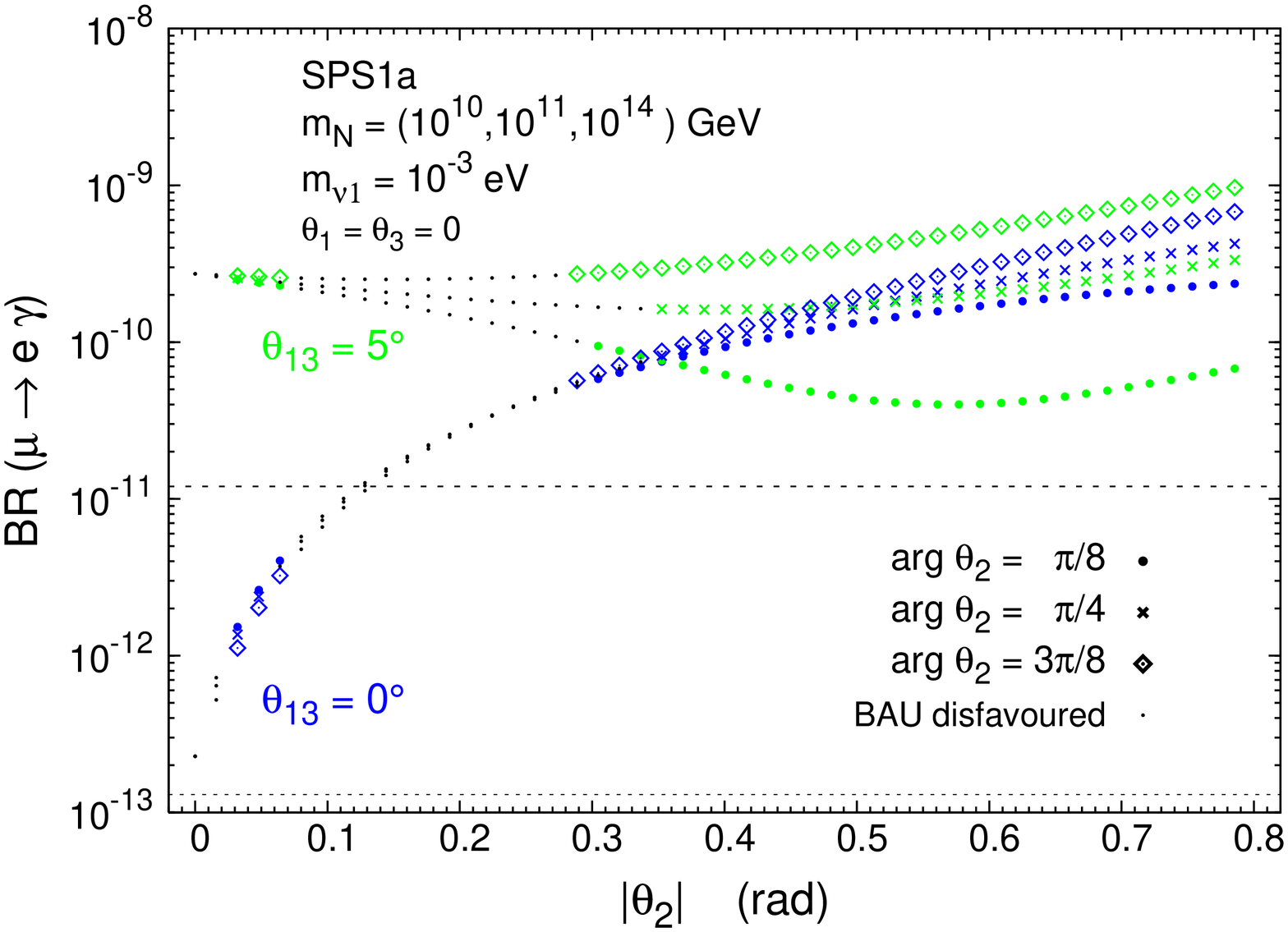,width=60mm,angle=270,clip=}
    \end{tabular}
    \caption{BR($\mu \to e\, \gamma$) as a function
      of $|\theta_2|$, for $\arg
      \theta_2\,=\,\{\pi/8\,,\,\pi/4\,,\,3\pi/8\}$ (dots, times,
      diamonds, respectively) and
      $\theta_{13}=0^\circ$, 5$^\circ$ (blue/darker, green/lighter lines).
      We take $m_{\nu_1}\,=\,10^{-5}$
      ($10^{-3}$) eV, on the left (right) panel. 
      In all cases black dots
      represent points associated with a disfavoured BAU scenario and 
      a dashed (dotted) horizontal line denotes the present 
      experimental bound (future sensitivity).} 
    \label{fig:modt2:argt2:1214}
  \end{center}
\end{figure}

In Fig.~\ref{fig:modt2:argt2:1214}, we display the numerical results,
considering $m_{\nu_1}\,=\,10^{-5}$ eV and $m_{\nu_1}\,=\,10^{-3}$~eV, 
while for the heavy neutrino masses we 
take $m_{N}\, =\, (10^{10},\,10^{11},\,10^{14})$~GeV.
There are several important conclusions to be drawn from
Fig.~\ref{fig:modt2:argt2:1214}. Let us first discuss the case 
$m_{\nu_1}\,=\,10^{-5}$~eV. As previously 
mentioned, one can obtain a baryon asymmetry 
in the range $10^{-10}$ to $10^{-9}$
for a considerable region of the analysed $|\theta_2|$ range. In
particular, a deviation from the $R = \mathbbm{1}$ case as small as
for instance, 
$\theta_2\,=\,0.05\,e^{\pi/8\,i}$ can account for an amount of BAU
close to the WMAP value. A wide region with 
larger values of $|\theta_2|$ ($0.3 \lesssim |\theta_2| \lesssim 0.8$) 
can also
accommodate a viable baryon asymmetry, as can be seen from 
Fig.~\ref{fig:modt2:argt2:1214}.
Notice also that there is a clear separation between the predictions
of $\theta_{13}=0^\circ$ and $\theta_{13}=5^\circ$, with the latter
well above the present experimental bound. At present, this would imply an
experimental impact of $\theta_{13}$, in the sense that the BR predictions 
become potentially
detectable for this non-vanishing $\theta_{13}$ value. 
With the planned
MEG sensitivity~\cite{mue:Ritt}, both cases would be within experimental reach.
However, 
this statement is strongly dependent on the
assumed parameters, in particular $m_{\nu_1}$.
For instance, a larger value of $m_{\nu_1}=10^{-3}$~eV, illustrated on
the right panel of Fig.~\ref{fig:modt2:argt2:1214}, leads to a very
distinct situation regarding the sensitivity to $\theta_{13}$.
While for smaller values of $|\theta_2|$
the branching ratio displays a clear sensitivity to having
$\theta_{13}$ equal or different from zero (a separation larger than two orders of
magnitude for $|\theta_2| \lesssim 0.05$), the effect of $\theta_{13}$ is
diluted for increasing values of $|\theta_2|$. For $|\theta_2| \gtrsim 0.3$ 
the BR($\mu \to e\, \gamma$)
associated with $\theta_{13}\,=\,5^\circ$ can be even smaller than for 
$\theta_{13}\,=\,0^\circ$. This implies that in this case, 
a potential measurement of BR($\mu \to e\, \gamma$) would not be
sensitive to $\theta_{13}$. 

Moreover, $m_{\nu_1}$ also affects the BAU-favoured regions. In
general, larger values of $m_{\nu_1}$ (still smaller than $10^{-3}$ eV) 
widen the range of $|\theta_2|$
for which a viable BAU can be obtained.
This can be understood from the fact that for very small (or zero) $\theta_2$ 
and $\theta_3$ (and with fixed $m_{N_1}$), $m_{\nu_1}$ controls the size
of the Yukawa couplings to the lightest right-handed neutrino, $N_1$.
On the other side, these are the Yukawa couplings 
governing the washout parameter $\widetilde m_1$ for thermal leptogenesis,
as introduced in Eq.~(\ref{Eq:m1tilde}). 
For very small $\theta_2$ and $\theta_3$, an optimal value 
$\widetilde m_1 \approx 10^{-3}$ eV can be reached for $m_{\nu_1}
\approx 10^{-3}$~eV 
(c.f.\ Eq.~(\ref{eq:BAU_R})), whereas smaller $m_{\nu_1}$ lead to 
suppressed leptogenesis in this case. 
For larger values of $\theta_2$ and/or $\theta_3$, 
which can be still consistent with 
leptogenesis, $m_{\nu_1}$ becomes less important, since the
other light neutrino masses $m_{\nu_2}$ and/or $m_{\nu_3}$ contribute to 
$\widetilde m_1$ as well. In most of the following analysis, we will use $m_{\nu_1}
\approx 10^{-3}$~eV and enable a successful thermal leptogenesis by
introducing a small $R$-matrix rotation angle $\theta_2$. 
In what concerns 
the sensitivity to $\theta_{13}$ via LFV, this is clearly a conservative choice
since, as previously mentioned, lower values of $m_{\nu_1}$
(e.g. $m_{\nu_1}= 10^{-5}$ eV) would lead to a
more favourable situation.    

Whether or not a BAU-compatible SPS 1a scenario would be disfavoured by
current experimental data on BR($\mu \to e\, \gamma$) requires a
careful weighting of several aspects. 
Even though Fig.~\ref{fig:modt2:argt2:1214} suggests that for this
particular choice of parameters only very small values of $\theta_2$
and $\theta_{13}$ would be in agreement with current experimental
data, a distinct choice of $m_{N_3}$
(e.g. $m_{N_3}=10^{13}$ GeV) would lead to a rescaling of the
estimated BRs by a factor of approximately $10^{-2}$.
Although we do not display 
the associated plots here, in the latter case 
nearly the entire $|\theta_2|$ range would be 
in agreement with experimental data (in fact the points which are
below the present MEGA bound on Fig.~\ref{fig:modt2:argt2:1214} would
then lie below the projected MEG sensitivity).

Regarding the other SPS points, which are not shown here, 
we find BRs for SPS 1b comparable to those of SPS 1a.
Smaller ratios are associated with SPS 2, 3 and 5, while larger (more
than one order of magnitude) BRs occur for SPS 4.

\begin{figure}[t]
  \begin{center} \hspace*{-10mm}
    \psfig{file=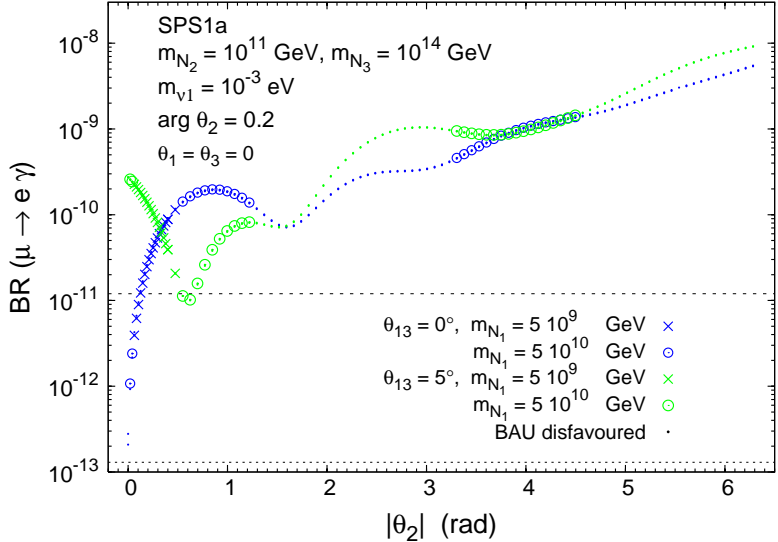,width=75mm,angle=270,clip=}
    \caption{BR($\mu \to e\, \gamma$) as a function
      of $|\theta_2|$, for SPS 1a, with $\arg
      \theta_2\,=\,0.2$. $\theta_{13}\,=\,0^\circ\,,5^\circ$ (blue/darker,
      green/lighter 
      lines, respectively), and $m_{N_1}= 5 \,\times\,10^{9}$ GeV, $5
      \,\times\,10^{10}$ GeV (crosses, circles, respectively). Dots
      represent points associated with a disfavoured BAU scenario for
      either $m_{N_1}= 5 \,\times\,10^{9}$ or $5
      \,\times\,10^{10}$ GeV and 
      a dashed (dotted) horizontal line denotes the present 
      experimental bound (future sensitivity).}
    \label{fig:modt2:argt2:SSW13:1214}
  \end{center}
\end{figure}

Let us now consider how the value of
$m_{N_1}$ affects the amount of BAU, and thus indirectly the branching ratio
associated to a given choice of $\theta_2$ that accounts for a viable BAU
scenario. 
In Fig.~\ref{fig:modt2:argt2:SSW13:1214} we present the BR($\mu \to e\,
\gamma$) as a function of $|\theta_2|$ 
for two distinct heavy neutrino
spectra: $m_{N}\, =\, (5 \,\times\,10^{9},\,10^{11},\,10^{14})$ GeV
and $m_{N}\, =\, (5 \,\times\,10^{10},\,10^{11},\,10^{14})$ GeV (values for
$m_{N_1}$ respectively smaller and larger than what was previously
considered). 
Regarding $\arg \theta_2$, we have chosen an example which
represents a minimal deviation from
the real case, $\arg \theta_2 \,=\,0.2$, and set $\theta_1\,=\theta_3\,=0$.
We consider SPS 1a, and again show both cases associated with
$\theta_{13}\,=\,0^\circ\,,5^\circ$.
From this figure, it can be seen that
in the case $m_{N_1}=5\,\times\, 10^{9}$~GeV, 
only one BAU-favoured window is opened, for small values of $\theta_2$ 
($0<\theta_2\lesssim \pi/4$).
In contrast, for $m_{N_1}=5\,\times\, 10^{10}$ GeV,
a second window opens, corresponding to the $\mod \pi$ periodicity
evidenced in Fig.~\ref{fig:bau:modt2:argt2} (also some additional
points at very small $|\theta_2|$ are allowed). The width of the
$|\theta_2|$ interval for this second window shrinks
with decreasing  $m_{N_1}$. In particular, for $m_{N_1}=10^{10}$ GeV
(not displayed) this interval becomes extremely small. 
The latter effect
can be understood from the interplay of $\theta_2$
and $m_{N_1}$ on the relevant BAU parameters of Eq.~(\ref{eq:BAU_R}). 
While $\widetilde m_1$ is unchanged and as long as $m_{N_1} \lesssim
T_\mathrm{RH}$, the produced  
baryon asymmetry increases with $m_{N_1}$. 
For a given value of  $m_{N_1}$, the disappearance of the 
second window associated with larger values of $|\theta_2|$ ($\pi
\lesssim |\theta_2| \lesssim 3\,\pi/2$), is due to a stronger washout, 
which leads to values of  $n_\text{B}/n_\gamma$ below the viable BAU range
of Eq.~(\ref{BAUeffrange}). 

Finally, let us notice that the BAU-favoured ranges of $\theta_2$ imply
very distinct predictions for both the BRs, and the associated
$\theta_{13}$ sensitivity. Even though the BRs arising from 
the second $\theta_2$ window are significantly larger, 
in this case the sensitivity to $\theta_{13}$ is considerably reduced,
as is clearly manifest in Fig.~\ref{fig:modt2:argt2:SSW13:1214}.
All the previous facts taken into account, 
we will often
rely on the choice $m_{N_1}=10^{10}$ GeV and 
$\theta_2\,=\,0.05\,e^{0.2\,i}$ as a means of
ensuring a viable BAU scenario via a minimal deviation
from the $R = \mathbbm{1}$ case.

\subsubsection{Influence of $\theta_1$}
\begin{figure}[t]
  \begin{center} \hspace*{-15mm}
    \psfig{file=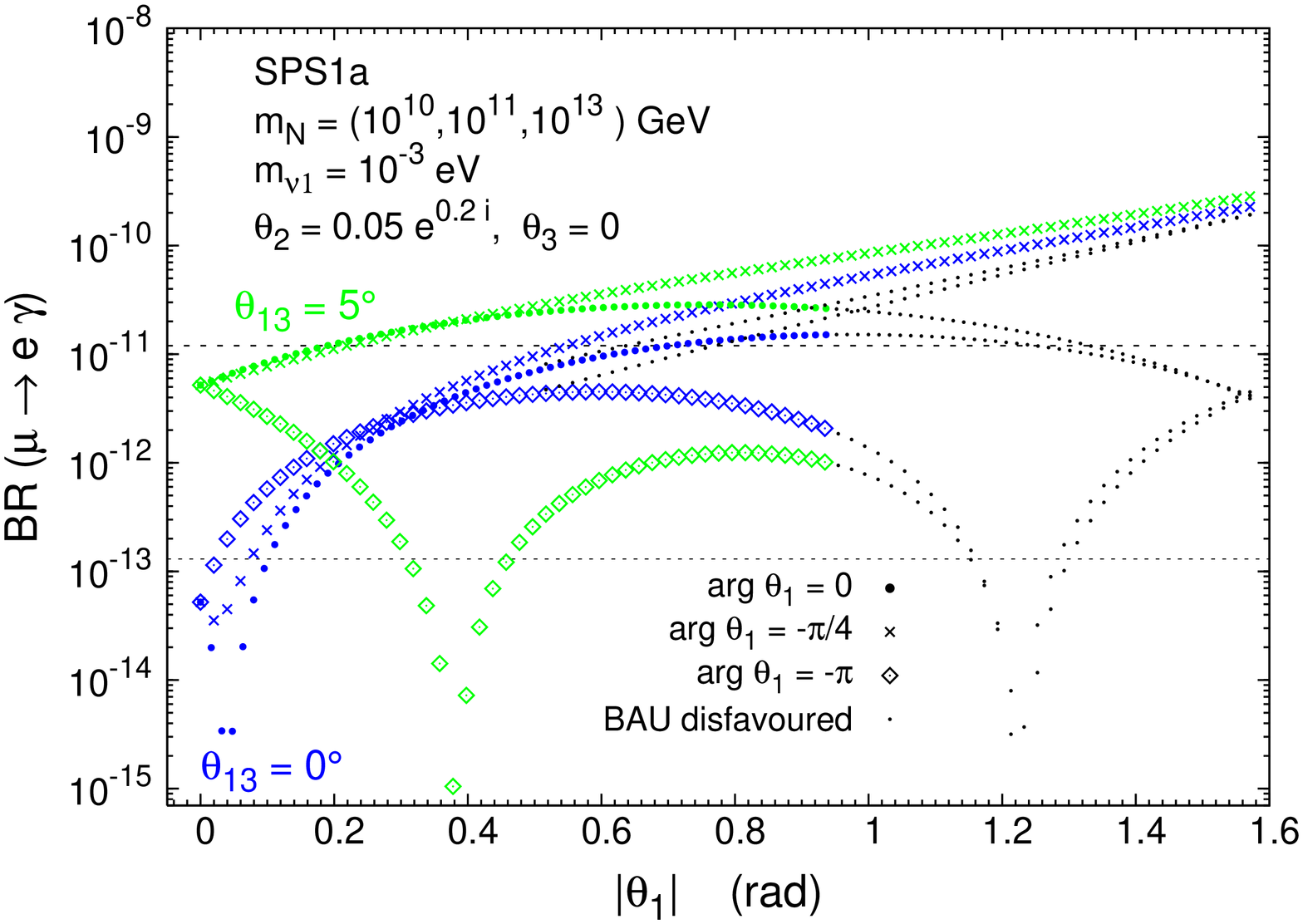,width=70mm,angle=270,clip=}
    \caption{BR($\mu \to e\, \gamma$) as a function
      of $|\theta_1|$, for $\arg
      \theta_1\,=\,\{0\,,\,-\pi/4\,,-\pi\}$ (dots, times, diamonds,
      respectively) and $\theta_{13}=0^\circ\,,5^\circ$ (blue/darker,
      green/lighter lines). 
      BAU is enabled by the choice 
      $\theta_2\,=0.05\,e^{0.2\,i}$ ($\theta_3\,=0$). 
      In all cases black dots
      represent points associated with a disfavoured BAU scenario and
      a dashed(dotted) 
      horizontal line denotes the present experimental bound (future
      sensitivity).}  
    \label{fig:modt1:argt1:1214}
  \end{center}
\end{figure}

\begin{figure}[t]
  \begin{center} \hspace*{-15mm}
    \psfig{file=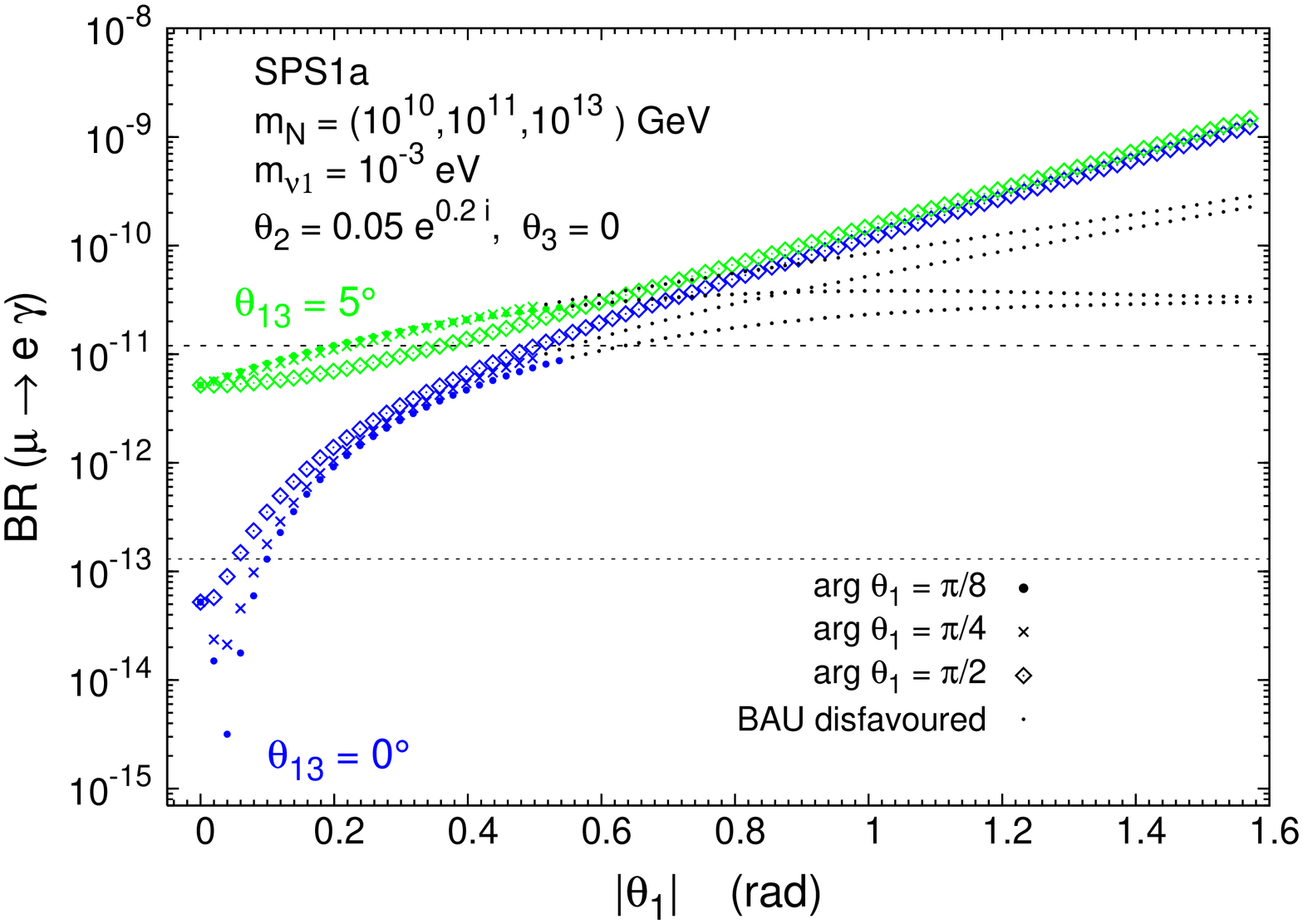,width=70mm,angle=270,clip=}
    \caption{BR($\mu \to e\, \gamma$) as a function
      of $|\theta_1|$, for $\arg
      \theta_1\,=\,\{\pi/8\,,\,\pi/4\,,\pi/2\}$ (dots, times, diamonds,
      respectively) and $\theta_{13}=0^\circ\,,5^\circ$ (blue/darker,
      green/lighter lines). 
      BAU is enabled by the choice 
      $\theta_2\,=0.05\,e^{0.2\,i}$ ($\theta_3\,=0$). 
      In all cases black dots
      represent points associated with a disfavoured BAU scenario and
      a dashed(dotted) 
      horizontal line denotes the present experimental bound (future
      sensitivity).}  
    \label{fig:modt1:argt1:1214:pos}
  \end{center}
\end{figure}

It has become clear from the previous analysis that a departure from
the $R = \mathbbm{1}$ case via non-vanishing values of $\theta_2$ can
significantly affect the BR sensitivity to $\theta_{13}$. 
Here we will show that $\theta_1$ plays an equally important role on
the present discussion.
In Figs.~\ref{fig:modt1:argt1:1214} and~\ref{fig:modt1:argt1:1214:pos} 
we display the BR($\mu \to e\, \gamma$) as a function of 
$|\theta_1|$, for different values of its argument. 

As already observed
in~\cite{Arganda:2005ji}, the effect of departing from the case $R =
\mathbbm{1}$ by varying $\theta_1$ leads to important 
additional contributions to the considered LFV decays.
Here, we have only presented the case $m_{N_3}=10^{13}$ GeV,
since for $m_{N_3}=10^{14}$ GeV the
experimental exclusion line is already crossed for very small values
of $\theta_1$ ($|\theta_1| \approx 0.1$). 
Opposed to the $\theta_2$ case, and as expected
from the analytical estimates, there is little dependence of the 
BR on the choice of the lightest neutrino mass\footnote{This
dependence is only manifest for $\theta_{13}\approx 0^\circ$ and
appears in terms proportional to $m_{N_2}$, so that it is considerably
suppressed. On the other hand, 
and as it occurred for $\theta_2$, larger values of 
$m_{\nu_1}$ widen the range of $\theta_1$
for which a viable BAU scenario can be obtained.}.
Considering the other SPS scenarios leads to analogous results, the
only difference lying in a global rescaling of the BR($\mu \to
e\, \gamma$), and the discussion is similar to that regarding
$\theta_2$. 

In the case of negative arguments, the influence of $\theta_1$ is
shown in Fig.~\ref{fig:modt1:argt1:1214}.
Notice that in all cases, for extremely small values of $|\theta_1|$
($|\theta_1| \lesssim 0.1$), we again recover for $\theta_{13}=5^\circ$ 
BRs which are larger, and clearly distinguishable
from the $\theta_{13}=0^\circ$ case.
In contrast, for a large (negative) 
$\arg \theta_1$, the situation is reversed, and 
the predictions for BR($\mu \to e\, \gamma$)
associated to $\theta_{13}=5^\circ$ are actually smaller than for 
$\theta_{13}=0^\circ$. 
This becomes manifest when $\arg \theta_1\lesssim -\pi/2$, a regime
for which the BR starts decreasing with increasing $|\theta_1|$. 
For real (and negative)
$\theta_1$, (i.e. $\arg \theta_1=-\pi$) the effect is such that 
for $\theta_{13} =5^\circ$ two local minima of the BR are present
(although one disfavoured by BAU), both with an associated value of
the BR below the planned MEG sensitivity
(for this specific choice of SPS point and seesaw parameters).
These ``dips'' reflect a
cancellation between the terms proportional to $m_{\nu_2}$ and
$m_{\nu_3}$ (see Eq.~(\ref{Y21:LLog})),
which in fact is also present for
$\theta_{13} \approx 0^\circ$, albeit only for the second, BAU-disfavoured,
$|\theta_1|$ value. 
It is worth pointing out that this apparent accidental cancellation
for a specific choice of the $R$-matrix parameters could correspond to the
occurrence of texture zeros in the neutrino Yukawa couplings 
(motivated, for instance, from some flavour symmetry, or arising within
specific seesaw models\footnote{
  At this point, we find it interesting to mention the connection between the 
  sensitivity (or lack thereof) of the BR$(\mu \to e \,\gamma)$ to
  $\theta_{13}$ in terms of the neutrino Yukawa couplings, as predicted 
  within the framework of Sequential 
  Dominance~\cite{King:2002nf} models.
  For the case of ``Heavy Sequential Dominance'', 
  $\theta_{13}$ is predicted as 
  $\theta_{13} \approx |(Y_\nu)_{31}| /\sqrt{2}|(Y_\nu)_{32}| +
  f((Y_\nu)_{2i})$, where $f$ is a function of the $(Y_\nu)_{2i}$
  couplings.
  If the predicted $\theta_{13}$ is driven by 
  the first term, and since the BR$(\mu
  \to e\, \gamma) \sim (Y^\dagger_\nu)_{23}  
  m_{N_3} L_{33} (Y_\nu)_{31}$, then there is a direct connection between
  $\theta_{13}$ and BR$(\mu \to e\, \gamma)$. This is, for
  example, what happens if we choose $R=\mathbbm{1}$. In
  contrast, if the second term dominates the predicted $\theta_{13}$ value,
  the connection of BR$(\mu \to e\, \gamma)$ to $\theta_{13}$ is 
  lost~\cite{Blazek:2002wq}. This corresponds to the ``dip'' of the BR$(\mu
  \to e\, \gamma)$ in Fig.~\ref{fig:modt1:argt1:1214}.}). 
Although not stable under RGE effects, these zeros effectively translate into 
very small entries in the Yukawa couplings, 
which can account for the observed suppression of the
BR~\cite{Arganda:2005ji} corresponding to the ``dips'' in 
Fig.~\ref{fig:modt1:argt1:1214}. We would like to remark that, generically,
the position and depth of these ``dips'' depend on the chosen values of 
all the seesaw parameters.

In Fig.~\ref{fig:modt1:argt1:1214:pos} we present a few examples of
$\arg \theta_1 >0$. In this case, the discussion of the BRs and
sensitivity to $\theta_{13}$ is very similar to that conducted for 
small negative arguments. That is, for small $|\theta_1|$ values, the
predictions for the 
two $\theta_{13}$ cases are clearly distinguishable. On the other
hand, and irrespective of the argument (positive or negative),
for sufficiently large $|\theta_1|$, the lines corresponding to the
cases $\theta_{13}=0^\circ$ and 5$^\circ$ eventually meet, and thus 
for this choice of parameters the 
sensitivity of the BR to $\theta_{13}$ is lost. 

Another relevant aspect to be inferred from
Figs.~\ref{fig:modt1:argt1:1214} and~\ref{fig:modt1:argt1:1214:pos} 
is how $\theta_1$
affects the BAU predictions enabled by $\theta_2$. 
Unlike what occurs for $\theta_2$ and $\theta_3$, the role of $\theta_1$
in accounting for the observed BAU is somewhat more indirect. In
particular, and as 
mentioned in Section~\ref{bau-theory}, $\theta_1$ 
essentially deforms the favoured BAU areas in the
$\theta_2-\theta_3$ plane. 
For instance, and for the chosen BAU-enabling $\theta_2$ value in
Fig.~\ref{fig:modt1:argt1:1214}, a real value of 
$\theta_1$ larger than 0.9 leads to an 
estimated $n_\text{B}/n_\gamma$  which is no longer within the 
viable BAU range of Eq.~(\ref{BAUeffrange}).
A distinct situation occurs for the cases  
$\arg \theta_1=-\pi/4$ and $\pi/2$, where the entire $|\theta_1|$ range
successfully accounts for $n_\text{B}/n_\gamma$ within
$[10^{-10},10^{-9}]$.

\vspace*{8mm}
To conclude this subsection, let us add two further comments.
Regarding the influence of 
$\theta_3$ it suffices to mention that although relevant
with respect to BAU (see Fig.~\ref{fig:bau:modt3:argt3}), 
we have not found a significant BR($\mu \to e\, \gamma$) dependence on
the latter parameter. 
This is a consequence of having 
the Yukawa couplings to the heaviest right-handed neutrino dominating,
since a $\theta_3$ $R$-matrix rotation leaves  unchanged the
couplings $(Y_\nu)_{i3}$. 
In this case, the sensitivity to $\theta_{13}$ is very similar to what
was found for the $R=\mathbbm{1}$ case. In the remaining analysis we will
fix $\theta_3=0$.

Concerning the EDMs, which are clearly non-vanishing in the presence
of complex $\theta_i$, we have numerically checked that for all the
explored parameter space, the predicted values for the electron, muon
and tau EDMs are well below the experimental bounds given in
Eq.(\ref{EDM:exp}).

\subsection{Dependence on other relevant parameters: 
$\pmb{m_{N_3}}$ and $\pmb{\tan \beta}$}

\begin{figure}[t]
  \begin{center} \hspace*{-10mm}
     \psfig{file=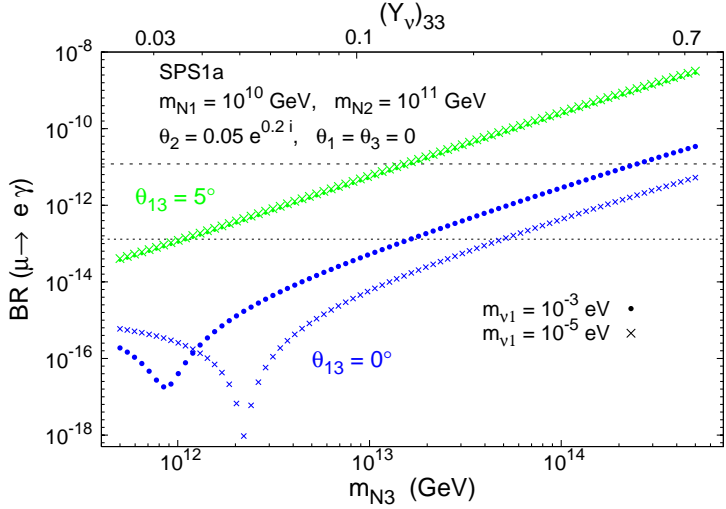,width=70mm,angle=270,clip=}
  \caption{BR($\mu \to e\, \gamma$) as a function of 
      $m_{N_3}$ for SPS 1a, with
      $m_{\nu_1}\,=\,10^{-5}$~eV and $m_{\nu_1}\,=\,10^{-3}$~eV (times, dots,
      respectively), and
      $\theta_{13}=0^\circ,\,5^\circ$ (blue/darker, green/lighter lines). 
      BAU is enabled by the choice 
      $\theta_2\,=0.05\,e^{0.2\,i}$ ($\theta_1=\theta_3=0$). 
      On the upper horizontal axis we display the associated value of
      $(Y_\nu)_{33}$. A dashed (dotted)
      horizontal line denotes the present experimental bound (future
      sensitivity).}  
    \label{fig:MN3:MN2:SPS1a}
  \end{center}
\end{figure}

Throughout the discussion regarding the dependence of the branching ratios
on the $R$-matrix complex angles, it has often been stressed that the
leading contributions to the BRs were those proportional to the mass
of the heaviest right-handed neutrino, $m_{N_3}$. This is indeed the most
relevant parameter. 
Here, and to briefly summarise the effect of $m_{N_3}$,
let us present the predictions for BR($\mu \to e\, \gamma$) as a function
of the latter mass, while keeping $m_{N_1}$ and $m_{N_2}$ fixed.
We have checked that the BRs do not significantly depend on 
$m_{N_1}$ and $m_{N_2}$, apart from one exception for 
$m_{N_2}$, which we will later comment.
The results for SPS 1a are displayed in Fig.~\ref{fig:MN3:MN2:SPS1a}.
For completeness, we have included in the upper horizontal axis
the associated value of $(Y_\nu)_{33}$ (with similar values being obtained
for $(Y_\nu)_{32}$).

We find from Fig.~\ref{fig:MN3:MN2:SPS1a} that the full RGE result
grows with $m_{N_3}$ in a very similar fashion to that 
predicted by the LLog approximation, i.e.
$m_{N_3}^2\,\log^2 m_{N_3}$.
It is clear that without a
predictive theoretical framework for $m_{N_3}$ (e.g. GUT models) or
indirect experimental evidence for the scale of the seesaw mechanism,
there is a large uncertainty regarding the value of $m_{N_3}$. 
Within our chosen scenario of hierarchical heavy neutrinos 
($m_{N_1}\ll m_{N_2} \ll m_{N_3}$), assuming that the observed BAU is 
generated via a mechanism of thermal leptogenesis (with $m_{N_1}
\gtrsim 10^{9}$ GeV), and given the gauge coupling unification 
scale\footnote{The possibility of larger LFV effects arising
  from the existence of a higher energy scale,
  e.g. $M_\text{Planck}$, has been addressed by other authors. See, for
  instance~\cite{Illana:2003pj}.} ($M_X \approx 2 \times
10^{16}$ GeV), 
the natural choice for $m_{N_3}$ would lie in the range
[$10^{10}$ GeV, $10^{15}$ GeV].
It is obvious from Fig.~\ref{fig:MN3:MN2:SPS1a} 
that such an uncertainty in $m_{N_3}$ translates into
predictions for the BR ranging over many orders of magnitude.
Hence, one can at most extract an upper bound on $m_{N_3}$ for the chosen set of
input parameters. For instance, in Fig.~\ref{fig:MN3:MN2:SPS1a}, 
$m_{N_3}\, \gtrsim\, 10^{13}\, (10^{14})$ GeV 
is not allowed by the present experimental bounds on the 
BR($\mu \to e\, \gamma$) for $\theta_{13}=5^\circ \,(0^\circ)$.
Notice that, although the sensitivity to $\theta_{13}$ is clearly 
displayed in Fig.~\ref{fig:MN3:MN2:SPS1a} 
(with more than two orders of magnitude separation of the 
$\theta_{13}=0^\circ$ and $5^\circ$ lines), without
additional knowledge of $m_{N_3}$ it will be 
very difficult to disentangle the several $\theta_{13}$ cases.
However, this argument can be reversed. This strong dependence on
$m_{N_3}$ could indeed be used to derive hints on $m_{N_3}$ 
from a potential BR measurement. We will return to this type of
considerations in the following section.

It is also worth commenting on the local minima appearing in
Fig.~\ref{fig:MN3:MN2:SPS1a} for the lines
associated with $\theta_{13}=0^\circ$. As mentioned before, these ``dips'' 
are induced by the effect of the running of $\theta_{13}$, shifting it
from zero to a negative value.
In the LLog approximation, 
the ``dips'' can be understood from Eq.~(\ref{Y21:LLog}) as
a cancellation between the terms proportional to $m_{N_3}\,L_{33}$ and 
$m_{N_2}\,L_{22}$ in the limit $\theta_{13}(m_M) \to 0^-$ 
(with $\theta_1=\theta_3=0$).
The depth of the minimum is larger for smaller $m_{\nu_1}$,
as visible in Fig.~\ref{fig:MN3:MN2:SPS1a}. We have also checked that
an analogous effect 
takes place when one investigates the dependence of BR($\mu \to e\,
\gamma$) on $m_{N_2}$. It is only in this limit $\theta_{13}(m_M) \to 0^-$,
and in the vicinity of the ``dip'', that $m_{N_2}$ can visibly
affect the BRs.

Regarding the other SPS points, with the exception of SPS 3 and 5,
the results from the full RGE computation (not displayed here)
are also in good agreement with the LLog approximation.
The predicted BRs for SPS 3 are found to be larger than those of the 
LLog by a factor of approximately 3.
This divergence is due to the fact that in the
LLog approximation the effects of $M_{1/2}$ in the running of the
soft-breaking parameters of Eq.~(\ref{misalignment_sleptons}) 
are not taken into account. Therefore, for low $M_0$ and large
$M_{1/2}$ (as is the case of SPS 3), 
there is a significant difference between the results of
the full and approximate computations, as previously noted 
by~\cite{Petcov:2003zb,Chankowski:2004jc}. 
Moreover, this difference becomes more evident 
for low values of $\tan \beta$.

\begin{figure}[t]
  \begin{center} \hspace*{-10mm} 
\begin{tabular}{cc}
	\psfig{file=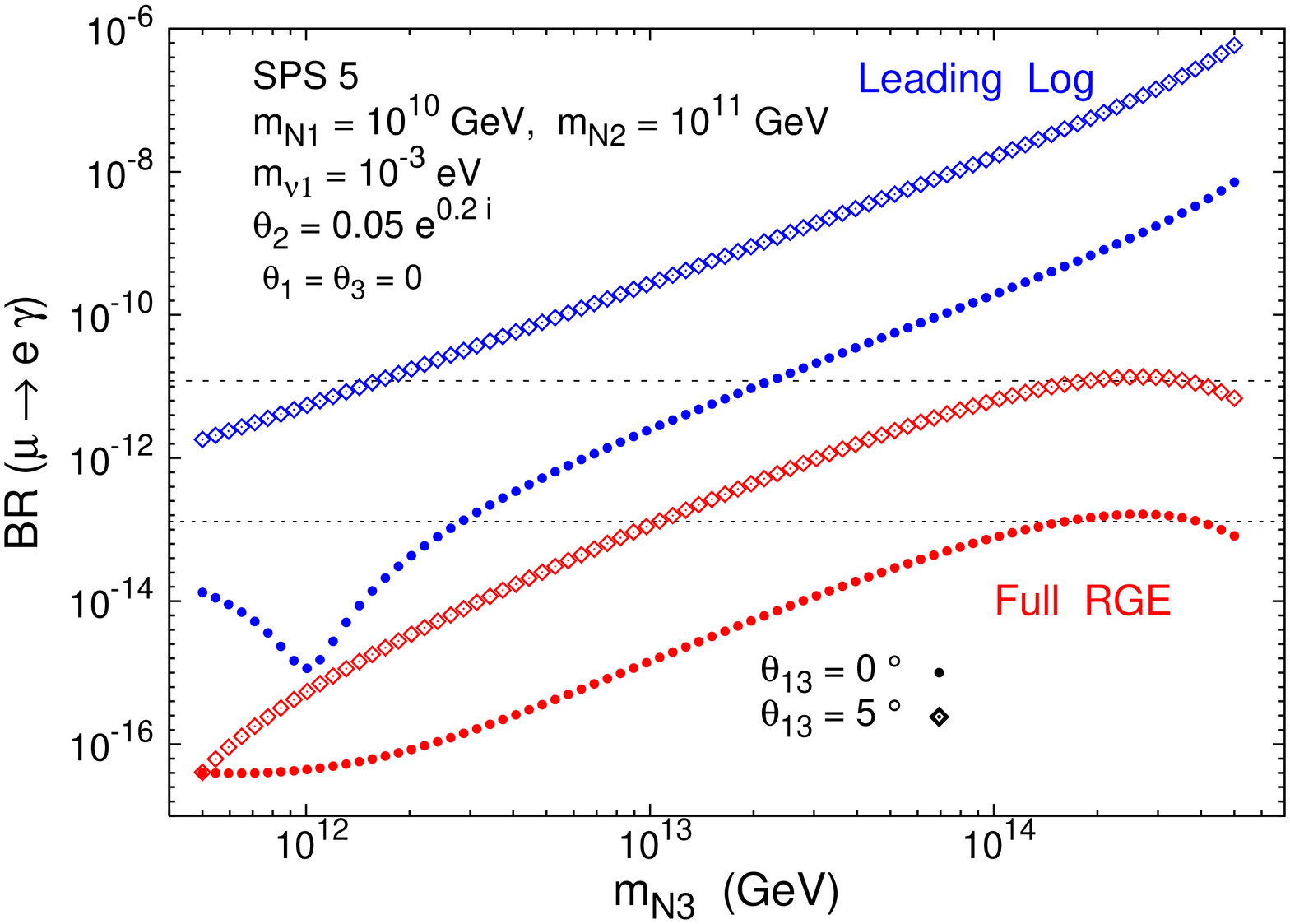,width=60mm,angle=270,clip=}
	&
\psfig{file=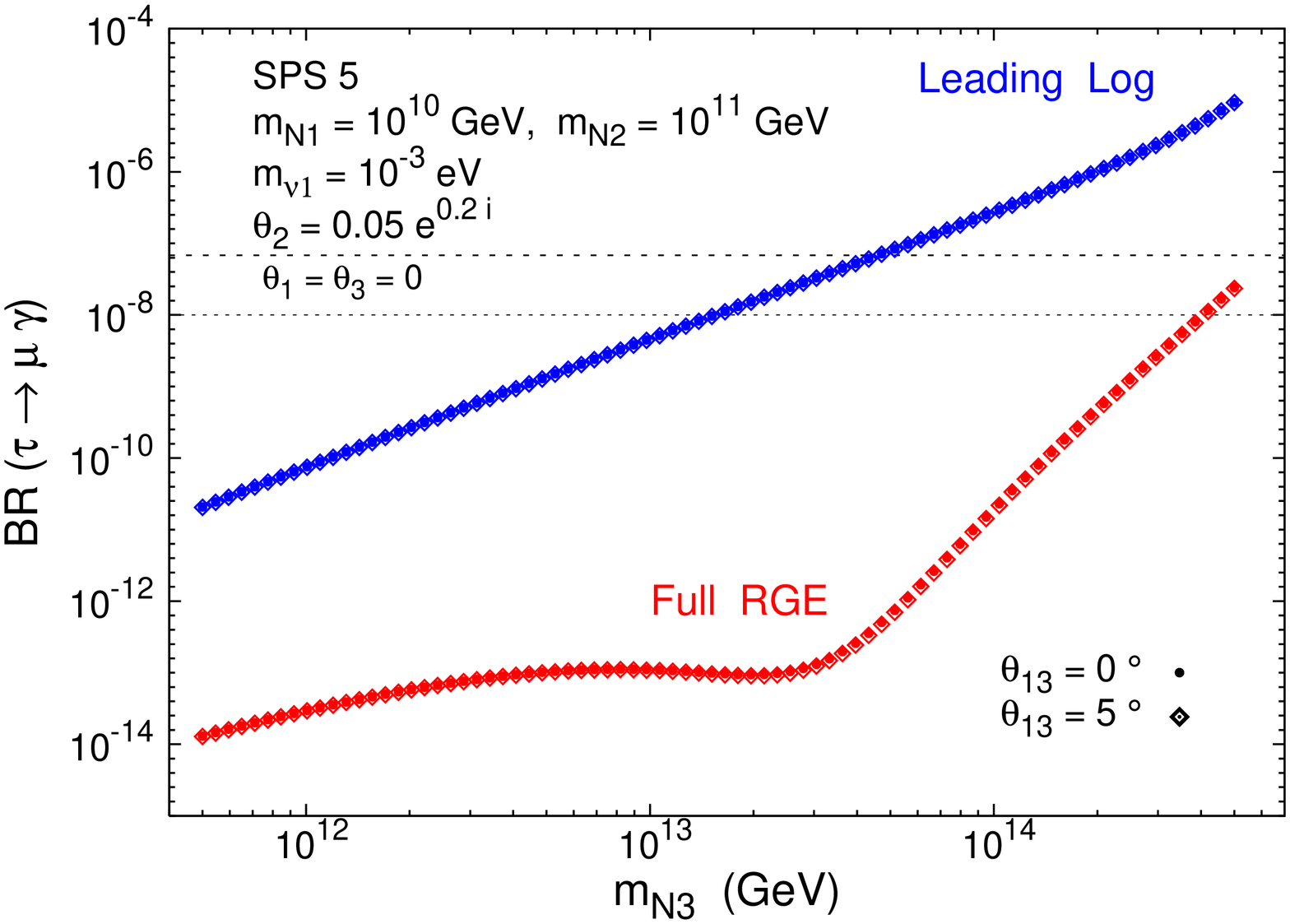,width=60mm,angle=270,clip=}
\end{tabular}
    \caption{Prediction for BR($\mu \to e\, \gamma$) 
      and BR($\tau \to \mu\, \gamma$) as a function of 
      $m_{N_3}$ for SPS 5, using the LLog approximation (upper blue
      lines) and full RGE (lower red lines).
      $\theta_{13}=0^\circ,\,5^\circ$ (dots, diamonds, respectively). 
      BAU is enabled by the choice 
      $\theta_2\,=0.05\,e^{0.2\,i}$ ($\theta_1=\theta_3=0$).  A dashed
      (dotted)
      horizontal line denotes the present experimental bound (future
      sensitivity).} 
    \label{fig:MN3:LLog:SPS5}
  \end{center}
\end{figure}
The divergence of the two computations is more dramatic for SPS 5.
This is shown in Fig.~\ref{fig:MN3:LLog:SPS5},
where we compare the dependence of the BR($\mu \to e\,
\gamma$) and BR($\tau \to \mu\,\gamma$) on 
$m_{N_3}$, as given from the full computation,
and in the leading log approximation. The latter approximation 
over-estimates by more than four orders of magnitude the
values of the BR($\mu \to e\,\gamma$). The full RGE and LLog results
diverge even more regarding the BR($\tau \to \mu\,\gamma$), with a
separation that can be as large as five orders of magnitude.
It is also manifest from Fig.~\ref{fig:MN3:LLog:SPS5} 
that the qualitative behaviour of the
full results with
respect to $m_{N_3}$ is no longer given by $m_{N_3}^2\,\log^2 m_{N_3}$.
The reason for this divergence is associated to the large negative
value of the trilinear coupling\footnote{The effect of the sign of
  $A_0$ in the failure of the LLog approximation has already been
  discussed in~\cite{Petcov:2003zb}.}, $A_0$. We considered other large
negative values of $A_0$, in all cases leading to the same conclusion.
Taking large positive $A_0$ also leads to an important, albeit not as large,
separation (for instance, three orders of magnitude for $A_0=1000$ GeV). 

\begin{figure}[t]
  \begin{center} \hspace*{-10mm}
    \begin{tabular}{cc}
      \psfig{file=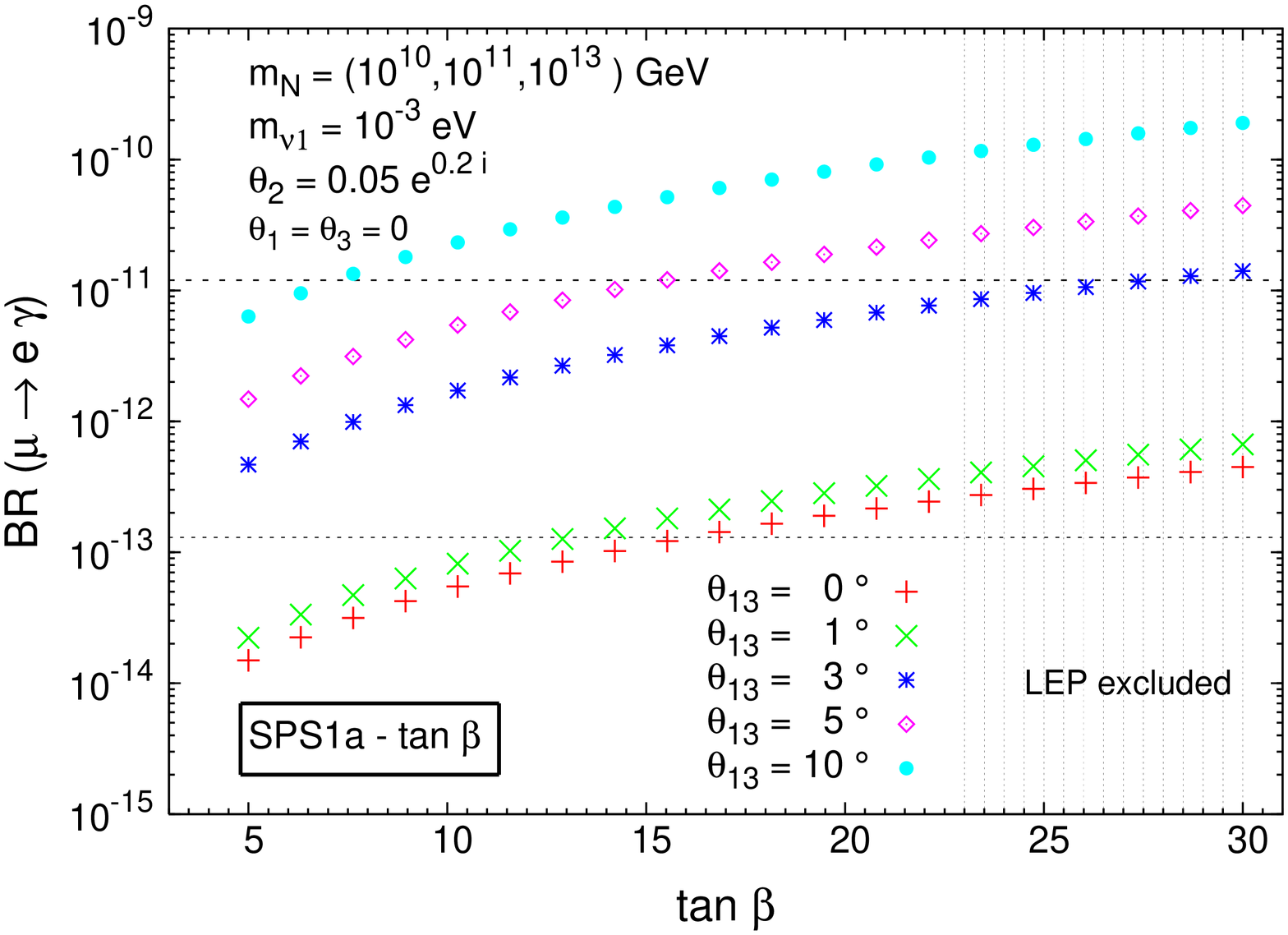,width=60mm,angle=270,clip=} &
      \psfig{file=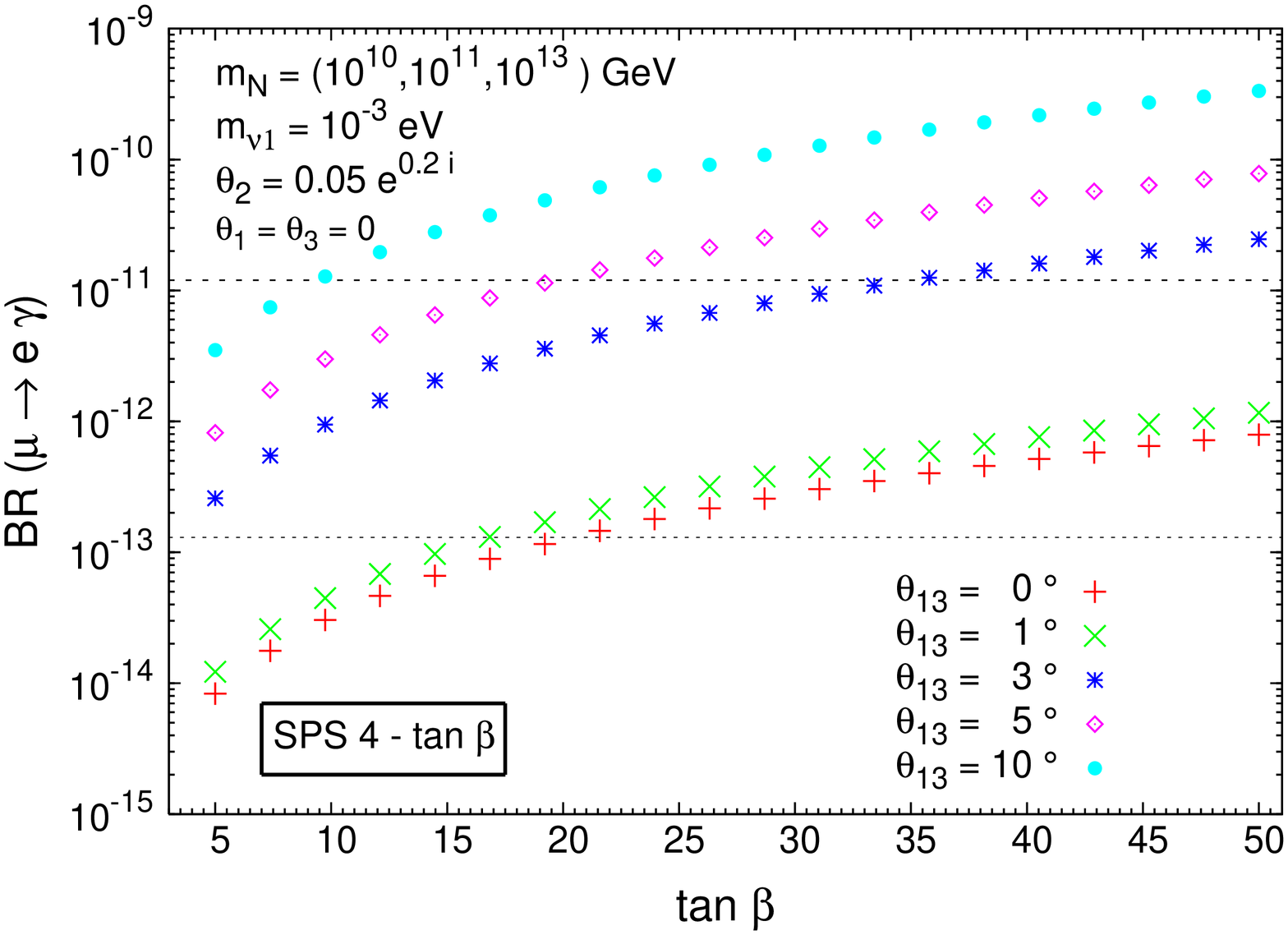,width=60mm,angle=270,clip=} 
    \end{tabular}
    \caption{BR($\mu \to e\, \gamma$) as a function of 
      $\tan \beta$, for  $m_{N}\, =\, (10^{10},\,10^{11},\,10^{13})$ GeV, 
      for SPS 1a (left) and SPS 4
      (right). $\theta_{13}= 0^\circ,1^\circ,3^\circ,5^\circ$, and $10^\circ$ 
      (crosses, times, asterisks,
      diamonds and dots, respectively).
      BAU is enabled by the choice 
      $\theta_2\,=0.05\,e^{0.2\,i}$ ($\theta_1=\theta_3=0$). A dashed (dotted)
      horizontal line denotes the experimental bound (future
      sensitivity). Vertical shaded regions correspond to regions
      with spectra excluded by LEP data.} 
    \label{fig:tanbeta:SPS1a4}
  \end{center}
\end{figure}

Finally, we briefly comment on the BR dependence on $\tan \beta$. 
As it is well known, the BRs approximately 
grow as $\tan^2 \beta$, and therefore this is also a relevant parameter. 
In Fig.~\ref{fig:tanbeta:SPS1a4}, we
plot a generalisation of the SPS points 1a and 4 (defined by 
$M_0$, $M_{1/2}$, $A_0$ and $\text{sign\,} \mu$) with free 
$\tan \beta$, and present the
sensitivity of the branching ratios to distinct values of $\theta_{13}$. 
Again, as can be seen in Fig.~\ref{fig:tanbeta:SPS1a4}, the sensitivity
to $\theta_{13}$ is clearly manifest, in the sense that for a given
$\tan \beta$ the predictions for the BRs are very distinct for different
$\theta_{13}$ values.
However, the $\tan \beta$ dependence is so important that two $\tan
\beta$ values, for instance 10 and 20, 
lead to predictions of the BR that diverge as much as
those one obtains from the comparison of $\theta_{13}=3^\circ$ and
$5^\circ$ (for a fixed value of $\tan \beta$).
This implies that unless the experimental range for $\tan \beta$
is far more constrained than at present,
we cannot conclude about the allowed/disallowed $\theta_{13}$ values 
from the present $\mu \to e\, \gamma$ bounds.
Just like as argued for $m_{N_3}$, the strong BR dependence 
on $\tan \beta$ can be constructively used to further constrain $\tan \beta$ 
from a potential BR($\mu \to e\, \gamma$) measurement. We will address
this topic in the following section.

\section{Experimental prospects: hints on SUSY and Seesaw parameters
  from measuring $\theta_{13}$ and BRs}\label{prospects:hints}

In the previous section, we analysed how the several free parameters
of the SUSY seesaw scenario affect the predictions for the 
BR($\mu \to e\, \gamma$). We also emphasised how the sensitivity of
the latter ratios to $\theta_{13}$ 
can be altered by the uncertainty introduced from the indetermination of
$\theta_i$, $\tan \beta$ and, most of all, $m_{N_3}$. The question we
aim to address in this section is whether a joint measurement of the
BRs and $\theta_{13}$ can shed some light on apparently unreachable
parameters, like $m_{N_3}$. 

The expected improvement in the experimental sensitivity to the LFV
ratios (see Table~\ref{LFV:bounds:future}) support the possibility that 
a BR be measured in the future, thus
providing the first experimental evidence for new
physics, even before its discovery at the LHC.
The prospects are especially encouraging regarding $\mu \to e\,
\gamma$, where the sensitivity will improve by at least two
orders of magnitude.
Moreover, and given the impressive
effort on experimental neutrino physics, a measurement
of $\theta_{13}$ will likely also occur in the 
future~\cite{Ables:MINOS:2004,Komatsu:2002sz,Migliozzi:2003pw,Huber:2006vr,Itow:2001ee,Blondel:2006su,Huber:2006wb,Burguet-Castell:2005pa,Campagne:2006yx}.  
In what follows, let us envisage a future ``toy''-like scenario,
where we will assume the following hypothesis: 
(i) measurement of BR($\mu \to e\, \gamma$);
(ii) measurement of $\theta_{13}$;
(iii) discovery of SUSY at the LHC, with a given spectrum.
Furthermore, we assume that BAU is explained via thermal
leptogenesis, with a hierarchical heavy-neutrino spectrum. 

\begin{figure}[t]
  \begin{center} \hspace*{-10mm}
    \begin{tabular}{cc}
	\psfig{file=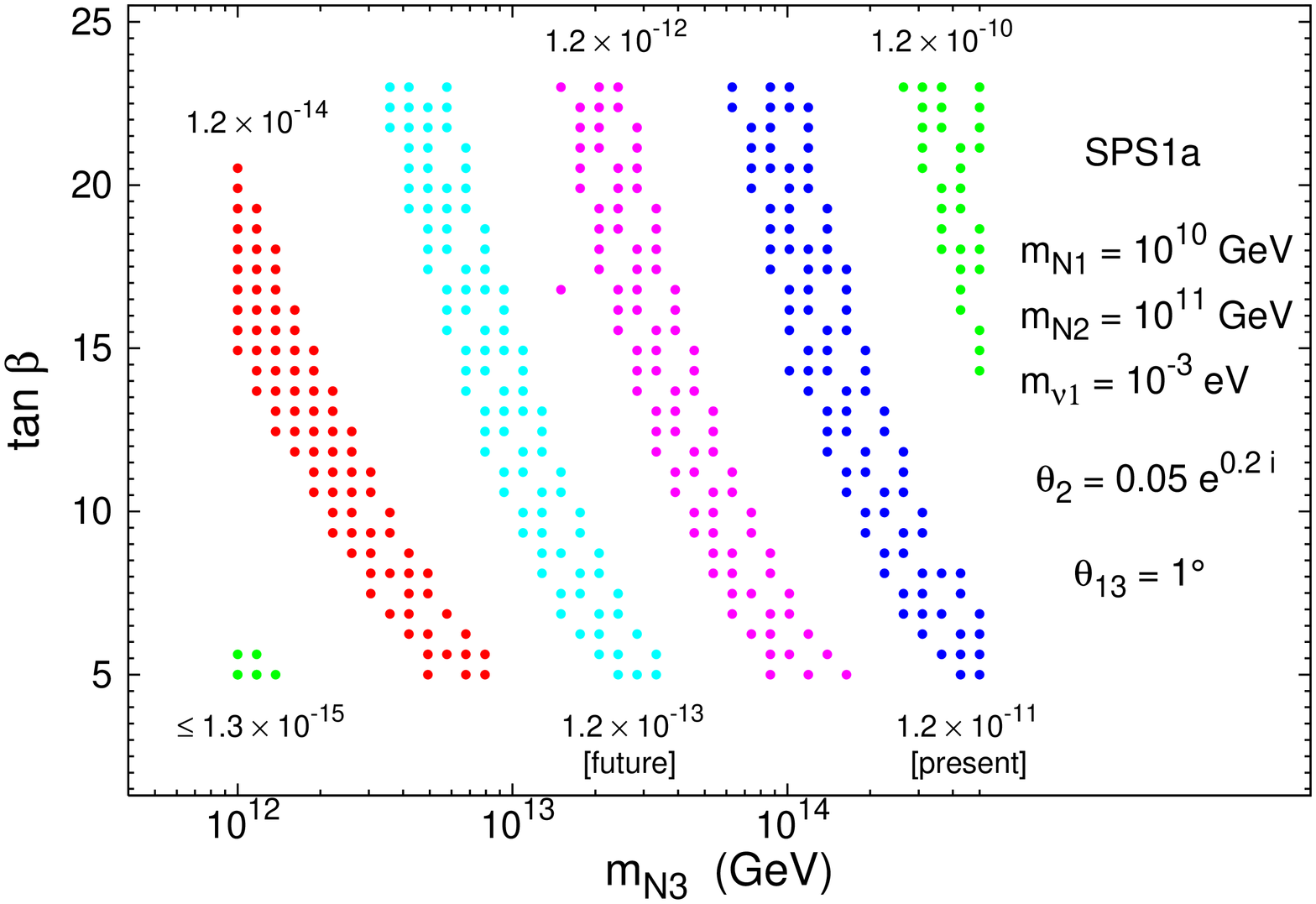,width=60mm,angle=270,clip=} &
	\psfig{file=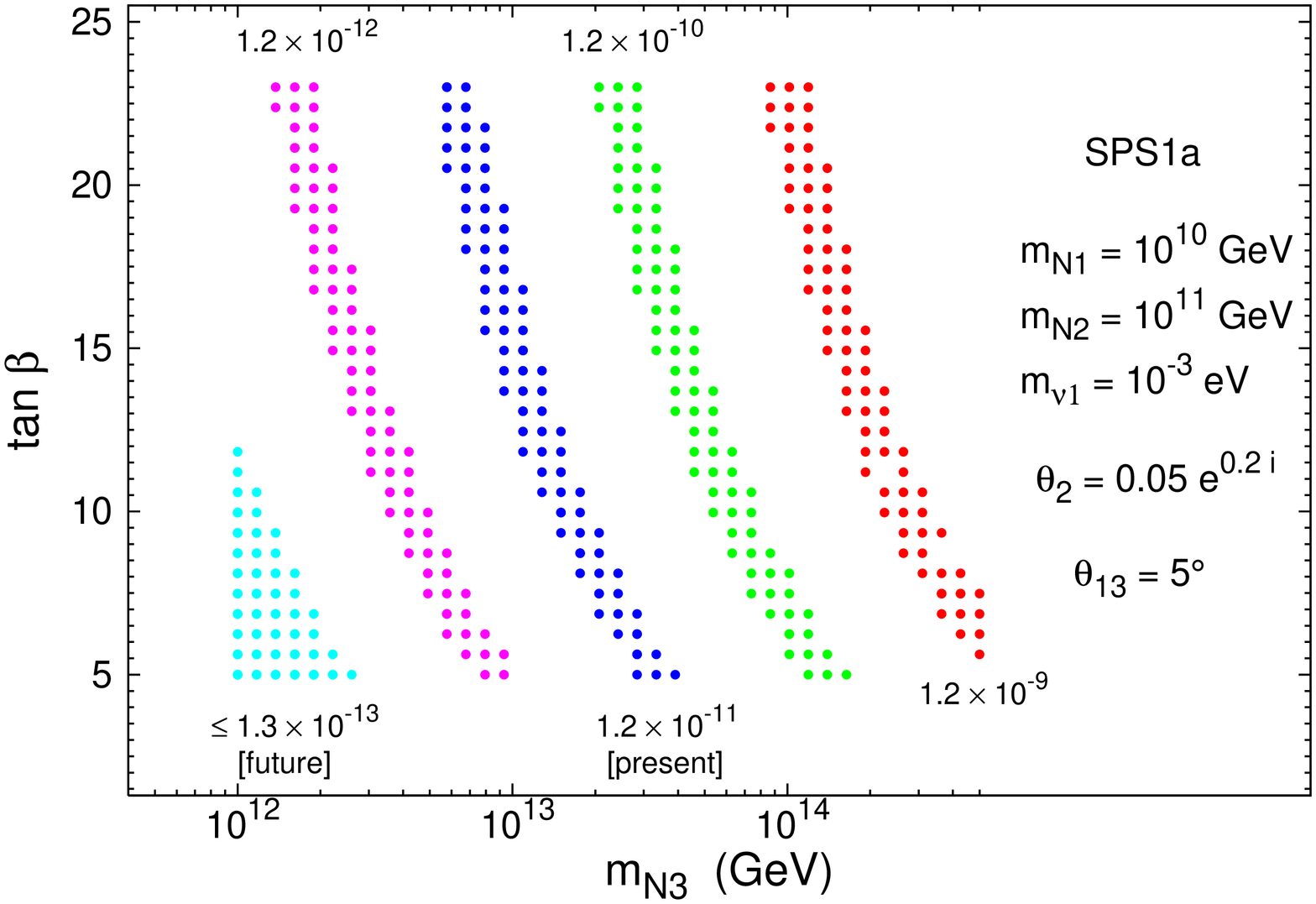,width=60mm,angle=270,clip=} 
    \end{tabular}
    \caption{Contours of BR($\mu \to e\, \gamma$)$= (1.2 \pm
      0.1) \times 10^{-n}$, with  $n=10,\dots, 15$ in the $m_{N_3} - \tan \beta$ plane, 
      for a generalised SPS 1a scenario. 
      We assume $\theta_{13}=1^\circ \pm
    0.1^\circ\, (5^\circ \pm0.5^\circ)$ 
    on the left (right) panel. BAU is enabled by the choice 
      $\theta_2\,=0.05\,e^{0.2\,i}$ ($\theta_1=\theta_3=0$). The
    current experimental bound is associated with the darkest (blue)
    surface, while the future sensitivity is represented by the
    lightest (cyan) one.} 
    \label{fig:M3:tanb:BR:SPS1a:t2:5d}
  \end{center}
\end{figure}

Under the above conditions, let us conduct the following exercise.
First, choosing SPS~1a, $m_{N_1}=10^{10}$ GeV, $m_{N_2}=10^{11}$ GeV,
$m_{\nu_1}=10^{-3}$ eV, $\theta_2\,=0.05\,e^{0.2\,i}$ (a minimal
BAU-enabling deviation from the $R=\mathbbm{1}$ case), 
and with $\theta_{13}$ set to $1^\circ$ ($\pm
0.1^\circ$) and to $5^\circ$ ($\pm
0.5^\circ$), we predict the BRs as a function of $\tan \beta$
and $m_{N_3}$. We then plot the contour lines for
constant BR values in the $m_{N_3} - \tan \beta$ plane. In
Fig.~\ref{fig:M3:tanb:BR:SPS1a:t2:5d} we display the corresponding 
contours for the central values of  
$1.2\times 10^{-n}$ with $n=10,\dots, 15$, allowing for a 10\%
spread-out around these values. 
The predicted contours should be compared with the 
present bound and future sensitivity of
$1.2\times 10^{-11}$~\cite{Brooks:1999pu} and $1.3\times
10^{-13}$~\cite{mue:Ritt}, respectively.  

Given a potential SUSY discovery, 
the implications of a measurement of BR($\mu \to e\, \gamma$) and
$\theta_{13}$ are clearly manifest in Fig.~\ref{fig:M3:tanb:BR:SPS1a:t2:5d}.
From this figure we first learn that, even in the absence of an experimental
determination of $\tan \beta$, a potential measurement of BR($\mu \to
e\, \gamma$) and $\theta_{13}$ will allow to constrain $m_{N_3}$.   
For example, an hypothetical measurement of BR($\mu \to e\, \gamma$)$
\approx 1.2 (\pm 0.1)\times 10^{-12}$ 
would point towards the following allowed ranges of $m_{N_3}$: 
\begin{align}\label{BR:t13:mn3}
&\theta_{13}\,\approx \,1^\circ\, \Rightarrow\,\,\,
2\, \times\,10^{13}\,\,\text{GeV}\,\lesssim \, m_{N_3}
\lesssim \,2\, \times\,10^{14}\,\,\text{GeV}\,, \nonumber\\
&\theta_{13}\,\approx \,5^\circ\, \Rightarrow\,\,\,
1.5\, \times\,10^{12}\,\,\text{GeV}\,\lesssim \, m_{N_3}
\lesssim \,10^{13}\,\,\text{GeV}\,.
\end{align}
Other assumptions for the BRs would equally lead to an order of
magnitude interval for the constrained values of $m_{N_3}$.
If in addition to the s-spectrum, we assume that 
$\tan \beta$ is experimentally 
determined, then the intervals for $m_{N_3}$ presented in
Eq.~(\ref{BR:t13:mn3}) can be significantly reduced. For instance,
assuming that SPS 1a is indeed reconstructed (that is, $\tan
\beta=10$), then we would find  
\begin{align}\label{BR:t13:tanb:mn3}
&\theta_{13}\,\approx \,1^\circ\, \Rightarrow\,\,\,
4\, \times\,10^{13}\,\,\text{GeV}\,\lesssim \, m_{N_3}
\lesssim \,7\, \times\,10^{13}\,\,\text{GeV}\,, \nonumber\\
&\theta_{13}\,\approx \,5^\circ\, \Rightarrow\,\,\,
3\, \times\,10^{12}\,\,\text{GeV}\,\lesssim \, m_{N_3}
\lesssim 5\, \times\,10^{12}\,\,\text{GeV}\,. 
\end{align}

\begin{figure}[t]
  \begin{center} \hspace*{-10mm}
    \begin{tabular}{cc}
	\psfig{file=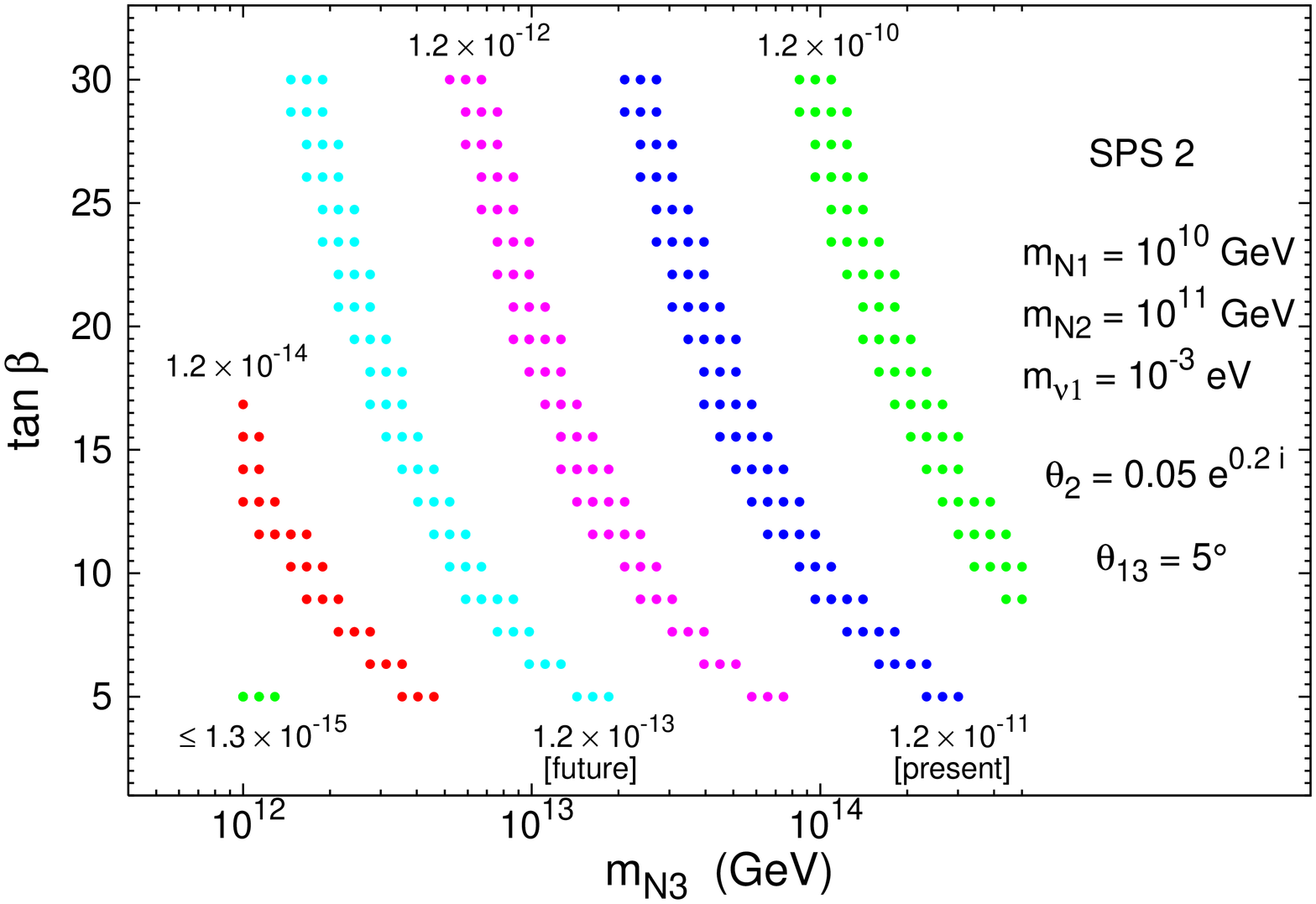,width=60mm,angle=270,clip=} &
	\psfig{file=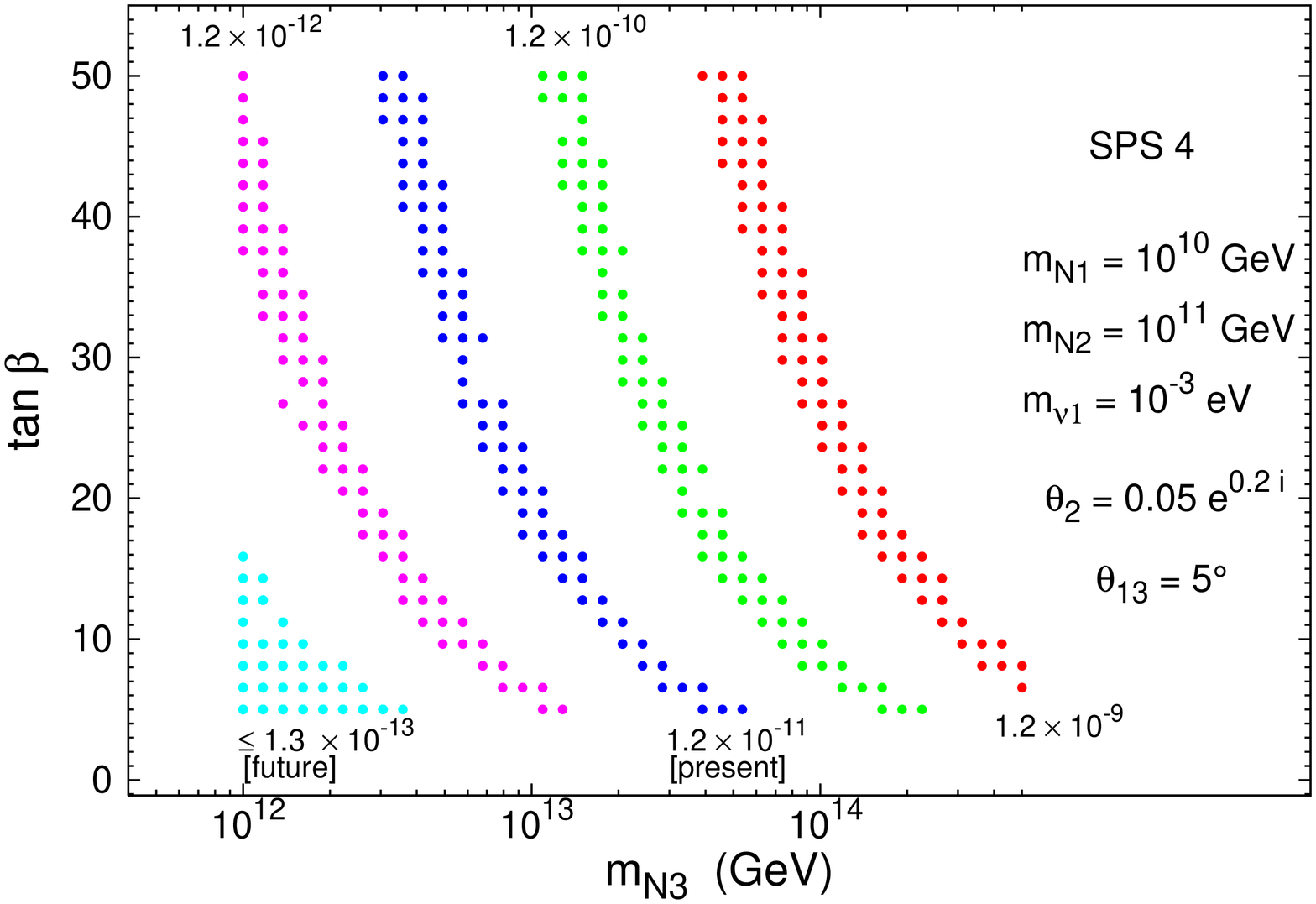,width=60mm,angle=270,clip=} 
    \end{tabular}
    \caption{Contours of BR($\mu \to e\, \gamma$)$= (1.2 \pm
      0.1) \times 10^{-n}$, with  $n=10,\dots, 15$ in the $m_{N_3} - \tan
      \beta$
      plane for a generalised 
      SPS 2 (left) and SPS 4 (right) scenarios. 
      We assume $\theta_{13}=5^\circ \pm 0.5^\circ$.
      BAU is enabled by the choice 
      $\theta_2\,=0.05\,e^{0.2\,i}$ ($\theta_1=\theta_3=0$).}
    \label{fig:M3:tanb:BR:SPS4:t2:5d}
  \end{center}
\end{figure}

The hypothetical reconstruction of any other SPS-like scenario would
lead to similar one order of magnitude intervals for $m_{N_3}$ 
but with distinct $m_{N_3}$ central values.  
As expected, for the same BR and $\theta_{13}$ measurements, 
SPS 2, 3 and 5 lead to larger values of $m_{N_3}$, the contrary
occurring for SPS 1b and 4. This can be seen in 
Fig.~\ref{fig:M3:tanb:BR:SPS4:t2:5d}, where
we present an analogous study to that of
Fig.~\ref{fig:M3:tanb:BR:SPS1a:t2:5d}, but focusing on SPS 2 and SPS 4,
and only considering $\theta_{13}= 5^\circ \pm 0.5$.

Concerning the comparison with current experimental bounds, one can
also draw some conclusions regarding the excluded regions of the 
$m_{N_3}$-$\tan \beta$ plane.
From both Figs.~\ref{fig:M3:tanb:BR:SPS1a:t2:5d} and 
~\ref{fig:M3:tanb:BR:SPS4:t2:5d}, for $\theta_{13}=5^\circ$, and 
for the chosen set of input parameters, we infer that in all cases the
upper-right regions of the $m_{N_3}$-$\tan \beta$ plane are 
clearly disfavoured.
For instance, for SPS~1a, $m_{N_3} \gtrsim 10^{14}$ GeV 
would be excluded for any value of $\tan \beta$. In the case of 
SPS 2, the exclusion region would be delimited by
$\tan \beta \gtrsim 10$ and $m_{N_3} \gtrsim 10^{14}$ GeV. The
most pronounced exclusion region is for SPS 4, and is given by
$\tan \beta \gtrsim 20$, $m_{N_3} \gtrsim 10^{13}$ GeV. With the
expected future sensitivity, these exclusion regions will be 
significantly enlarged.

A potential caveat to the previous discussion is the fact that, as seen in
Section~\ref{results}, there is a very important dependence of the BRs
on the $R$-matrix parameters $\theta_i$. Not only will this have implications
on how accurate the indirect estimates of $m_{N_3}$ are, but will also
affect any judgement regarding 
the experimental viability of a SUSY seesaw scenario.
We recall that, as shown in Section~\ref{results}, 
other choices of $\theta_2$ (and $\theta_1$) 
can lead to substantially smaller or larger BRs, 
therefore modifying the exclusion regions of
Figs.~\ref{fig:M3:tanb:BR:SPS1a:t2:5d} and ~\ref{fig:M3:tanb:BR:SPS4:t2:5d}.

\begin{figure}[t]
  \begin{center} \hspace*{-10mm}
	\psfig{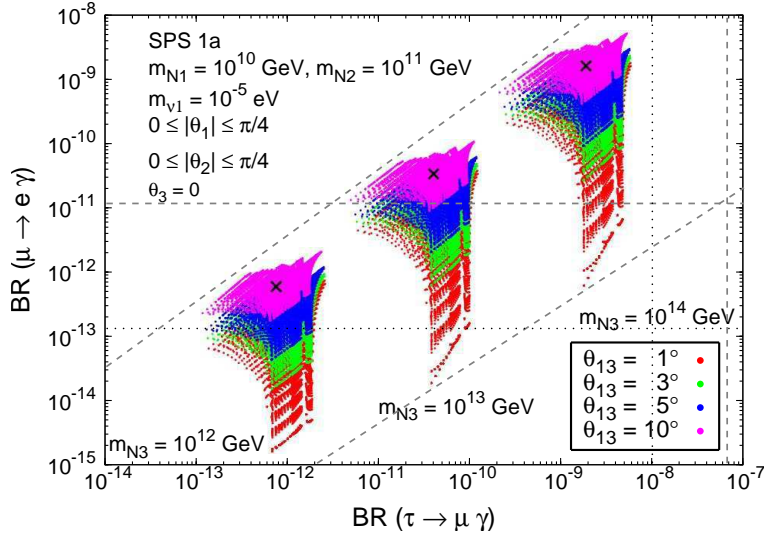} 
    \caption{Correlation between BR($\mu \to e\,\gamma$) and 
      BR($\tau \to \mu\,\gamma$) as a function of $m_{N_3}$, for SPS
      1a. The areas displayed represent the scan over $\theta_i$ 
      as given in Eq.~(\ref{doubleBR:input}). From bottom to top, 
      the coloured regions correspond to 
      $\theta_{13}=1^\circ$, $3^\circ$, $5^\circ$ and $10^\circ$ (red,
      green, blue and pink, respectively). Horizontal and vertical 
      dashed (dotted) lines denote the experimental bounds (future
      sensitivities). } 
    \label{fig:doubleBR}
  \end{center}
\end{figure}

To take into account the strong $R$-matrix dependence, 
let us conduct in what follows
a more comprehensive survey of the parameter space.
For SPS 1a, and for distinct choices of the heaviest neutrino mass, 
we scan over the BAU-enabling $R$-matrix angles (setting $\theta_3$ to
zero) as 
\begin{align}\label{doubleBR:input}
& 0\, \lesssim \,|\theta_1|\,\lesssim \, \pi/4 \,, \quad \quad
-\pi/4\, \lesssim \,\arg \theta_1\,\lesssim \, \pi/4 \,, \nonumber \\
& 0\, \lesssim \,|\theta_2|\,\lesssim \, \pi/4 \,, \quad \quad
\quad \,\,\,\,\,\,
0\, \lesssim \,\arg \theta_2\,\lesssim \, \pi/4 \,, \nonumber \\
& m_{N_3}\,=\,10^{12}\,,\,10^{13}\,,\,10^{14}\,\text{GeV}\,.
\end{align}
Given that, as previously emphasised, $\mu \to e\,\gamma$ is very
sensitive to $\theta_{13}$, whereas this is not the case for 
BR($\tau \to \mu\,\gamma$), 
and that both BRs display the same approximate behaviour with 
$m_{N_3}$ and $\tan \beta$, we now propose to study the correlation between
these two observables. This optimises the impact of a 
$\theta_{13}$ measurement, since it allows to minimise the uncertainty
introduced from not knowing $\tan \beta$ and $m_{N_3}$, and at the
same time offers a better illustration of the uncertainty associated
with the $R$-matrix angles.
In this case, the correlation of the BRs with respect to $m_{N_3}$
means that, for a fixed set of parameters, varying $m_{N_3}$ implies
that the predicted point 
(BR($\tau \to \mu\,\gamma$),~BR($\mu \to e \, \gamma$)) 
moves along a line with approximately constant slope in the 
BR($\tau \to \mu\,\gamma$)-BR($\mu \to e \, \gamma$) plane.
On the other hand, varying $\theta_{13}$ leads to a 
displacement of the point along the vertical axis.
In Fig.~\ref{fig:doubleBR}, we illustrate this correlation for SPS
1a, and for the previously selected $m_{N_3}$ and $\theta_{1,2}$
ranges (c.f. Eq.~(\ref{doubleBR:input})). We consider the following values,
$\theta_{13}=1^\circ$, $3^\circ$, $5^\circ$ and $10^\circ$, and only
include the BR predictions allowing for a favourable BAU. In addition, and as
done throughout our analysis,
we have verified that all the points in this figure lead to charged lepton
EDM predictions which are compatible with present experimental bounds.
More specifically, we have obtained
values for the EDMs lying in the following ranges (in units of e.cm):
\begin{equation}
10^{-39}\, \lesssim \, |d_e| \, \lesssim \,2\times10^{-35}\,,
\,
6\times10^{-37}\, \lesssim \, |d_\mu| \, \lesssim \,1.5
\times10^{-32}\,,
\,
10^{-34}\, \lesssim \, |d_\tau| \, \lesssim \,4\times\,10^{-31}\, .
\end{equation}

For a fixed value of $m_{N_3}$, and for a given value of $\theta_{13}$, the
dispersion arising from a $\theta_1$ and $\theta_2$ variation produces a small
area rather than a point in the 
BR($\tau \to\mu\,\gamma$)-BR($\mu \to e \, \gamma$) plane.
The dispersion along the BR($\tau \to \mu\,\gamma$) axis is of
approximately one order of magnitude for all $\theta_{13}$. 
In contrast, the dispersion along the BR($\mu \to e\,\gamma$) axis
increases with decreasing $\theta_{13}$ (in
agreement with the findings of Section~\ref{results}),
ranging from 
an order of magnitude for $\theta_{13}=10^\circ$,
to over three orders of magnitude for the case of small $\theta_{13}$
($1^\circ$). 
From Fig.~\ref{fig:doubleBR} 
we can also infer that other choices of $m_{N_3}$ (for $\theta_{13}
\in [1^\circ, 10^\circ]$) would lead to BR
predictions which would roughly lie within the diagonal lines depicted
in the plot. Comparing
these predictions for the shaded areas along the expected diagonal
``corridor'', with the allowed experimental region, allows to conclude
about the impact of a $\theta_{13}$ measurement on the allowed/excluded 
$m_{N_3}$ values.

The most important conclusion from Fig.~\ref{fig:doubleBR} is that for
SPS~1a, and for the parameter space defined in Eq.~(\ref{doubleBR:input}), 
an hypothetical $\theta_{13}$ measurement larger than $1^\circ$, together 
with the present experimental bound on the BR($\mu \to e\,\gamma$),
will have the impact of excluding values of $m_{N_3} \gtrsim 10^{14}$
GeV. This lends support to the hints already drawn from 
Fig.~\ref{fig:M3:tanb:BR:SPS1a:t2:5d}. Moreover, with the planned MEG
sensitivity, the same $\theta_{13}$ measurement can further constrain 
$m_{N_3} \lesssim 3\times 10^{12}$~GeV.
The impact of any other $\theta_{13}$ measurement can be analogously
extracted from Fig.~\ref{fig:doubleBR}.

\begin{figure}[t]
\begin{center} \hspace*{-10mm}
\begin{tabular}{cc}
	\psfig{file=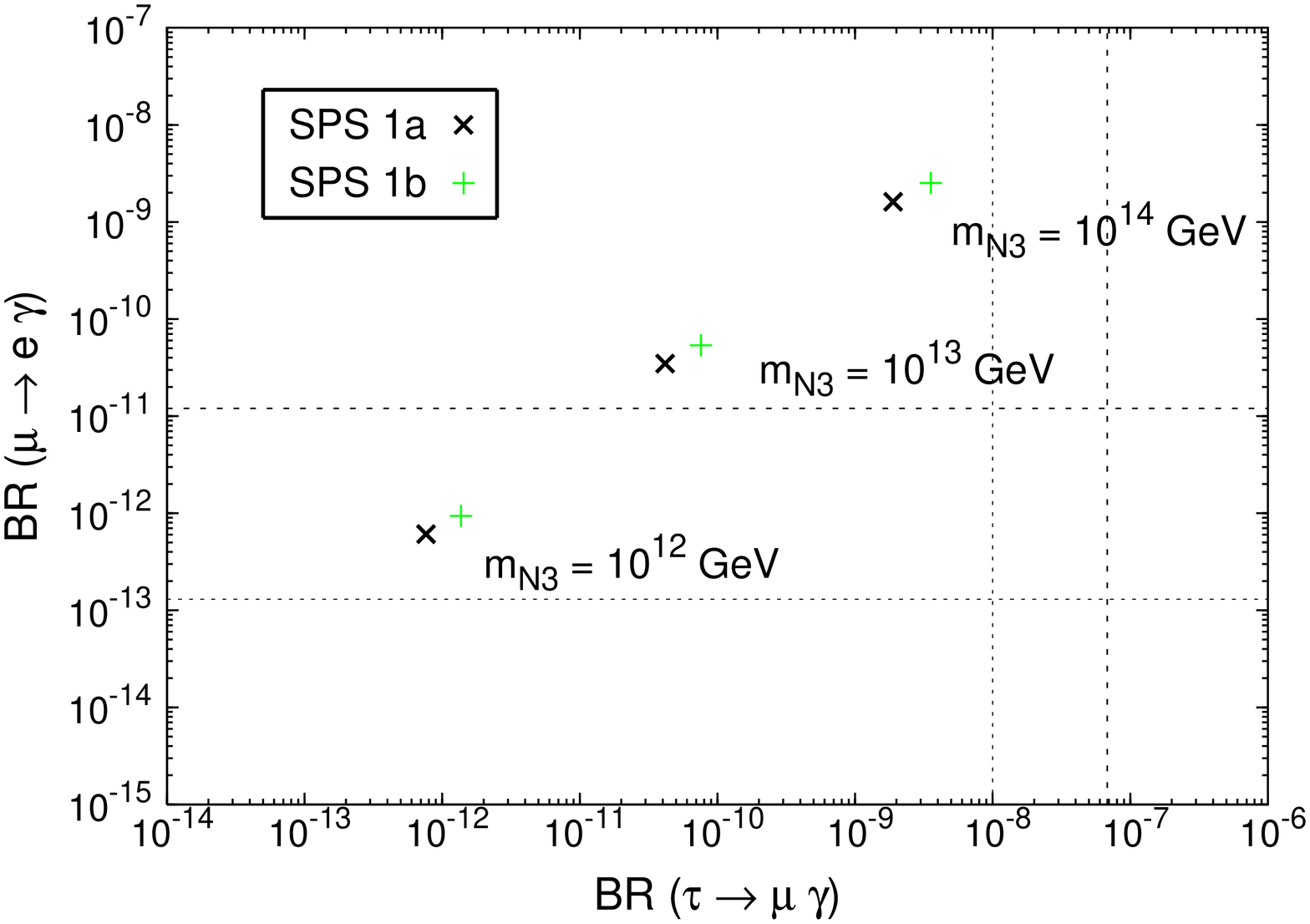,width=50mm,angle=270,clip=} &
	\psfig{file=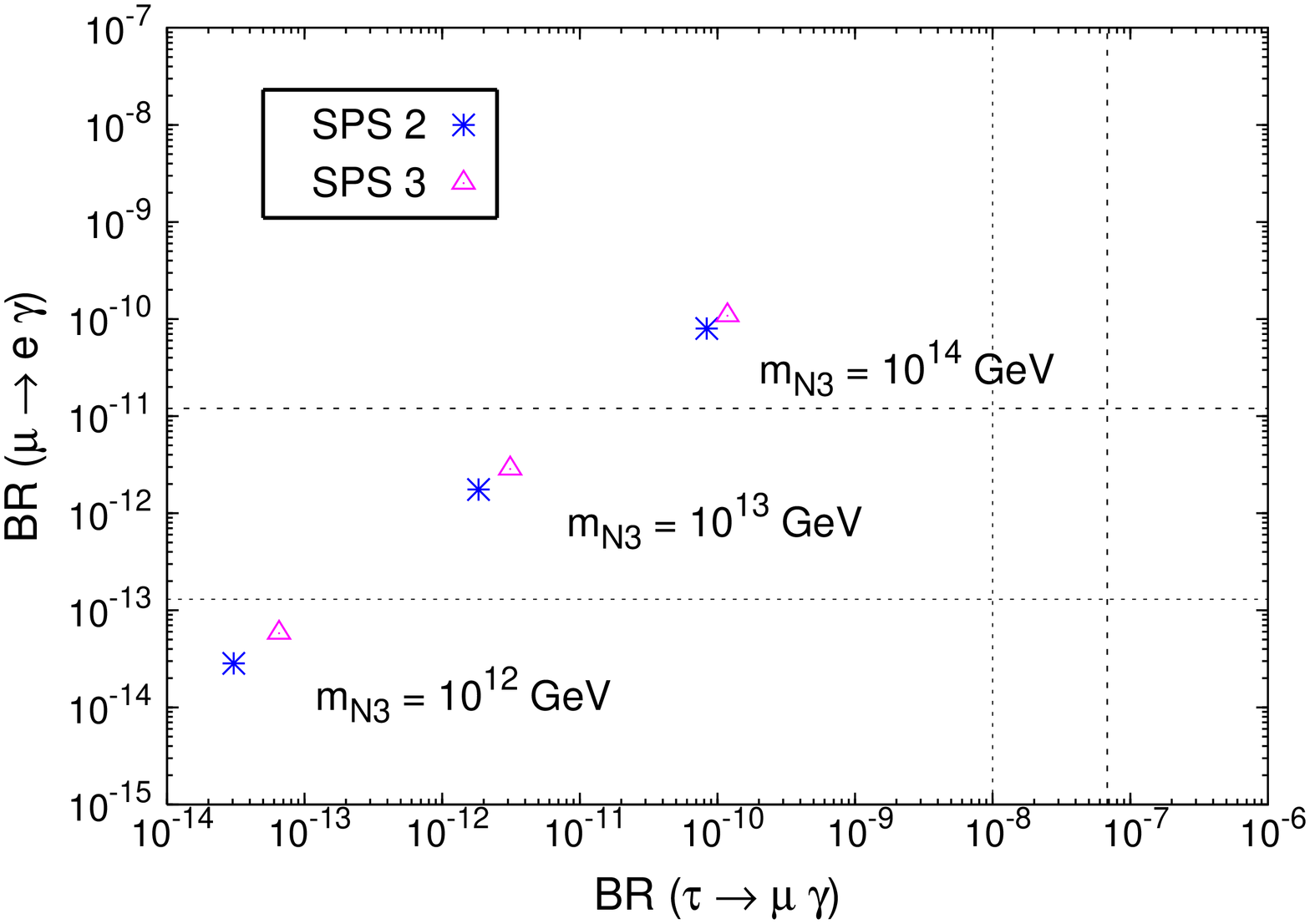,width=50mm,angle=270,clip=} \\
	\psfig{file=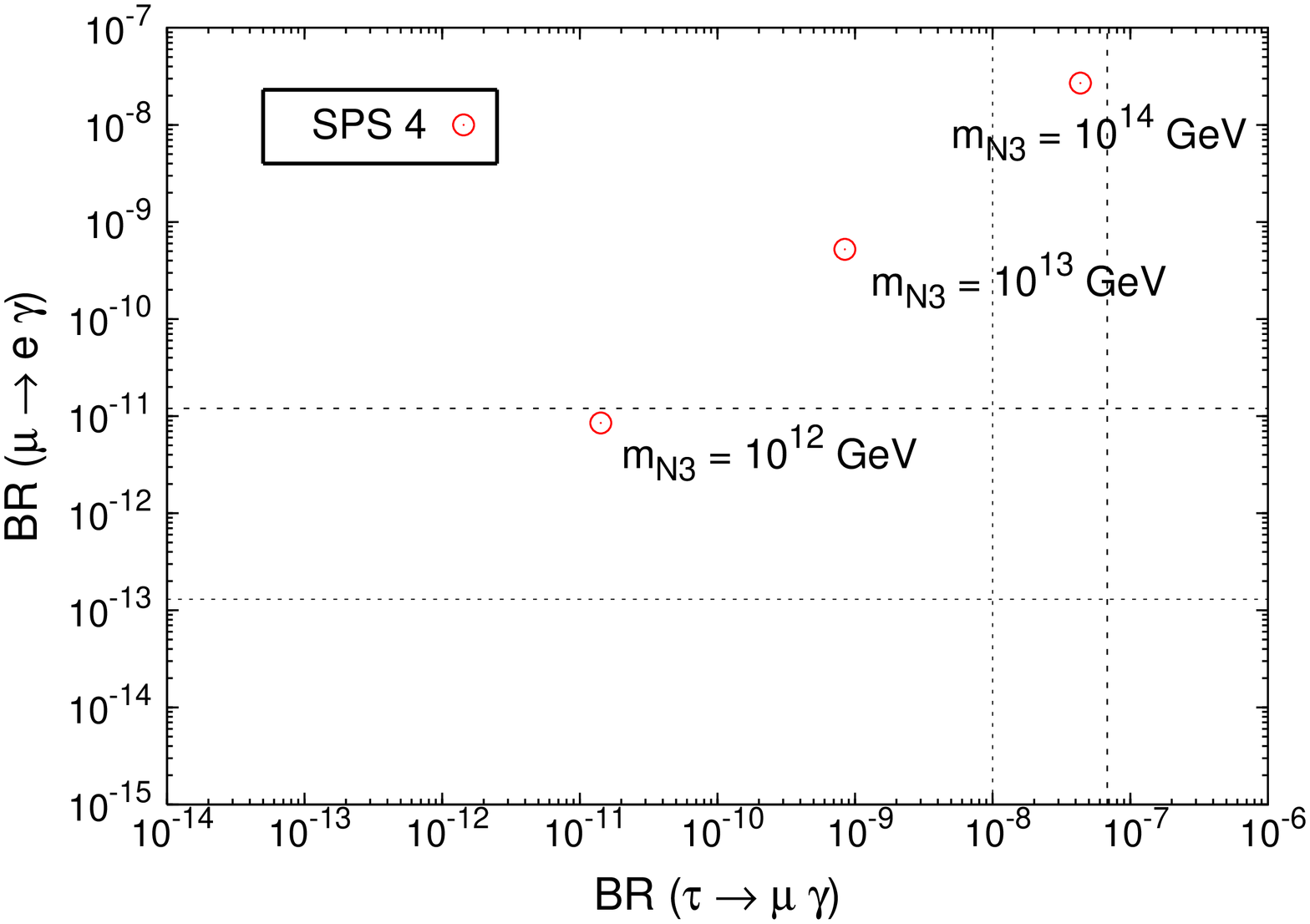,width=50mm,angle=270,clip=} &
	\psfig{file=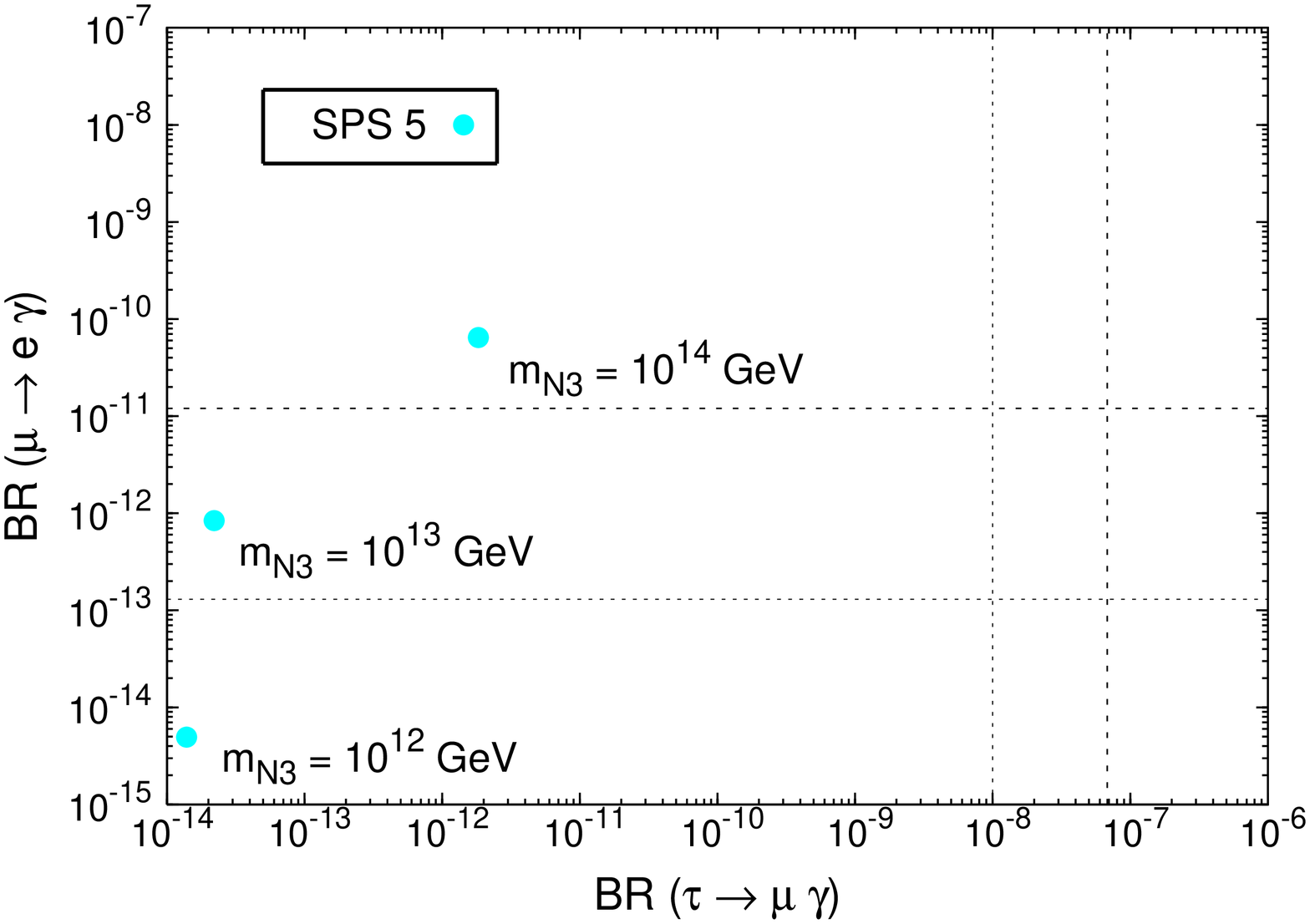,width=50mm,angle=270,clip=}
    \end{tabular}
    \caption{Correlation between BR($\mu \to e\,\gamma$) and 
      BR($\tau \to \mu\,\gamma$) as a function of $m_{N_3}$. 
      The points displayed mimic the behaviour of the central 
      $\theta_{13}=10^\circ$ marked point in Fig.~\ref{fig:doubleBR},
      for SPS points 1a, 1b, 2, 3, 4 and 5. Horizontal and vertical 
      dashed (dotted) lines denote the experimental bounds (future
      sensitivities).}
    \label{fig:doubleBR:1.5}
  \end{center}
\end{figure}

Similar conclusions can be reached for the other SPS points, as seen
in Fig.~\ref{fig:doubleBR:1.5}, 
where we only display the predictions corresponding to the point
marked with a cross in the centre of the $\theta_{13}=10^\circ$
shaded area of Fig.~\ref{fig:doubleBR} (taking into account
all $\theta_{13}$ values would lead to replications of the shaded
areas observed in Fig.~\ref{fig:doubleBR}).
Regarding SPS 1b, the discussion is very similar to that of SPS 1a,
and the inferred constrains on $m_{N_3}$ are almost identical. 
SPS 2 and SPS 3 offer very close predictions, and when compared to SPS
1a, for the same $\theta_{13}$ measurement,
allow to extract slightly weaker bounds on $m_{N_3}$. 
On the other hand, SPS 4 clearly provides the most
stringent scenario and a measurement of $\theta_{13}=10^\circ$ 
is only compatible with $m_{N_3} \lesssim 10^{12}$~GeV.
Notice also that this is the only case where the present experimental
bound from BR($\tau \to \mu\,\gamma$) plays a relevant role.
SPS 5 provides the weakest bounds on $m_{N_3}$ but nevertheless 
still allows to exclude $m_{N_3} \gtrsim 10^{14}$ GeV from a measurement
of $\theta_{13}=10^\circ$. Finally, it is interesting to notice that
the observed correlations for SPS 5 are manifestly different from the 
other cases, in agreement with the findings of Section~\ref{results}.
In this case, varying $m_{N_3}$ leads to predictions of the BR($\mu
\to e\,\gamma$) and BR($\tau \to \mu\,\gamma$) 
which are not linearly correlated, 
opposed to what would be expected from the LLog approximation.

\section{Conclusions}\label{conclusions}
In this work, we have investigated lepton flavour violating muon 
and tau decays in the CMSSM extended by three right-handed (s)neutrinos, 
and used a type-I seesaw mechanism to explain the smallness of the
neutrino masses.  
As typical examples of an mSUGRA-like scenario,  
several SPS SUSY benchmark points were considered.
We have parameterised the solutions to the seesaw
equation in terms of a complex orthogonal matrix $R$ and of
the right-handed neutrino masses, requiring compatibility with low-energy
data. We have considered scenarios of hierarchical light and heavy neutrinos.
In addition, we imposed consistency with present bounds on charged
lepton EDMs and baryogenesis via thermal leptogenesis taking into
account constraints on the reheating temperature from non-thermal LSP
production by gravitino decay. 
We have studied in great detail the sensitivity of the BRs to
$\theta_{13}$, giving special emphasis to the $\mu \to e\,\gamma$
decay channel.

In a first stage, we have considered the simple case $R=\mathbbm{1}$ where
there are no additional neutrino mixings other than those in the $U_\text{MNS}$. 
We have found a very pronounced sensitivity to $\theta_{13}$ 
in the decay channels  
$\mu \to e\,\gamma$, $\mu \to 3\,e$, $\tau \to e\,\gamma$ and 
$\tau \to 3\,e$. 
Varying $\theta_{13}$ from $0^\circ$ to $10^\circ$, the branching 
ratios for the above processes increase by several orders of magnitude. 
In view of the present experimental bounds and the expected future
sensitivity, $\mu \to e\,\gamma$ is by far the most promising channel to
study the sensitivity to $\theta_{13}$ in LFV processes. We would like to 
notice that $\mu \to 3\,e$ may also offer interesting additional 
information. 
We have presented the predictions for the branching 
ratios for various SPS SUSY benchmark points. We further  
emphasised the importance of a full numerical 
computation, which we have found to differ significantly from the LLog 
approximation in some cases.

We have then explored how the sensitivity of BR($\mu \to e\,\gamma$)
to $\theta_{13}$ is altered when we take into account the remaining
SUSY seesaw parameters. In this sense, we have found that the most 
relevant parameters are $\theta_1$, $\theta_2$, $m_{N_3}$ and $\tan \beta$  
and we have systematically studied their influence on the impact of 
$\theta_{13}$ on BR($\mu \to e\,\gamma$).
We have also noticed that the sensitivity to $\theta_{13}$ improves
for lower values of $m_{\nu_1}$ ($m_{\nu_1} \lesssim 10^{-3}$ eV).

Compared to the special case $R=\mathbbm{1}$, non-vanishing $\theta_i$ 
can have important consequences.  
In particular, the sensitivity to $\theta_{13}$ is considerably reduced 
for large values of $|\theta_1|$ and $|\theta_2|$.
However, thermal leptogenesis severely constrains $\theta_{2}$ and 
$\theta_{3}$ (but not generically $\theta_1$). 
In fact, the requirement of successful thermal leptogenesis with 
constraints on the reheating temperature from non-thermal LSP
production by gravitino decay, suggests a region 
$|\theta_{2,3}| \lesssim 1$.
In these ranges, we have studied the sensitivity of the BR to 
$\theta_{13}$.
Generically, the separation between the BR predictions for 
distinct $\theta_{13}$ is reduced when we move from  
$R=\mathbbm{1}$ to $R\neq \mathbbm{1}$, and one could be led to the
conclusion that the BR sensitivity to $\theta_{13}$ would be reduced.
However, we have also found cases of $R\neq \mathbbm{1}$  
where, although this separation is reduced, the BR predictions are
now larger (and can be above the experimental bounds)  
and different $\theta_{13}$ values can be distinguished 
even more efficiently than in the $R=\mathbbm{1}$ case.

Regarding the right-handed neutrino masses, the most relevant one for
the LFV BRs is clearly $m_{N_3}$ (with a marginal role being played by 
$m_{N_2}$). Even though $m_{N_1}$ does not directly affect the BRs,
it nevertheless plays a relevant role with respect to baryogenesis. 
This, together with the assumption of hierarchical right-handed
neutrinos, leads furthermore to an indirect lower bound for $m_{N_3}$. 
For a given choice of $\theta_{13}$,
the dependence on $m_{N_3}$ is so pronounced that for the
investigated range $[10^{11}\,\text{GeV}, 10^{15}\,\text{GeV}]$, 
the BRs change by over six orders of magnitude. 
Thus, and even though the sensitivity to $\theta_{13}$ is clearly 
manifest (for instance, more than two orders of
magnitude separation between the BR predictions of
$\theta_{13}=1^\circ$ and $5^\circ$, for a given value of $\theta_2$)
without additional knowledge of $m_{N_3}$ it will be very difficult to
disentangle the several $\theta_{13}$ cases. 

In a similar fashion, the sensitivity of the BRs to $\theta_{13}$
can be altered by the uncertainty introduced from the indetermination
of $\tan \beta$.
The study of the generalised SPS points shows that changing $\tan \beta$
from 5 to 50 translates in BR($\mu \to e\,\gamma$) 
predictions which differ by two orders of magnitude, so that unless
there is an experimental determination of $\tan \beta$, it will also 
be hard to distinguish the distinct $\theta_{13}$ predictions.
Moreover, we have emphasised that this strong dependence on $m_{N_3}$ 
and $\tan \beta$ can be constructively used as
a means of extracting information on these parameters from a 
potential joint measurement of $\theta_{13}$ and BR($\mu \to e\,\gamma$).

Remarkably, within a particular SUSY scenario and scanning 
over specific $\theta_1$ and $\theta_2$ BAU-enabling ranges for various 
values of $\theta_{13}$, the comparison of the
theoretical predictions for BR($\mu \to e\,\gamma$) and 
BR($\tau \to \mu\,\gamma$) with the present experimental bounds allows 
to set $\theta_{13}$-dependent upper bounds on $m_{N_3}$. 
Together with the indirect lower bound arising from leptogenesis 
considerations, this clearly provides interesting hints on the value of the 
seesaw parameter $m_{N_3}$. For instance, in the SUSY scenario SPS1a and for 
values of $\theta_{13}$ in the present experimental allowed range, the
present MEGA constraint on BR$(\mu \to e \gamma)$ already sets an upper bound
on $m_{N_3}$, $m_{N_3} \lesssim 10^{13}$ GeV for $\theta_{13} \gtrsim 10^{\circ}$ and $m_{N_3}
\lesssim 6 \times 10^{13}$ GeV for $\theta_{13} \gtrsim 3^{\circ}$, as inferred from
Fig.~\ref{fig:doubleBR}. These bounds are even more stringent for the case of
SPS4 (see Fig.~\ref{fig:doubleBR:1.5}) where the present contraint on
BR$(\mu \to e \gamma)$ sets an upper bound of $m_{N_3}
\lesssim 3 \times 10^{12}$ GeV for $\theta_{13} \gtrsim 10^{\circ}$ and $m_{N_3}
\lesssim 10^{13}$ GeV for $\theta_{13} \gtrsim 3^{\circ}$. With the planned
future sensitivities, these bounds would further improve by approximately one order of magnitude.

Ultimately, a joint measurement of the LFV branching ratios, 
$\theta_{13}$ and the sparticle spectrum would be a powerful tool for 
shedding some light on otherwise unreachable SUSY seesaw parameters.

\section*{Acknowledgements}
We are grateful to M.~Raidal for providing tables with the  
numerical results for thermal leptogenesis in the MSSM, including effects of
reheating. We are also indebted to A.~Donini and
E.~Fern\'andez-Mart\'{\i}nez for providing relevant information regarding the 
experimental sensitivity to $\theta_{13}$.
E.~Arganda and M.~J.~Herrero acknowledge A.~Rossi for useful comments
regarding the BR($l_j \to 3\,l_i$) computation.
A.~M.~Teixeira is grateful to F.~R.~Joaquim for his valuable suggestions.
The work of S.~Antusch was partially supported by the EU 6$^\text{th}$
Framework Program MRTN-CT-2004-503369 ``The Quest for Unification:
Theory Confronts Experiment''.  
E.~Arganda acknowledges the Spanish MEC for financial support under the grant
AP2003-3776. 
The work of A.~M.~Teixeira has been supported by ``Funda\c c\~ao para
a Ci\^encia e Tecnologia'', grant SFRH/BPD/11509/2002 and by HEPHACOS
`` Fenomenolog\'{\i}a de las Interacciones Fundamentales: Campos,
Cuerdas y Cosmolog\'{\i}a'' P-ESP-00346. 
This work was also supported by the Spanish MEC under project FPA2003-04597.

\end{document}